\documentclass{aa}  
\usepackage[varg]{txfonts}

\usepackage{amsmath}
\usepackage{amssymb}
\usepackage{graphicx}
\usepackage{color}
\usepackage{comment}
\usepackage[normalem]{ulem}
\usepackage{enumitem}
\usepackage{xspace}

\usepackage{xcolor}
\definecolor{xlinkcolor}{cmyk}{1,1,0,0}
\usepackage[breaklinks=true,
     bookmarks=false,         
     pdfnewwindow=true,      
     colorlinks=true,    
     linkcolor=xlinkcolor,     
     citecolor=xlinkcolor,     
     filecolor=xlinkcolor,  
     urlcolor=xlinkcolor,      
     final=true]{hyperref}

\usepackage{pifont}
\newcommand{\cmark}{\ding{51}}%
\newcommand{\xmark}{\ding{55}}%

\bibpunct{(}{)}{;}{a}{}{,} 

\begin{document}

   \title{Steady-state accretion in magnetized protoplanetary disks}

   \author{Timmy~N.~Delage\inst{1},
          Satoshi~Okuzumi\inst{2},
          Mario~Flock\inst{1,3},
          Paola~Pinilla\inst{1,4},
          \and Natalia~Dzyurkevich\inst{5}}

   \institute{Max-Planck-Institut f\"{u}r Astronomie, K\"{o}nigstuhl 17, 69117, Heidelberg, Germany, \email{delage@mpia.de}
    \and Department of Earth and Planetary Sciences, Tokyo Institute of Technology, Meguro, Tokyo 152-8551, Japan
    \and Jet Propulsion Laboratory, California Institute of Technology, Pasadena, CA 91109, USA
    \and Mullard Space Science Laboratory, University College London, Holmbury St Mary, Dorking, Surrey RH5 6NT, UK 
    \and Institut f\"{u}r Theoretische Astrophysik, University of Heidelberg, Albert-\"{U}berle Str. 2, Heidelberg, D- 69120, Germany}

   \date{Received date / 
   Accepted date}

\titlerunning{Steady-state accretion in magnetized protoplanetary disks}
\authorrunning{Delage~et~al.~}

 
  \abstract
   {The transition between magnetorotational (MRI)-active and magnetically-dead regions corresponds to a sharp change in the disk turbulence level where pressure maxima may form; hence potentially trapping dust particles, and explaining some of the observed disk sub-structures.}
   {We aim to bring the first building blocks toward a self-consistent approach to assess the dead zone outer edge as a viable location for dust trapping, under the framework of viscously-driven accretion.}
   {We present a 1+1D global magnetically-driven disk accretion model that captures the essence of the MRI-driven accretion, without resorting to 3D global non-ideal MHD simulations. The gas dynamics is assumed to be solely controlled by the MRI and hydrodynamic instabilities. For given stellar and disk parameters, the Shakura-Sunyaev viscosity parameter $\alpha$ is determined self-consistently under the adopted framework from detailed considerations of the MRI with non-ideal MHD effects (Ohmic resistivity and ambipolar diffusion), accounting for disk heating by stellar irradiation, non-thermal sources of ionization, and dust effects on the ionization chemistry. Additionally, the magnetic field strength is constrained and adopted to maximize the MRI activity.}
   {We demonstrate the use of our framework by investigating steady-state MRI-driven accretion in a fiducial protoplanetary disk model around a solar-type star. We find that the equilibrium solution displays no pressure maximum at the dead zone outer edge, except if a sufficient amount of dust particles have accumulated there before the disk reaches a steady-state accretion regime. Furthermore, the steady-state accretion solution describes a disk that displays a spatially extended long-lived inner disk gas reservoir (the dead zone) accreting a few $10^{-9}\, M_{\odot}.\rm{yr}^{-1}$. By conducting a detailed parameter study, we find that the extend to which the MRI can drive efficient accretion is primarily determined by the total disk gas mass, the representative grain size, the vertically-integrated dust-to-gas mass ratio, and the stellar X-ray luminosity.}
   {A self-consistent time-dependent coupling between gas, dust, stellar evolution models and our general framework on million-year timescales is required to fully understand the formation of dead zones and their potential to trap dust particles.}

    
   \maketitle
%

\section{Introduction} \label{sect:intro}

The advent of state of the art telescopes such as the Atacama Large Millimiter/sub-millimiter Array (ALMA) or Spectro-Polarimetric High-contrast Exoplanet Research at the Very Large Telescope (SPHERE/VLT) has revealed astonishing sub-structures in protoplanetary disks \citep[e.g.,][]{2015ApJ...808L...3A, Nomura_2016, P_rez_2016, Ginski_2016, de_Boer_2016, Andrews_2016, 2016arXiv160805123I, Kurtovic_2021}. One of the preferred explanations for those sub-structures are local pressure maxima where dust particles radially drift toward and are trapped. 

Such local pressure maxima can occur in various ways. Embedded massive planets can easily form them and explain the diverse multi-wavelength observations \citep[e.g,][]{Zhu_2012, Pinilla_2012, Ataiee_2013, de_Juan_Ovelar_2013, 2015ApJ...809L...5D, Pohl_2015, 2015A&A...573A...9P}. However, there has been detection and confirmation of such planets in only one protoplanetary disk so far \citep[PDS70,][]{Keppler_2018, Christiaens_2019}, although the kinematic signatures of potential planets in different systems are encouraging \citep[e.g. HD 163296,][]{2018ApJ...860L..12T,2019Natur.574..378T}. Furthermore, analysis of current observational capabilities suggest that several of the proposed planets that explain some of the sub-structures could have been already detected \citep[e.g.,][]{asensiotorres2021perturbers}. This could indicate that planets may not be the universal origin for disk sub-structures, especially in the case of younger systems. 

Protoplanetary disks can rapidly accrete their gas content to feed the growth of the central forming star \citep[see e.g.,][]{doi:10.1146/annurev-astro-081915-023347}. To sustain such an overall inflow of material, the angular momentum must also be transported \citep[see e.g.,][]{Armitage_2011}. Different sources of angular momentum transport have been suggested to play a role in the disk evolution, including hydrodynamic-driven mechanisms such as gravitational instability \citep[e.g.,][]{10.1093/mnras/225.3.607, Lodato_2004, Vorobyov_2009}, baroclinic instabilities \citep[e.g.,][]{Klahr_2003, Raettig_2013}, vertical shear instability \citep[VSI, e.g.,][]{10.1111/j.1365-8711.1998.01118.x, Nelson_2013, Lin_2015, Manger_2020, 2020ApJ...897..155F}; or magnetohydrodynamical (MHD)-driven mechanisms such as MHD winds \citep[e.g.,][]{10.1093/mnras/199.4.883, Suzuki_2009, Bai_2016a, Bai_2016b}, and the magnetorotational instability \citep[MRI, e.g.,][]{1991ApJ...376..214B, 1998RvMP...70....1B, 1995ApJ...440..742H}. Another way of forming pressure maxima can thus be by the unsteady overall gas inflow, with material piling-up where the inflow becomes slower than average \citep[see e.g.,][for formation of rings and gaps by variable magnetized disk winds]{2017MNRAS.468.3850S, 2018MNRAS.477.1239S, 2019MNRAS.484..107S}. 

The MRI-driven accretion is of particular interest because it can be substantially quenched or suppressed at some locations in the disk by non-ideal MHD effects such as the Ohmic resistivity, the Hall effect and ambipolar diffusion; hence causing an unsteady overall gas inflow. These non-ideal MHD effects arise from the weak level of ionization in the disk \citep[e.g.,][]{1996ApJ...457..355G, 2000ApJ...530..464F, 2002ApJ...570..314S, 2002ApJ...577..534S, 2003ApJ...585..908F, 2005ApJ...628L.155I,2006A&A...445..223I, 2007ApJ...659..729T, 2010ApJ...708..188T, 2008ApJ...679L.131T, 2011ApJ...735....8P,2011ApJ...739...50B}, or poor coupling between the field and the gas bulk motion at low densities and high magnetic field strengths \citep[e.g.,][]{2011ApJ...736..144B}. As a result, magnetically-dead zones arise. In general, how much and where the MRI is suppressed in protoplanetary disks is a complex problem that significantly depends on the dust properties and its abundance, the magnetic field strength and its configuration \citep[e.g.,][]{2015MNRAS.454.1117S}, the complex gas and dust chemistry \citep[e.g.,][]{2011ApJ...735....8P}, and the various ionization sources \citep[e.g.,][]{2013ApJ...772....5C}. These are crucial ingredients setting the non-ideal MHD strength terms, hence the MRI activity. Overall, the rate of gas flow decreases in magnetically-dead zones \citep[e.g.,][]{2010A&A...515A..70D}. Therefore, the gas can accumulate in the transition from a MRI-active to non-active region, hence potentially forming a local pressure maximum.

There have been several works studying pressure traps created by a sharp change in the disk turbulence level (disk viscosity), particularly at the inner and outer edge of the dead zone \citep[e.g.,][]{2006A&A...446L..13V, 2007ApJ...664L..55K, 2008A&A...487L...1B, 2009ApJ...690..407K, 2010ApJ...721.1585K, 2013A&A...556A..37D, 2016A&A...596A..81P, 2018ApJ...861..144M, 2021MNRAS.504..280J}. In the case of the outer edge of the dead zone, some of those studies show that the pressure bump is capable of trapping dust particles, and stops the rapid inward migration of the larger pebbles \citep[e.g.,][]{1977MNRAS.180...57W}. All of these used the Shakura-Sunyaev $\alpha$-prescription \citep{1973A&A....24..337S}, wherein the quantity $\alpha$ encodes the disk turbulence level. Crucially, however, they have all adopted ad-hoc prescriptions of $\alpha$ without accounting for the detailed physics of the MRI. Conversely, several works have investigated the detailed behavior of MRI-active and non-active regions by performing 3D local shearing box or global simulations \citep[e.g.,][]{2010A&A...515A..70D, 2010ApJ...708..188T, 2011ApJ...736..144B, 2011ApJ...739...50B, 2011ApJ...742...65O, 2011ApJ...735..122F, 2015A&A...574A..68F}. However, these studies do not implement gas and dust evolution (growth processes included) on million-year timescales, since it is computationally unfeasible due to the very different timescales at play. For example, turbulent eddies grow and decay on a timescale of one orbital period \citep[e.g,][]{2006A&A...452..751F}, while dust particles grow to macroscopic sizes and settle to the mid-plane on a timescale of a few orbital periods depending on the grain size \citep[e.g.,][]{1981Icar...45..517N, 2005A&A...434..971D, 2008A&A...487L...1B}. 

Assessing the dead zone mechanism as a possible explanation for the observed disk sub-structures requires a self-consistent coupling between non-ideal MHD calculations and gas/dust evolution models on million-year timescales, as we will show. To date, such a coupling still remains to be done. One way to achieve it could be by building a "trade-off" model where a 1D viscous disk model and non-ideal MHD calculations are combined: the $\alpha$-parameter is determined by the MRI-driven turbulence, accounting for non-ideal MHD effects and detailed modeling of the gas ionization level in the protoplanetary disk. Doing so, this model would capture the essence of the MRI and be computationally cheap enough to make the implementation into 1D gas/dust evolution models feasible on million-year timescales.

In this paper, we aim to present such a magnetically-driven disk accretion model, where the local mass and angular momentum transport are assumed to be solely controlled by the MRI and hydrodynamic instabilities. We include the key physical processes relevant for the outer regions ($r \gtrsim 1\,$au, where $r$ is the distance from the central star) of class II protoplanetary disks: (1) disk heating by stellar irradiation; (2) dust settling; (3) non-thermal ionization from stellar X-rays, galactic cosmic rays and the decay of short/long-lived radionuclides; (4) ionization chemistry based on a semi-analytical chemical model that includes the effect of dust. In order to know where the MRI can operate in the disk under the framework of viscously-driven accretion, the general methodology is to carefully model the gas ionization degree, compute the magnetic diffusivites of the non-ideal MHD effects as well as their corresponding Elsasser numbers, and apply a set of conditions for active MRI derived from 3D numerical simulations. Some previous studies attempted to make such "trade-off" models \citep[e.g.,][]{2008ApJ...689..532T,2012ApJ...753L...8O,2013ApJ...771...44O,2013ApJ...765..114D}. However, they are either a more parametrized version than ours (e.g., their $\alpha$ value in the MRI-active layer is set arbitrarily), or they do not include all the necessary physics to appreciate the MRI-driven turbulence in protoplanetary disks (e.g., ambipolar diffusion is omitted). To focus on the roles of the MRI and hydrodynamic instabilities, we ignore accretion driven by magnetic winds. The role of wind-driven accretion will be investigated in a future study.

We demonstrate the use of our framework by investigating the specific case of steady-state MRI-driven accretion of a fiducial disk around a solar-type star, under the assumption that the MRI activity is maximally efficient permitted by the non-ideal MHD effects considered. Particularly, we are interested in describing its overall structure (gas surface density, ionization level, viscosity, magnetic field strength required for the MRI activity to be maximally efficient, accretion rate, and the dead zone outer edge location), and finding what model parameters are crucial to MRI-driven accretion in protoplanetary disks. In this context, our framework solves for the gas surface density through an iterative process required to satisfy the steady-state accretion assumption. \footnote{We emphasize that our framework can also be used with e.g. the self-similar solution \citep[][]{1974MNRAS.168..603L} as for the gas surface density. In this case though, the derived accretion rate would be radially variable; implying that some regions of the disk can accrete more than others. This scenario would thus not describe a steady-state accretion state for the disk.}

The layout of the paper is as follows. In Sects.~\ref{sect:model} and \ref{sect:MRI}, we explain the key steps to determine self-consistently the Shakura-Sunyaev viscosity parameter $\alpha$ from detailed considerations of the MRI with non-ideal MHD effects, under the framework of viscously-driven accretion. In Sect.~\ref{sect:methodology}, we present the methodology employed to study steady-state accretion in our framework. In Sect.~\ref{sect:results-fiducial model}, we present the results for the fiducial protoplanetary disk model around a solar-type star. In Sect.~\ref{sect:results-parameter study}, we perform an exhaustive parameter study to determine the key parameters shaping the previous equilibrium solution. In Sect.~\ref{sect:discussion}, we discuss the implications of our results as well as the main limitations. Finally, Sect.~\ref{sect:summary and conclusions} summarizes our conclusions.

\section{Disk Model} \label{sect:model}

By assuming the protoplanetary disk to be geometrically thin ($H_{\rm{gas}} \ll r$), the vertical and radial dimensions can be decoupled into a 1+1D $(r,z)$ framework, where each radial grid-point contains an independent vertical grid. Furthermore, we assume the disk to be axisymmetric and symmetric about the mid-plane, implying that computing the domain $z \geqslant 0$ is enough to obtain the full solution. In our disk model, the radial grid is computed from $r_{\rm{min}}$ to $r_{\rm{max}}$, with $N_{r}$ cells logarithmically spaced. For every radial grid-point $r \in [r_{\rm{min}},r_{\rm{max}}]$, the corresponding vertical grid is computed from the disk mid-plane ($z = 0$) to $z_{\rm{max}}$, with $N_{z}$ cells linearly spaced. In all our simulations, we choose the following general setup: $r_{\rm{min}} = 1 \,$au, $r_{\rm{max}} = 200 \,$au, $N_{r} = 256$ cells, $z_{\rm{max}} = 5 \, H_{\rm{gas}}(r)$ for every radial grid-points $r \in [r_{\rm{min}},r_{\rm{max}}]$, and $N_{z} = 512$ cells. Here $H_{\rm{gas}}$ corresponds to the disk gas scale height defined as
\begin{equation}
    H_{\rm{gas}} \equiv \frac{c_s}{\Omega_K},
    \label{eq:gas scale height}
\end{equation}

\noindent where the isothermal sound speed $c_s$ and the Keplerian angular velocity $\Omega_K$ are given by $c_s = \sqrt{\frac{k_B T}{\mu m_{\rm H}}}$ and $\Omega_K = \sqrt{\frac{G M_\star}{r^3}}$ ; with $k_B$ the Boltzmann constant, $T$ the gas temperature, $\mu = 2.34$ the mean molecular weight (assuming solar abundances), $m_{\rm H}$ the atomic mass of hydrogen, $G$ the gravitational constant, and $M_{\star}$ the stellar mass.

Table~\ref{tab:Table1} summarizes all the fiducial parameters considered in our disk model.

\subsection{The Shakura-Sunyaev prescription}

In the context of the Shakura-Sunyaev $\alpha$-disk model, the local mass and radial angular momentum transport are determined by the local turbulent parameter $\alpha$ (also called viscosity parameter). In our disk model, we assume the transport to be solely controlled by both the MRI and hydrodynamic instabilities, meaning that the origin of $\alpha$ is the accretion driven by the MRI and hydrodynamic instabilities. The local turbulent parameter $\alpha$ is defined by the relation \citep{1973A&A....24..337S}
\begin{equation}
    \nu \equiv \alpha \frac{c_s^{2}}{\Omega_K},
\end{equation}

\noindent where $\nu$ is the kinematic viscosity arising from turbulence within the disk, provided that the turbulence can be described as a local process.

It is important to note that assuming the MRI to be the dominant magnetically-controlled mechanism for turbulence means that we need to investigate the MRI activity in the $r-z$ plane, which requires the solution for vertical stratification since $\alpha$ changes with disk height. Consequently, the key for a successful 1D description of vertically-layered accretion within the Shakura-Sunyaev viscous $\alpha$-disk model is to consider the effective turbulent parameter $\bar{\alpha}$, defined as the pressure-weighted vertical average of the local turbulent parameter $\alpha$
\begin{equation}
    \bar{\alpha}(r) \equiv \frac{\int_{-\infty}^{+\infty} \alpha(r,z) \: P_{\rm{gas}}(r,z) \: dz}{\int_{-\infty}^{+\infty} P_{\rm{gas}}(r,z) \: dz},
    \label{eq:alpha}
\end{equation}

\noindent where $r$ is the distance from the star, $z$ is the height from the mid-plane, and  $P_{\rm{gas}}$ is the gas pressure. 

In order to obtain an appropriate $\bar{\alpha}$ from the detailed physics of the MRI, we need to specify several ingredients. First, we need to set the gas and dust properties (Sect.~\ref{sect:disk properties}). Second, we need to implement the ionization sources in order to compute the disk ionization level (Sect.~\ref{sect:ionization sources}). Third, we need ionization chemistry to obtain the number densities of all charged particles initiated by the ionization process (Sect.~\ref{sect:chemistry}). Finally, we need to compute the magnetic diffusivities in order to derive where the MRI can operate as well as the turbulence strength generated (Sect.~\ref{sect:MRI}).
\begin{table}
    \begin{center}
    \resizebox{\columnwidth}{!}{
    \begin{tabular}{ |p{5.1cm}||p{2.3cm}|p{1.8cm}|}
    \hline
    \hline
    \small{\textbf{FIDUCIAL PARAMETERS}} & \small{\textbf{Symbol [Units]}} & \small{\textbf{Value}}\\
    \hline
    Inner radial boundary & $r_{\rm{min}} \,$[au] & $1$ \\
    \hline
    Outer radial boundary & $r_{\rm{max}} \,$[au] & $200$ \\
    \hline
    Radial grid resolution & $N_{r}$ & $256$ \\
    \hline
    Disk vertical extent & $z_{\rm{max}} \,[H(r)]$ & $5$ \\
    \hline
    Vertical grid resolution & $N_{z}$ & $512$ \\
    \hline
    Stellar mass & $M_\star \,[M_{\odot}]$ & $1$ \\
    \hline
    Bolometric luminosity & $L_\star \,[L_{\odot}]$ & $2$ \\
    \hline
    Total disk gas mass & $M_{\rm{disk}} \,[M_\star]$ & $0.05$ \\
    \hline
    Mean molecular weight & $\mu$ & $2.34$ \\
    \hline
    Mean molecular mass & $m_{\rm{neutral}} \,$[g] & $\mu \times m_{H}$ \\
    \hline
    Grain size & $a_{\rm{dust}} \,[\mu$m] & $1$ \\
    \hline
    Grain bulk density & $\rho_{\rm{bulk}} \,$[g.cm$^{-3}$] & $1.4$ \\
    \hline
    Dust-to-gas mass ratio & $f_{\rm{dg}}$ & $10^{-2}$ \\
    \hline
    Stellar X-ray luminosity & $L_{\rm{XR}} \,$[erg.s$^{-1}$] & $10^{-3.5} \times L_\star$ \\
    \hline
    Direct X-rays amplitude & $\zeta_{\rm{1,XR}} \,$[s$^{-1}$] & $6 \times 10^{-12}$\\
    \hline
    Direct X-rays penetration depth & $\Sigma_{\rm{1,XR}} \,$[g.cm$^{-2}$] & $0.0035$ \\
    \hline
    Scattered X-rays amplitude & $\zeta_{\rm{2,XR}} \,$[s$^{-1}$] & $10^{-15}$ \\
    \hline
    Scattered X-rays penetration depth & $\Sigma_{\rm{2,XR}} \,$[g.cm$^{-2}$] & $1.64$ \\
    \hline
    Cosmic rays amplitude & $\zeta_{\rm{CR,ISM}} \,$[s$^{-1}$] & $10^{-17}$ \\ 
    \hline
    Cosmic rays penetration depth & $\Sigma_{\rm{CR}} \,$[g.cm$^{-2}$] & $96$ \\
    \hline 
    Radionuclides amplitude & $\zeta_{\rm{RA}, \, 0} \,$[s$^{-1}$]  & $7.6 \times 10^{-19}$\\
    \hline
    Ion mass ($\rm{\rm{HCO}}^{+}$) & $m_{i} \,$[g] & $29 \times m_{H}$ \\
    \hline
    Ion sticking coefficient & $s_{i}$ & $1$ \\
    \hline
    Electron sticking coefficient & $s_{e}$ & $0.3$ \\
    \hline
    Hydrodynamic turbulent parameter & $\alpha_{\rm{hydro}}$ & $10^{-4}$ \\
    \hline
    \hline
\end{tabular}}
\caption{Summary of the parameters for our fiducial disk model. $m_{\rm H}$ corresponds to the atomic mass of Hydrogen.}
\label{tab:Table1}
\end{center}
\end{table}

\subsection{Gas and Dust Properties} \label{sect:disk properties}

\paragraph{Gas.} We consider a central star of mass $M_\star$ and bolometric luminosity $L_\star$. Additionally, we assume that the envelop has dispersed to reveal a gravitationally stable and viscously accreting disk with gas surface density $\Sigma_{\rm{gas}}$. In general, $\Sigma_{\rm{gas}}$ would be an input profile of the model such as e.g., a power-law or the self-similar solution. In this present study, we are specifically interested in the steady-state accretion solution. Hence, $\Sigma_{\rm{gas}}$ is given by the equilibrium solution described in Sect.~\ref{sect:methodology} (Eq.~\eqref{eq:gas surface density}). We further assume that the total disk gas mass varies as $M_{\rm{disk}} \propto M_\star$ \citep{2006ApJ...645.1498S, 2013ApJ...773..168M}. If not stated otherwise, we fiducially adopt $M_\star = 1 \, M_{\odot}$, $L_\star = 2 \, L_{\odot}$ and $M_{\rm{disk}} = 0.05 \, M_\star$. 

The disk is assumed vertically isothermal with a radial temperature profile set by absorbing stellar irradiation:
\begin{equation}
T(r) = \left[T^{4}_{\rm{1 \,au}} \left(\frac{r}{1 \, \rm{au}}\right)^{-2} \left(\frac{L_\star}{L_{\odot}}\right) + T^{4}_{\rm{bkg}}\right]^{\frac{1}{4}},
\label{eq:Temperature}
\end{equation}

\noindent where $T_{\rm{1 \,au}} = 280 \,$K is the gas temperature at $1 \,$au for $L_\star = L_{\odot}$, and $T_{\rm{bkg}} = 10 \,$K is the background gas temperature corresponding to the primordial temperature of the cloud prior to the collapse. We note that any extra projection factors arising from the relative inclination between the disk surface and incident stellar irradiation are not included. Adopting this gas temperature radial profile  means that we assume the disk to be optically thin to stellar irradiation. We explore how the equilibrium solution may depend on this choice by discussing the dependence on the gas temperature model in Appendix~\ref{sect:effect temperature model}.

Assuming hydrostatic equilibrium in the vertical direction gives the gas volume density distribution
\begin{equation}
    \rho_{\rm{gas}}(r,z) = \frac{\Sigma_{\rm{gas}}(r)}{\sqrt{2 \pi} H_{\rm{gas}}(r)} \exp{\left(-\frac{z^2}{2 H^{2}_{\rm{gas}}(r)}\right)}.
    \label{eq:volume gas density}
\end{equation}

\noindent The total number density of gas particles $n_{\rm{gas}}$ is then given by $n_{\rm{gas}} = \frac{\rho_{\rm{gas}}}{m_{\rm{neutral}}}$, where $m_{\rm{neutral}} = \mu m_{\rm H}$ is the mean molecular mass.

We can reformulate Eq.~\eqref{eq:alpha} for the effective turbulent parameter $\bar{\alpha}$ as:
\begin{equation}
\bar{\alpha}(r) = \frac{\int_{-\infty}^{+\infty} \alpha(r,z) \: \rho_{\rm{gas}}(r,z) \: dz}{\int_{-\infty}^{+\infty} \rho_{\rm{gas}}(r,z) \: dz} = 2 \: \frac{\int_{0}^{+\infty} \alpha(r,z) \: \rho_{\rm{gas}}(r,z) \: dz}{\Sigma_{\rm{gas}}(r)},
\label{eq:alpha bar}
\end{equation}

\noindent where the first equality (derived using $P_{\rm{gas}} = \rho_{\rm{gas}} c_s^{2}$) holds only if the disk is vertically isothermal, and the second equality uses the assumption that the disk is axisymmetric and symmetric about the mid-plane. If not stated otherwise, $\bar{\alpha}$ is always computed using the second equality of Eq.~\eqref{eq:alpha bar}.

\paragraph{Dust.} We consider dust particles as a mono-disperse population of perfect compact spheres of radius $a_{\rm{dust}}$, intrinsic volume density $\rho_{\rm{bulk}} = 1.4 \,$g.cm$^{-3}$ \citep[consistent with the solar abundance when H$_2$O ice is included in grains, see][]{1994ApJ...421..615P} and mass $m_{\rm{dust}} = \frac{4}{3} \pi \rho_{\rm{bulk}} a_{\rm{dust}}^{3}$. If not stated otherwise, we choose the fiducial value $a_{\rm{dust}} = 1 \, \mu$m.

Most of the transport and dynamics of dust particles in protoplanetary disks are regulated by the interactions between themselves and the gas. A way to quantify the importance of the drag forces on the dynamics of a dust particle is by its Stokes number, defined as the dimensionless version of the stopping time of that particle. In general, dust particles with a radius smaller than roughly $1-10 \,$mm can reside in the so-called Epstein drag regime where $a_{\rm{dust}}$ is smaller than the mean-free-path of gas particles. In this case, the Stokes number near the mid-plane is given by \citep[e.g.,][]{2008A&A...487L...1B, 2010A&A...513A..79B}
\begin{equation}
    \rm {St} = \frac{\pi \rho_{\rm{bulk}} a_{\rm{dust}}}{2 \Sigma_{\rm{gas}}}.
    \label{eq:stockes number}
\end{equation}

In the vertical direction of the disk structure, dust models predict that grains tend to settle toward the mid-plane as they grow in size \citep{1995Icar..114..237D}, and so even for micron-sized dust particles \citep{2016ApJ...822..111K}. Additionally, dust settling has been confirmed to be at play by ALMA observations \cite[see e.g.,][]{2019A&A...624A...7V, 2020A&A...642A.164V}. As a result, we can expect the number of dust particles to drop significantly above a dust scale height $H_{\rm{dust}}$ that can be much smaller than the gas scale height $H_{\rm{gas}}$. Assuming dust stirring to be induced by the MRI-driven turbulence, we can relate $H_{\rm{dust}}$, $H_{\rm{gas}}$, St and $\bar{\alpha}$ by \citep[e.g.,][]{1995Icar..114..237D, 2016SSRv..205...41B}: 
\begin{equation}
    H_{\rm{dust}} \equiv H_{\rm{gas}} \sqrt{\frac{\bar{\alpha}}{\bar{\alpha} + \rm{St}}}.
    \label{eq:dust scale height}
\end{equation}
\noindent This expression assumes that ${\rm St}/D_{\rm{gas}}$ is independent of $z$, where $D_{\rm{gas}}$ is the gas diffusion coefficient. We also implicitly approximate $D_{\rm{gas}}$ by the height independent quantity $\bar{\nu} = \bar{\alpha} c_s H_{\rm gas}$ (i.e., we use the common approximation that the gas diffusivity equals the gas kinematic viscosity).

Finally, we assume that the volume dust density follows a Gaussian distribution in the vertical direction 
\begin{equation}
    \rho_{\rm{dust}}(r,z) = \frac{\Sigma_{\rm{dust}}(r)}{\sqrt{2 \pi} H_{\rm{dust}}(r)} \exp{\left(-\frac{z^2}{2 H^{2}_{\rm{dust}}(r)}\right)},
    \label{eq:volume dust density}
\end{equation}

\noindent where $H_{\rm{dust}}$ is given by Eq.~\eqref{eq:dust scale height} and $\Sigma_{\rm{dust}}$ is the dust surface density given by $\Sigma_{\rm{dust}} = f_{\rm{dg}} \Sigma_{\rm{gas}}$, with $f_{\rm{dg}}$ the vertically-integrated dust-to-gas mass ratio. If not stated otherwise, we adopt the standard ISM value $f_{\rm{dg}} = 10^{-2}$. The total number density of all dust particles $n_{\rm{dust}}$ is then given by $n_{\rm{dust}} = \frac{\rho_{\rm{dust}}}{m_{\rm{dust}}}$.

\subsection{Ionization Sources} \label{sect:ionization sources}

In this study, the non-thermal ionization sources include stellar X-rays, galactic cosmic rays, and the decay of short/long-lived radionuclides. Charged particles are created primarily by ionization of molecular hydrogen (H$_2$) and helium (He). The ionization rate for He is related to the one for H$_2$ by $\zeta^{(He)} = 0.84 \times \zeta^{(H_2)}$ \citep[see][]{1990MNRAS.243..103U, 2009ApJ...690...69U}. It is thus enough to know $\zeta^{(H_2)}$ to compute the total ionization rate. The total ionization rate $\zeta$ is given by $\zeta = \zeta^{(H_2)} x_{H_2} + \zeta^{(He)} x_{He}$, where $x_{H_2} = \frac{n_{H_2}}{n_{\rm{gas}}}$ and $x_{He} = \frac{n_{He}}{n_{\rm{gas}}}$ are the factional abundances of H$_2$ and He, respectively. We calculate $x_{H_2}$ and $x_{He}$ from the solar system abundance by \citet{1989GeCoA..53..197A}: $x_{H_2} = \frac{n_{H_2}}{n_{\rm{gas}}} = \frac{n_{H_2}}{n_{H}} \times \frac{n_{H}}{n_{\rm{gas}}}$, with $\frac{n_{H_2}}{n_{H}} = 0.5$ (all H nuclei are in the form of H$_2$), and $x_{He} = \frac{n_{He}}{n_{\rm{gas}}} = \frac{n_{He}}{n_{H}} \times \frac{n_{H}}{n_{\rm{gas}}}$, with $\frac{n_{He}}{n_{H}} = 0.0975$. Here, $n_{\rm H}$ is the number density of hydrogen nucleus which can be estimated as $n_{\rm H} = \rho_{\rm gas}/(1.4m_{\rm H})$ for the solar abundance \citep[e.g.,][]{1999ApJ...518..848I}. In total, we have
\begin{equation}
    \zeta (r,z) = 0.97 \times \zeta^{(H_2)}(r,z).
    \label{eq:total ionization rate}
\end{equation}
Besides, we follow standard prescriptions where the ionization rates are given as a function of vertical gas column densities: $\Sigma_{\rm{gas}}^{+}(r,z) = \int_{z}^{+\infty} \rho_{\rm{gas}}(r,z') dz' $ and $\Sigma_{\rm{gas}}^{-}(r,z) = \Sigma_{\rm{gas}}(r) - \Sigma_{\rm{gas}}^{+}(r,z)$; where $\Sigma_{\rm{gas}}^{+}(r,z)$ and $\Sigma_{\rm{gas}}^{-}(r,z)$ represent the vertical gas column density measured from above and below a height of interest $z$, respectively, at a given radius $r$.

\paragraph{X-rays ionization rate $\zeta^{(H_2)}_{\rm{XR}}$ for H$_2$.} We adopt the fitting formula of \citet{2009ApJ...701..737B}, based on the Monte Carlo simulations from \citet{1999ApJ...518..848I}. We estimate the total stellar X-ray luminosity $L_{\rm{XR}}$ from the empirical median relationship $L_{\rm{XR}} \approx 10^{-3.5} \times L_\star$ \citep{2007A&A...468..353G}, and use the fitting coefficients at X-ray temperature $T_{\rm{XR}} = 3 \,$keV, which gives 
\begin{equation}
    \zeta^{(H_2)}_{\rm{XR}} = \zeta^{(H_2)}_{\rm{XR,\,direct}} + \zeta^{(H_2)}_{\rm{XR,\,scattered}},
    \label{eq:X-ray ionisation rate}
\end{equation}

\noindent where
\begin{multline}
    \zeta^{(H_2)}_{\rm{XR,\,direct}}(r,z) = \left(\frac{L_{\rm{XR}}}{10^{29} \, \rm{erg.s^{-1}}}\right) \left(\frac{r}{1 \, \rm{au}}\right)^{-2.2} \times \\
     \zeta_{\rm{1,XR}} \left[ \exp{\left( -\left(\frac{\Sigma_{\rm{gas}}^{+}(r,z)}{\Sigma_{\rm{1,XR}}}\right)^{0.4} \right)} + \exp{\left( -\left(\frac{\Sigma_{\rm{gas}}^{-}(r,z)}{\Sigma_{\rm{1,XR}}}\right)^{0.4} \right)} \right]
     \label{eq:Direct X-ray ionisation rate},
\end{multline}

\noindent and
\begin{multline}
    \zeta^{(H_2)}_{\rm{XR,\,scattered}}(r,z) = \left(\frac{L_{\rm{XR}}}{10^{29} \, \rm{erg.s^{-1}}}\right) \left(\frac{r}{1 \, \rm{au}}\right)^{-2.2} \times \\
     \zeta_{\rm{2,XR}} \left[ \exp{\left( -\left(\frac{\Sigma_{\rm{gas}}^{+}(r,z)}{\Sigma_{\rm{2,XR}}}\right)^{0.65} \right)} + \exp{\left( -\left(\frac{\Sigma_{\rm{gas}}^{-}(r,z)}{\Sigma_{\rm{2,XR}}}\right)^{0.65} \right)} \right].
     \label{eq:Scattered X-ray ionisation rate}
\end{multline}

\noindent The first term of Eq.~\eqref{eq:X-ray ionisation rate} corresponds to the direct X-rays contribution of the total stellar X-rays ionization rate, whereas the second term corresponds to its scattered X-rays contribution. $\zeta^{(H_2)}_{\rm{XR,\,direct}}$ describes attenuation of X-rays photons by absorption, with the unattenuated coefficient $\zeta_{\rm{1,XR}} = 6 \times 10^{-12} \,$s$^{-1}$ and the penetration depth $\Sigma_{\rm{1,XR}} \approx 0.0035 \,$g.cm$^{-2}$. $\zeta^{(H_2)}_{\rm{XR,\,scattered}}$ describes a contribution from scattering, with the unattenuated coefficient $\zeta_{\rm{2,XR}} = 10^{-15} \,$s$^{-1}$ and the penetration depth $ \Sigma_{\rm{2,XR}} \approx 1.64 \,$g.cm$^{-2}$. Here we have re-calculated the penetration depths given by \citet{2009ApJ...701..737B} in terms of hydrogen nucleus into the corresponding gas surface density penetration depths ($\Sigma_{\rm{1,XR}}$ and $\Sigma_{\rm{2,XR}}$).

\paragraph{Cosmic rays ionization rate $\zeta^{(H_2)}_{\rm{CR}}$ for H$_2$.} We adopt the following fitting formula given by \citet{2009ApJ...690...69U}:
\begin{multline}
        \zeta^{(H_2)}_{\rm{CR}}(r,z) = \frac{\zeta_{\rm{CR,ISM}}}{2} \times \\
        \Bigg\{ \exp{\left(-\frac{\Sigma_{\rm{gas}}^{+}(r,z)}{\Sigma_{\rm{CR}}}\right)} \left[1+\left(\frac{\Sigma_{\rm{gas}}^{+}(r,z)}{\Sigma_{\rm{CR}}}\right)^{\frac{3}{4}}\right]^{-\frac{4}{3}} + \\ \exp{\left(-\frac{\Sigma_{\rm{gas}}^{-}(r,z)}{\Sigma_{\rm{CR}}}\right)} \left[1+\left(\frac{\Sigma_{\rm{gas}}^{-}(r,z)}{\Sigma_{\rm{CR}}}\right)^{\frac{3}{4}}\right]^{-\frac{4}{3}} \Bigg\},
    \label{eq:cosmic ray ionisation rate}
\end{multline}

\noindent where we choose the unattenuated cosmic rays ionization rate $\zeta_{\rm{CR,ISM}} = 10^{-17} \,$s$^{-1}$ and the cosmic ray penetration depth $\Sigma_{\rm{CR}} = 96 \,$g.cm$^{-2}$. Nevertheless, we note that the galactic cosmic rays ionization rate suffers from large uncertainties \citep[e.g.,][]{2003Natur.422..500M, 2013ApJ...772....5C}.

\paragraph{Radionuclides ionization rate $\zeta^{(H_2)}_{\rm{RA}}$ for H$_2$.} The combined effect of the decay of short and long-lived radionuclides is included in the following ionization rate
\begin{equation}
    \zeta^{(H_2)}_{\rm{RA}}(r,z) = \zeta_{\rm{RA}, \, 0} \times \frac{\left(\rho_{\rm{dust}}(r,z)/\rho_{\rm{gas}}(r,z)\right)}{10^{-2}},
    \label{eq:radionuclides ionization rate}
\end{equation}

\noindent where $\zeta_{\rm{RA}, \, 0} = 7.6 \times 10^{-19} \, \rm{s^{-1}}$. This constant is expected to be in the range $(1-10) \times 10^{-19} \,$s$^{-1}$, and is higher at smaller radii and early times \citep[e.g.,][]{2009ApJ...690...69U, 2013ApJ...777...28C}. We further scale the radionuclides ionization rate by the local dust-to-gas ratio $\rho_{\rm{dust}}/\rho_{\rm{gas}}$, normalized by the ISM value $10^{-2}$. Since radionuclides (e.g., $^{26}$Al) are refractory and locked into dust particles, a change in the local dust-to-gas ratio is expected to induce a local change in the number of radionuclides locally available to ionize the gas. We note that Eq.~\eqref{eq:radionuclides ionization rate} does not account for the escape of decay products, and therefore overestimates radionuclide ionization rates in low gas column density regions ($\Sigma_{\rm gas}\lesssim 10\,$g cm$^{-2}$). Nevertheless, radionuclide ionization does not dominate the total ionization rate in such regions of low gas surface densities reached in the outer disk regions. In practice, this inconsistency is thus not an issue.

We compute the ionization rate for H$_2$ $\zeta^{(H_2)}$ by summing the X-rays, cosmic rays, and radionuclides contributions such as $\zeta^{(H_2)}= \zeta^{(H_2)}_{\rm{XR}}+ \zeta^{(H_2)}_{\rm{CR}}+ \zeta^{(H_2)}_{\rm{RA}}$; and finally obtain the total ionization rate $\zeta$ (mean ionization rate for all gas molecules) using Eq.~\eqref{eq:total ionization rate}.

\subsection{Ionization Chemistry} \label{sect:chemistry}

By assuming that local ionization-recombination equilibrium is reached at every locations in the disk, we obtain the abundance of all charge carriers initiated by the ionization process. In this work, we do not implement a chemical reaction network \citep[see e.g.,][]{2009ApJ...701..737B, 2004A&A...417...93S, 2006A&A...445..223I, 2011ApJ...739...50B}. Instead, we implement a semi-analytical chemical model based on \citet{2009ApJ...698.1122O}. The main motivation behind our choice is the need of a chemical model that carefully captures the dust charge state as well as the gas ionization state, without greatly enhancing the computational complexity of the whole problem. The difference between the semi-analytical chemical model presented here and the one described in \citet{2009ApJ...698.1122O} is the fact that we consider compact rather than fluffy grains.

Similarly to \citet{2009ApJ...698.1122O}, the following fundamental approximations are made: (1) ions can be represented by a single dominant ion species; (2) the grain charge distribution can be approximated as a continuous function of $Z$. The key parameter in such an approach is the dimensionless parameter $\Gamma$ that measures the ratio of the electrostatic attraction or repulsion energy between a charged dust particle and an incident ion or free electron, respectively, to the thermal kinetic energy. It is defined as 
\begin{equation}
    \Gamma \equiv \frac{\left(-\left<Z\right> e^{2}/a_{\rm{dust}}\right)}{k_B T}= \frac{- \left<Z\right> \lambda_O}{a_{\rm{dust}}} = \frac{- \left<Z\right>}{\tau}. 
    \label{eq:Gamma-parameter}
\end{equation}

\noindent In the first equality, $\left<Z\right>$ is the mean charge carried by dust particles at a given location in the disk $(r,z)$, and $\lambda_O$ is given by $\lambda_O = \frac{e^2}{k_B T}$; with $e$ the elementary charge, and $T$ the gas temperature defined by Eq.~\eqref{eq:Temperature}. In the second equality, the dimensionless parameter $\tau$ is given by $\tau = \frac{a_{\rm{dust}}}{\lambda_O}$.

We need to further specify the forms of the effective collision cross-sections between ions/dust particles ($\sigma_{\rm{di}}(\Gamma)$) and free electrons/dust particles ($\sigma_{\rm{de}}(\Gamma)$) to properly account for grain surface chemistry. Without loss of generality, they can be written as $\sigma_{\rm{di}}(\Gamma) = s_i \sigma_{\rm{dust}} P_{i}(\Gamma)$ and $\sigma_{\rm{de}}(\Gamma) = s_e \sigma_{\rm{dust}} P_{e}(\Gamma)$, where $\sigma_{\rm{dust}} = \pi a_{\rm{dust}}^2$ is the grain geometric cross-section; $s_i = 1$ and $s_e = 0.3$ are the sticking coefficients for ions and free electrons, respectively; $P_{i}(\Gamma)$ and $P_{e}(\Gamma)$ are dimensionless factors determined by the electrostatic interactions between ions/dust particles and free electrons/dust particles, respectively. Since we consider compact dust particles, we account for their induced electric polarization forces using the results from \citet{1987ApJ...320..803D}. 
We assume $\left<Z\right> < 0$ (i.e., $\Gamma > 0$) because free electrons collide with grains much faster, hence more frequently than ions \citep[see e.g,][]{1991ApJ...368..181N, 2007Ap&SS.311...35W}. Consequently, $P_{i}(\Gamma)$ and $P_{e}(\Gamma)$ are functions of $\Gamma$ that can be written as
\begin{equation}
    P_{i}(\Gamma) = (1 + \Gamma) \left(1 +\sqrt{\frac{2}{\tau \left(1 + 2 \Gamma\right)}}\:\right) 
    \label{eq:normalized effective collision cross-sections between ions/dust particles}
\end{equation}
and
\begin{equation}
    P_{e}(\Gamma) = \exp{\left(-\frac{\Gamma}{1 + \left(\tau \Gamma\right)^{-\frac{1}{2}}}\right)} \left(1 + \sqrt{\frac{1}{\tau \left(4 + 3 \Gamma\right)}}\:\right)^{2},
    \label{eq:normalized effective collision cross-sections between free electrons/dust particles}
\end{equation}

\noindent where Eqs.~\eqref{eq:normalized effective collision cross-sections between ions/dust particles} and \eqref{eq:normalized effective collision cross-sections between free electrons/dust particles} are directly obtained from Eqs.~(3.4) and (3.5) of \citet{1987ApJ...320..803D}, respectively (see also Appendix~\ref{appendix:dZ2_ni_ne}).

For a successful implementation of ionization chemistry in our disk model, we demand that charge neutrality is reached at every locations in the disk
\begin{equation}
    n_i(r,z) - n_e(r,z) + \int_{Z}^{} Z \: n_{\rm{dust}}(r,z,Z) \: dZ = 0, 
    \label{eq:charge neutrality balance}
\end{equation}

\noindent where $n_i$ is the total number density of ions, $n_e$ is the total number density of free electrons; and $n_{\rm{dust}}(r,z,Z)$ is the number density of dust particles of net charge $Z$, located at $(r,z)$.

For $\tau \gtrsim 1$, the equilibrium grain charge distribution can be well approximated by a Gaussian distribution resulting in \citep[][see also Appendix \ref{appendix:dZ2_ni_ne}]{2009ApJ...698.1122O}
\begin{equation}
    n_{\rm{dust}}(r,z,Z) = \frac{n_{\rm{dust}}(r,z)}{\sqrt{2 \pi \left<\Delta Z^{2}\right>}} \exp{\left[-\frac{\left(Z - \left<Z\right> \right)^{2}}{2 \left<\Delta Z^{2}\right>}\right]},
    \label{eq:charge state distribution of dust particles}
\end{equation}

\noindent where $n_{\rm{dust}}(r,z) = \int_{Z}^{} n_{\rm{dust}}(r,z,Z) \: dZ$ is defined in Sect.~\ref{sect:disk properties}, the mean grain charge $\left<Z\right>$ is 
\begin{equation}
    \left<Z\right> \equiv \frac{1}{n_{\rm{dust}}(r,z)} \int_{Z}^{} Z \: n_{\rm{dust}}(r,z,Z) \: dZ = -\Gamma \tau, 
    \label{eq:mean charge}
\end{equation}
and the grain charge dispersion $\sqrt{\left<\Delta Z^{2}\right>}$ is defined as 
\begin{equation}
\begin{split}
    \left(\sqrt{\left<\Delta Z^{2}\right>}\right)^{2} &\equiv \frac{1}{n_{\rm{dust}}(r,z)} \int_{Z}^{} \left(Z-\left<Z\right>\right)^{2} \: n_{\rm{dust}}(r,z,Z) \: dZ \\
    &= \left[\frac{1}{\tau} \left( \frac{d\ln{P_i(\Gamma)}}{d \Gamma} - \frac{d \ln{P_e(\Gamma)}}{d \Gamma} \right)\right]^{-1} 
\end{split}
    \label{eq:charge dispersion}
\end{equation}
(see Appendix~\ref{appendix:dZ2_ni_ne} for derivation of Eq.~\eqref{eq:charge dispersion}, and a complete expression). 
 
Since the charge distribution is solely parametrized by $\Gamma$, $n_i$ and $n_e$ can be written as a function of that single parameter in the equilibrium state. Specifically, the balance between ionization and recombination yields (see \citealt{2009ApJ...698.1122O} and Appendix \ref{appendix:dZ2_ni_ne})
\begin{equation}
    n_i = \frac{2 \zeta n_{\rm{gas}}}{s_i u_i \sigma_{\rm{dust}} n_{\rm{dust}}} \frac{1}{P_{i}(\Gamma) \left(1 + \sqrt{1 + 2 g(\Gamma)}\right)}
    \label{eq:ions number density}
\end{equation}
\noindent and
\begin{equation}
    n_e = \frac{2 \zeta n_{\rm{gas}}}{s_e u_e \sigma_{\rm{dust}} n_{\rm{dust}}} \frac{1}{P_{e}(\Gamma) \left(1 + \sqrt{1 + 2 g(\Gamma)}\right)},
    \label{eq:free electrons number density}
\end{equation}

\noindent where 
\begin{equation}
    g(\Gamma) = \frac{2 k \zeta n_{\rm{gas}}}{s_i u_i s_e u_e \left(\sigma_{\rm{dust}} n_{\rm{dust}}\right)^{2}} \frac{1}{P_{i}(\Gamma) P_{e}(\Gamma)}.
    \label{eq:g(GAMMA)}
\end{equation}

\noindent If $g(\Gamma) \gtrsim 1$ the recombination process is dominated by gas-phase recombination, otherwise it is dominated by charge adsorption onto grains. $k$ is the rate coefficient for the gas-phase recombination defined below, $\zeta$ is the total ionization rate determined by Eq.~\eqref{eq:total ionization rate}, $n_{\rm{gas}}$ is the total number density of gas particles given in Sect.~\ref{sect:disk properties}, and $u_i$, $u_e$ are the mean thermal velocity of the dominant ion species and free electrons, respectively, given by $u_{i(e)} = \sqrt{\frac{8 k_{B} T}{\pi m_{i(e)}}}$ with $m_i$ being the the mass of the dominant ion species. In this study, we assume that the dominant ion species is $\rm{HCO}^+$ by setting  $m_i = 29 \,m_{\rm H}$ and $k = 2.4 \times 10^{-7} \left(\frac{T}{300 \, \rm{K}}\right)^{-0.69} \, \rm{cm}^{3}~\rm{s}^{-1}$ \citep[see][]{2013A&A...550A..36M}. The choice of the dominant ion species does not strongly affect our results as long as the dominant ions are molecular ions that recombine with free electrons dissociatively \citep[e.g.,][]{1974ApJ...192...29O}. 

The only remaining unknown of the problem is the key parameter $\Gamma$. Equation~\eqref{eq:charge neutrality balance} with Eqs.~\eqref{eq:mean charge}, \eqref{eq:ions number density} and \eqref{eq:free electrons number density} leads to the following equation whose root is $\Gamma$:
\begin{equation}
    \frac{1}{P_{i}(\Gamma)} - \left[ \frac{s_i u_i}{s_e u_e} \frac{1}{P_{e}(\Gamma)} + \frac{\Gamma}{\Theta} \frac{\left(1 + \sqrt{1 + 2 g(\Gamma)}\right)}{2} \right] = 0,
    \label{eq:equation for GAMMA}
\end{equation}

\noindent where $\Theta$ is a dimensionless parameter that quantifies which of free electrons and dust particles are the dominant carriers of negative charge. It is defined by 
\begin{equation}
    \Theta \equiv \frac{\zeta n_{\rm{gas}} \lambda_O}{s_i u_i \sigma_{\rm{dust}} a_{\rm{dust}} n_{\rm{dust}}^{2}}.
\end{equation}

To summarize, the abundance of all charge carriers are written as analytical functions of the single key parameter $\Gamma$. By numerically solving Eq.~\eqref{eq:equation for GAMMA}, we obtain $\Gamma$ and thus those abundances. In Appendix~\ref{appendix:comparison between the semi-analytical chemical model and a chemical reaction network}, we compare our semi-analytical approach to a chemical reaction network.

\section{Magnetorotational Instability} \label{sect:MRI}

The mass and angular momentum transport of weakly ionized protoplanetary disks is governed by MRI magnetic torques only if the coupling between the charged particles in the gas-phase and the magnetic field is sufficient; namely, the gas motion induced by those torques can generate magnetic stresses (due to orbital shear) faster than they can diffuse away due to non-ideal MHD effects. In order to obtain the disk turbulence level encoded into the Shakura-Sunyaev $\alpha$-parameter, we need to know where the MRI can operate (Sect.~\ref{sect:criteria for active MRI}); choose an appropriate magnetic field strength (see Sect.~\ref{sect:B field}); and link the MRI stress to the disk turbulence state (Sect.~\ref{sect:connecting MRI to alpha-disk model}).

\subsection{Criteria for Active MRI} \label{sect:criteria for active MRI}

In this study, the non-ideal MHD effects considered are Ohmic resistivity and ambipolar diffusion. Their corresponding Elsasser numbers are defined as 
\begin{equation}
    \Lambda \equiv \frac{\rm{v}_{\rm{Az}}^{2}}{\eta_O \Omega_K}
    \label{eq:omhic Elsasser number}
\end{equation}
and
\begin{equation}
    \rm{Am} \equiv \frac{\rm{v}_{A}^{2}}{\eta_{\rm{AD}} \Omega_K},
    \label{eq:ambipolar Elsasser number}
\end{equation}

\noindent where $\Lambda$ is the Ohmic Elssaser number, $\eta_O$ is the Ohmic magnetic diffusivity, Am is the ambipolar Elsasser number, $\eta_{\rm{AD}}$ is the ambipolar magnetic diffusivity, and $\rm{v}_{\rm{Az}}$ is the vertical component of the Alfv\'{e}n velocity $\rm{v}_{A}$ defined as $\rm{v}_{A} \equiv \frac{B}{\sqrt{4 \pi \rho_{\rm{gas}}}}$. We note the use of $\rm{v}_{A}$ instead of $\rm{v}_{\rm{Az}}$ in Eq.~\eqref{eq:ambipolar Elsasser number}, since we adopt the results of \citet{2011ApJ...736..144B} who use $\rm{v}_{A}$ to define $\rm{Am}$. Here $B_z$ is the vertical component of the magnetic field, and $B$ is the strength of the r.m.s. field. We assume that those two quantities are related by $B^{2} \approx 25 B_{z}^{2}$ \citep{2004ApJ...605..321S}. Our method to constrain $B$ is described in Sect.~\ref{sect:B field}.  

When Omhic resistivity dominates, it has been shown that the MRI can be active if $\Lambda > 1$ \citep[see e.g.,][]{2002ApJ...577..534S, 2007ApJ...659..729T}. When ambipolar diffusion is the dominant non-ideal MHD effect, and assuming the strong-coupling limit (see Appendix~\ref{sect:strong-coupling limit}), the MRI can operate even if $\rm{Am} < 1$, provided that the magnetic field strength is "sufficiently weak" \citep{2011ApJ...736..144B}. Here "sufficiently weak" means that the plasma $\beta$-parameter must exceed a minimum threshold $\beta_{\rm{min}}$ defined by Eq.~\eqref{eq:beta min}.

Consequently, we consider the following set of conditions for sustaining active MRI:
\begin{equation}
    \Lambda > 1,
    \label{eq:ohmic Elsasser number criterion for MRI}
\end{equation}
and
\begin{equation}
    \beta > \beta_{\rm{min}}(\rm{Am})
    \label{eq:magnetic field criterion for MRI}.
\end{equation}

\noindent The first condition will be referred in the following as the Ohmic condition, the second one as the ambipolar condition. Here the plasma $\beta$-parameter is defined as $\beta \equiv \frac{P_{\rm{gas}}}{P_{\rm{magnetic}}} = \frac{8 \pi \rho_{\rm{gas}} c_s^2}{B^2}$. The minimum threshold $\beta_{\rm{min}}$ is a function of the ambipolar Elsasser number $\rm{Am}$ given by \citep{2011ApJ...736..144B}
\begin{equation}
    \beta_{\rm{min}}(\rm{Am}) = \sqrt{\left(\frac{50}{\rm{Am}^{1.2}}\right)^2 + \left(\frac{8}{\rm{Am}^{0.3}} + 1\right)^2}.
    \label{eq:beta min}
\end{equation}

Whether the MRI is active or not depends on the magnetic diffusivities ($\eta_O$ and $\eta_{\rm{AD}}$), since they express the strength of their corresponding non-ideal MHD effects. They are principally determined by the ionization degree of the gas as well as the magnetic field strength. Armed with the equilibrium abundances of all charge carriers computed in Sect.~\ref{sect:chemistry} and adopting the magnetic field strength obtained in Sect.~\ref{sect:B field}, $\eta_O$ and $\eta_{\rm{AD}}$ are computed at every locations of the protoplanetary disk by following \citet{2007Ap&SS.311...35W} (see their Eqs.~(29) and (31)). To formally compute the various conductivites \citep[$\sigma_O$, $\sigma_H$ and $\sigma_P$ from][]{2007Ap&SS.311...35W}, we need to specify for the momentum rate coefficient $\left<\sigma v\right>_{j}$ \citep[appearing in the coefficient $\gamma_j$ of][see their Eq.~(21)]{2007Ap&SS.311...35W}  for ions ($j \equiv i$), free electrons ($j \equiv e$) and dust particles ($j \equiv dust$). We adopt the formulation from \citet{2011piim.book.....D} (see Table~2.1 and Eq.~(2.34)), \citet{2011ApJ...739...50B} (see Eq.~(15)), and \citet{1999MNRAS.303..239W} (see Eq.~(21)); respectively for $\left<\sigma v\right>_{i}$, $\left<\sigma v\right>_{e}$, and $\left<\sigma v\right>_{\rm{dust}}$
\begin{equation}
    \left<\sigma v\right>_{i} = 2.0 \times 10^{-9} \sqrt{\frac{m_{\rm H}}{\delta}} \, \rm{cm}^{3} \rm{s}^{-1},
    \label{eq:momentum rate coefficient for ions}
\end{equation}
\begin{equation}
    \left<\sigma v\right>_{e} = 8.3 \times 10^{-9} \times \max{\left[1,\sqrt{\frac{T}{100 \, \rm{K}}} \:\right]} \, \rm{cm}^{3}.\rm{s}^{-1}
    \label{eq:momentum rate coefficient for free electrons}
\end{equation}
and
\begin{equation}
    \left<\sigma v\right>_{\rm{dust}} = \sigma_{\rm{dust}} \sqrt{\frac{128 k_B T}{9 \pi m_{neutral}}},
    \label{eq:momentum rate coefficient for grains}
\end{equation}

\noindent where $\delta \equiv \frac{m_i m_{\rm{neutral}}}{m_i + m_{\rm{neutral}}}$ is the reduced mass in a typical ion-neutral collision, $T$ is the gas temperature defined by Eq.~\eqref{eq:Temperature}, and $\sigma_{\rm{dust}}$ is the grain geometric cross-section defined in Sect.~\ref{sect:chemistry}. 

An "MRI-active layer" is where both Eqs.~\eqref{eq:ohmic Elsasser number criterion for MRI} and \eqref{eq:magnetic field criterion for MRI} are satisfied; a "dead zone" is where the condition \eqref{eq:ohmic Elsasser number criterion for MRI} is not met, so that Omhic resistivity shuts off the MRI; and a "zombie zone" \citep[as termed by][]{2013ApJ...764...65M} is where the condition \eqref{eq:magnetic field criterion for MRI} is not fulfilled, so that ambipolar diffusion prohibits the MRI-driven turbulence with too strong magnetic fields. With these conditions for active MRI, the underlying assumption is that the MRI is either in a saturation level allowed by the non-ideal MHD effects or completely damped at any locations in the disk. Fundamentally, this assumption comes down to the fact that the MRI growth timescale is short (roughly $\propto \Omega_{K}^{-1}$) compared to viscous evolution timescales or dust evolution timescales.

The Hall effect has been ignored in the above criteria for active MRI, since quantifying its effect is beyond the scope of the present paper. Indeed, the inclusion of the Hall effect will further complicate the global picture; and it is still not clear how this non-ideal MHD effect impact on the MRI when it dominates. As a result, we use the above conditions for MRI even in the Hall-dominated regions of the protoplanetary disk. Nevertheless, we investigate where the Hall effect might be important and discuss the implications in Appendix~\ref{sect:the hall effect}.

\subsection{Magnetic Field Strength} \label{sect:B field}

The magnetic field strength plays a crucial role for the MRI, since the magnetic diffusivities depend on it.   

In our study, we make the assumption that if the MRI can operate then: (1) the MRI activity is maximally efficient given our conditions for active MRI; (2) the MRI-driven turbulence is in a saturation level allowed by the non-ideal MHD effects; and (3) the magnetic field is constant with height across the MRI-active layer. Assumptions (2) and (3) are motivated by various 3D stratified and unstratified shearing box and global simulations. They show that the MRI saturates on a state permitted by the non-ideal MHD effects \citep[e.g.,][]{1995ApJ...440..742H, 2004ApJ...605..321S, 2011ApJ...730...94S, 2012ApJ...761...95F, 2013ApJ...763...99P, 2013MNRAS.435.2281P}; and once saturated, turbulence stirring maintains roughly constant the field strength across the active layer \citep[e.g.,][]{2000ApJ...534..398M}. Adopting assumption (1) implies that we opt for an optimistic view of the MRI activity, which will generate the highest accretion rate possibly reachable given the set of conditions for active MRI.

To fulfill the maximum MRI activity assumption, we constrain and adopt the magnetic field strength so that it corresponds to the strongest field possible that would still allow for the MRI to operate (i.e., we want $\beta$ as low as possible while satisfying both Eqs.~\eqref{eq:ohmic Elsasser number criterion for MRI} and \eqref{eq:magnetic field criterion for MRI}). In practice, we do so by maximizing the component of the steady-state accretion rate that comes from the MRI-active layer ($\dot{M}_{\rm{acc,\,MRI}}$). Using Eq.~(24) of \citet{2011ApJ...739...50B}, it can be written as $\dot{M}_{\rm{acc,\,MRI}} \propto \frac{h_{\rm{MRI}} B^2}{\Omega_K}$, where $h_{\rm{MRI}}$ is the MRI-active layer thickness. This formula is derived assuming that $\alpha \propto \beta^{-1}$ (see Sect.~\ref{sect:connecting MRI to alpha-disk model}). Here we thus need to maximize the quantity $h_{\rm{MRI}} B^2$, where we want $B$ as strong as possibly allowed for the MRI to still operate. We note that $h_{\rm{MRI}}$ implicitly depends on $B$, hence maximizing $B$ alone formally leads to an infinitesimal $h_{\rm{MRI}}$, implying $\dot{M}_{\rm{acc,\,MRI}} \to 0$. Indeed, ambipolar diffusion prohibits the MRI-driven turbulence with too strong magnetic fields as indicated by Eqs.~\eqref{eq:magnetic field criterion for MRI} and \eqref{eq:beta min}. 

Additionally, the MRI activity can be considered as maximally efficient only if the most unstable modes (fastest growing modes) can develop in the MRI-active layer. In the upper layers of the MRI-active layer, the MRI behavior is marginally close to its ideal-MHD limit (Am $\approx 1$, see Fig.~\ref{fig:Elsasser_numbers}(a)). In this context, the wavelengths of the must unstable modes are given by \citep[e.g.,][]{1999ApJ...515..776S}
\begin{equation}
    \lambda_m = \frac{2 \pi \rm{v}_{\rm{Az}}}{\Omega_K},
    \label{eq:wavelength for the most unstable mode}
\end{equation}

\noindent where $\rm{v}_{\rm{Az}}$ is the vertical component of the Alfv\'{e}n velocity defined in Sect.~\ref{sect:criteria for active MRI}. Assuming that the MRI is maximally efficient ultimately comes down to the condition that the longest wavelength of the most unstable modes in the MRI-active layer must fit within both the disk gas scale height and the MRI-active layer thickness. Mathematically, this condition can be written as
\begin{equation}
    \lambda_{\rm{MRI}} \leq H_{\rm{gas}} \; \textnormal{and} \; \lambda_{\rm{MRI}} \leq h_{\rm{MRI}},
    \label{eq:highest wavelentgth of the most unstable mode fits within the disk gas scale height and the MRI-active layer thickness}
\end{equation}

\noindent where $\lambda_{\rm{MRI}}(r) = \max\limits_{\textnormal{z $\in$ MRI-active layer}} \Bigg\{\lambda_m(r,z) \Bigg\}$ is the longest wavelength of the most unstable modes in the MRI-active layer, at any radii $r$.
To summarize, the magnetic field strength $B$ is chosen such that the MRI can operate (both Eqs.~\eqref{eq:ohmic Elsasser number criterion for MRI} and \eqref{eq:magnetic field criterion for MRI} must be satisfied) and the MRI is maximally efficient (we find $B$ such that the quantity $h_{\rm{MRI}} B^2$ is maximized, and the condition \eqref{eq:highest wavelentgth of the most unstable mode fits within the disk gas scale height and the MRI-active layer thickness} is fulfilled).

To compute the magnetic field strength, we adopt a similar procedure as described in \citet{2013ApJ...764...65M}. For a specified disk radius $r$, we loop through a range of field strengths $B \in \left[10^{-5}-10^{3}\right] \,$Gauss, which covers the plausible range in stellar accretion disks. For a given value of that range, we determine the MRI-active layer thickness $h_{\rm{MRI}}$. We then check that the longest wavelength of the most unstable modes in the MRI-active layer fits within both the disk gas scale height and the MRI-active layer thickness. If it does, we compute the quantity $h_{\rm{MRI}} B^2$. If not, we automatically discard this magnetic field strength value by setting the quantity $h_{\rm{MRI}} B^2$ to $0$. We then reiterate the previous steps for the different field strength values in the range, and finally choose the final value $B$ for this specific disk radius $r$ such that the quantity $h_{\rm{MRI}} B^2$ is the highest. We repeat the above steps for a range of radii $r \in \left[r_{\rm{min}}-r_{\rm{max}}\right]$, until we obtain $B$ as a function of disk radius $r$. Once we have the radial profile, we finally reconstruct $B$ in the vertical direction by further assuming that it is vertically constant.

\subsection{From the MRI Accretion Formulation to turbulence} \label{sect:connecting MRI to alpha-disk model}

The last key ingredient missing in our disk model is the link between accretion stress produced by either the MRI or hydrodynamic instabilities and the effective turbulent parameter $\bar{\alpha}$ that goes into the Shakura-Sunyaev $\alpha$-disk model. In particular, we must specify the local turbulent parameter $\alpha$ as a function of location in the disk (see Eq.~\eqref{eq:alpha bar}). 

In non-MRI regions (dead or zombie zone), the source of viscosity is thought to be mainly from hydrodynamic instabilities that can produce non-MRI stresses. Since we do not run any hydrodynamic simulations, we assume the turbulence in all non-MRI regions to be encoded into a single constant turbulent parameter $\alpha_{\rm{hydro}}$. We further assume the hydrodynamic turbulent parameter $\alpha_{\rm{hydro}}$ to be constant across a given non-MRI zone, and taken as such of whether the region is a dead or zombie zone. Motivated by the results from 3D global hydrodynamic simulations of the VSI \citep[e.g.,][]{2020ApJ...897..155F, 2021arXiv210601159B}, we fiducially choose $\alpha_{\rm{hydro}} = 10^{-4}$.

In the regions of the disk where the MRI can operate (following our criteria in Sect.~\ref{sect:criteria for active MRI}), a variety of numerical simulations show that there exists a tight correlation between the normalized accretion stress and the $\beta$-plasma parameter \citep[see e.g.,][]{1995ApJ...440..742H, 2004ApJ...605..321S, 2010ApJ...716.1012P, 2011ApJ...736..144B}. The correlation can be represented by \citep[see e.g.,][their discussion in Appendix~B]{2018ApJ...861..144M}
\begin{equation}
    W_{r \phi, \rm{MRI}} \approx \frac{1}{2 \beta},
    \label{eq:accretion stress}
\end{equation}

\noindent where $\beta$ is the plasma $\beta$-parameter defined in Sect.~\ref{sect:criteria for active MRI}, corresponding to the optimal r.m.s. magnetic field strength obtained by the method described in Sect.~\ref{sect:B field}; and $W_{r \phi, \rm{MRI}}$ corresponds to the normalized accretion stress in the MRI-active layer. Here the total normalized accretion stress $W_{r \phi}$ is defined as the sum of the Reynolds and Maxwell stresses normalized by the gas pressure. 

In the MRI-active layer, we connect the local turbulent parameter $\alpha_{\rm{MRI}}$ to the normalized accretion stress $W_{r \phi, \rm{MRI}}$ by
\begin{equation}
    \alpha_{\rm{MRI}} = \left\{
                    \begin{array}{ll}
                         \frac{2}{3} W_{r \phi, \rm{MRI}} \approx \frac{1}{3 \beta} & \mbox{if} \; \frac{1}{3 \beta} > \alpha_{\rm{hydro}}  \\
                         \alpha_{\rm{hydro}} & \mbox{otherwise}
                    \end{array}
                \right.
    \label{eq:alpha MRI}
\end{equation}

\noindent The factor "2/3" in Eq.~\eqref{eq:alpha MRI} is explained in Sect.~2.2 of \citet{2017ApJ...845...31H} and in Appendix~B of \citet{2018ApJ...861..144M}. Also, we have added up the floor value $\alpha_{\rm{hydro}}$ because the MRI stress must dominate over the residual hydrodynamic stresses in MRI-active regions. If not, such regions are considered as dead in the sense that the turbulence is driven by hydrodynamic instabilities (even though the conditions for active MRI are satisfied).

To summarize, we calculate the appropriate effective turbulent parameter $\bar{\alpha}$ (Eq.~\eqref{eq:alpha bar}) by specifying the local turbulent parameter $\alpha$ at all locations of the protoplanetary disk:
\begin{equation}
    \alpha(r,z) = \left\{
                    \begin{array}{ll}
                         \alpha_{\rm{DZ}}(r,z) & \mbox{in the dead zone}  \\
                         \alpha_{\rm{AD}}(r,z) & \mbox{in the zombie zone} \\
                         \alpha_{\rm{MRI}}(r,z) & \mbox{in the MRI-active layer}
                    \end{array}
                \right.
    \label{eq:local turbulent parameter}
\end{equation}

\noindent where $\alpha_{\rm{DZ}}(r,z) = \alpha_{\rm{AD}}(r,z) =\alpha_{\rm{hydro}}$, and $\alpha_{\rm{MRI}}(r,z)$ is given by Eq.~\eqref{eq:alpha MRI}.

\section{Application: Steady-state Accretion} \label{sect:methodology}

\subsection{Approach} \label{sect:approach}

In this study, we apply our 1+1D global MRI-driven disk accretion model (described in Sects.~\ref{sect:model} and \ref{sect:MRI}) to investigate the steady-state MRI-driven accretion regime of a fiducial disk around a solar-type star. To do so, we self-consistently solve for the gas surface density through an iterative process (see Sect.~\ref{sect:the equilibrium solution}), alongside the appropriate effective turbulent parameter $\bar{\alpha}$, to ensure the accretion rate $\dot{M}_{\rm{acc}}$ to be radially constant (required for steady-state accretion) in the outer regions of the disk ($r \gtrsim 1\,$au). In practice, for each step of the iteration process, we  solve the coupled set of equations for both the gas surface density and the MRI. These equations are coupled because: (1) $\bar{\alpha}$ and $\dot{M}_{\rm{acc}}$ are constrained by the MRI accretion, which depends on the disk structure; (2) the disk structure in steady-state accretion is determined by $\bar{\alpha}$, $\dot{M}_{\rm{acc}}$, and stellar parameters. Although the detailed procedure is different, we note that our overall methodology is similar to the one adopted in \citet{2018ApJ...861..144M}. The authors focus on regions around the dead zone inner edge ($r < 1\,$au) by presenting a semi-analytical steady-state dust-free model in which the disk structure, thermal ionization and the viscosity due to the MRI are determined self-consistently. Among the key differences, our numerical model focuses on regions beyond $1 \,$au (including the dead zone outer edge), include the relevant non-thermal ionization sources at those locations in the disk, and employ a semi-analytical chemical model to capture the charge state of the disk dust-gas mixture.

In our disk model, we choose $M_{\rm{disk}}$ as a free input parameter, and self-consistently obtain the corresponding $\dot{M}_{\rm{acc}}$; rather than choosing $\dot{M}_{\rm{acc}}$ as the free parameter. This choice is particularly interesting if one wants to input the total disk gas mass. For example, this is useful for comparing the effective turbulent parameter $\bar{\alpha}$ obtained for a steady-state accretion disk with a disk that would be out of equilibrium but with the same total disk gas mass; or if one wants to compare the equilibrium solutions obtained for different stellar masses, where the total disk gas mass is fixed. We note that setting either $\dot{M}_{\rm{acc}}$ or $M_{\rm{disk}}$ as the free input parameter of the model is equivalent, and lead to the same results. 

While the steady-state solution well describes the inner inward-accreting region ($r \lesssim R_{\rm t}$) of viscously evolving disks, it cannot treat the outer viscously-expanding region ($r \gtrsim R_{\rm t}$). Here $R_{\rm t}$ corresponds to the transition radius at which the gas motion changes from inward to outward in a viscously evolving disk \citep[e.g.,][]{1998ApJ...495..385H}. Ignoring this outer region, our steady model underestimates the real total disk gas mass. However, if we regard the outer boundary $r_{\rm{max}}$ of our disk model as the transition radius $R_{\rm t}$, the neglected mass is at worst comparable to the mass within the model. \footnote{For instance, the self-similar $\alpha$-constant disk model has a gas mass distribution given by $M_<(r) = M_{\rm{tot,self-similar}}[1-\exp(-r/(2R_{\rm t}))]$; where $M_<(r)$ is the disk gas mass within disk radius $r$, and $M_{\rm{tot,self-similar}}$ is the total gas mass of the whole self-similar disk \citep[see e.g.,][]{1998ApJ...495..385H}. For this specific model, the inner inward-accreting region ($r \lesssim R_{\rm t}$) comprises $\approx 40\%$ of the total disk gas mass. Neglecting the gas mass within the outer viscously-expanding region ($r \gtrsim R_{\rm t}$) thus underestimates the total disk gas mass by only $\approx 60\%$.}

\subsection{The Equilibrium Solution} \label{sect:the equilibrium solution}

In the classical 1D viscous disk model, the steady-state accretion rate can be written as \citep[for a derivation see e.g.,][their Appendix~B]{2018ApJ...861..144M}:
\begin{equation}
	\dot{M}_{\rm{acc}} \equiv \frac{3 \pi c_{s}^{2} \bar{\alpha} \Sigma_{\rm{gas}}}{f_r\Omega_{K}} \approx \frac{3 \pi c_{s}^{2} \bar{\alpha} \Sigma_{\rm{gas}}}{\Omega_{K}}. 
  	\label{eq:Accretion rate1}
\end{equation}

\noindent where $f_r \equiv \left(1-\sqrt{\frac{R_{\rm{in}}}{r}} \:\right)$ comes from the thin boundary-layer condition at the inner edge of the disk, assuming a zero-torque inner boundary condition \citep[e.g.,][]{2019SAAS...45....1A}. In the $\alpha$-disk model, $R_{\rm{in}}$ is by definition the location where the Keplerian angular velocity $\Omega_K$ plateaus and turns over. For instance, if the disk extends up to the stellar surface, $R_{\rm{in}} = R_\star$ with $R_\star$ the stellar radius. In any case $R_{\rm{in}} \ll 1\,$au, justifying the second equality in Eq.~\eqref{eq:Accretion rate1} that we shall use in this paper, since we study the structure of the outer disk ($r \gtrsim 1\,$au).

Solving for $\Sigma_{\rm{gas}}$ by re-arranging Eq.~\eqref{eq:Accretion rate1}, we obtain the self-consistent steady-state gas surface density under the framework of viscously-driven accretion
\begin{equation}
    \Sigma_{\rm{gas}} \approx \frac{\dot{M}_{\rm{acc}} \Omega_{K}}{3 \pi c_{s}^{2} \bar{\alpha}}.
    \label{eq:gas surface density}
\end{equation}

When multiplying Eq.~\eqref{eq:gas surface density} by $2 \pi r$, integrating over radius and using the fact that $\dot{M}_{\rm{acc}}$ is radially constant for steady-state accretion, we can find the relation between the accretion rate and the total disk gas mass $M_{\rm{disk}}$ 
\begin{equation}
    \dot{M}_{\rm{acc}} \approx \frac{3 M_{\rm{disk}}}{2 \int_{r_{\rm{min}}}^{r_{\rm{max}}} \frac{r \Omega_K}{\bar{\alpha} c_s^{2}} \: dr},
    \label{eq:Accretion rate2}
\end{equation}

\noindent where $r_{\rm{min}}$ and $r_{\rm{max}}$ are the inner and outer boundary of our radial grid, respectively.
Eq.~\eqref{eq:Accretion rate2} is another way of expressing the dependence of the accretion rate on the disk structure. It allows us to compute $\dot{M}_{\rm{acc}}$ accordingly to the chosen free input parameter $M_{\rm{disk}}$, and ensure that the calculated $\Sigma_{\rm{gas}}$ is such that $M_{\rm{disk}} = \int_{r_{\rm{min}}}^{r_{\rm{max}}} 2 \pi r \Sigma_{\rm{gas}}(r) \: dr$. Here we emphasize that the total disk gas mass $M_{\rm{disk}}$ is defined as the gas mass enclosed between $r_{\rm{min}}$ and $r_{\rm{max}}$, and does not account for the mass residing beyond the outer boundary $r_{\rm{max}}$ ($M_{\rm{disk}}$ is equal to the real total disk gas mass within a factor of $2$, see Sect.~\ref{sect:approach}).

Once the free input parameter $M_{\rm{disk}}$ is chosen, we always compute $\dot{M}_{\rm{acc}}$ using Eq.~\eqref{eq:Accretion rate2} and $\Sigma_{\rm{gas}}$ using Eq.~\eqref{eq:gas surface density}, if not stated otherwise.

The steady-state $\Sigma_{\rm{gas}}$ and $\bar{\alpha}$ are obtained by iterating through the following steps (those two quantities utterly determine the equilibrium solution in our disk model):
\begin{itemize}
    \item Step 1. Compute the accretion rate $\dot{M}_{\rm{acc}}$ (Eq.~\eqref{eq:Accretion rate2}) for a given fixed total disk gas mass $M_{\rm{disk}}$.
    \item Step 2. Derive the steady-state gas structure by computing the gas surface density $\Sigma_{\rm{gas}}$ (Eq.~\eqref{eq:gas surface density}), and reconstructing the gas volume density distribution $\rho_{\rm{gas}}$ under the hydrostatic equilibrium assumption.
    \item Step 3. Compute the total ionization rate $\zeta$ as in Sect.~\ref{sect:ionization sources}.
    \item Step 4. Derive the number densities of free electrons, ions and charged dust particles by using the semi-analytical chemical model described in Sect.~\ref{sect:chemistry}.
    \item Step 5. Derive the Ohmic and ambipolar magnetic diffusivities ($\eta_O$ and $\eta_{\rm{AD}}$), and obtain their corresponding Elsasser numbers from which a set of conditions for active MRI can be derived as in Sect.~\ref{sect:criteria for active MRI}.
    \item Step 6. Derive the local turbulent parameter $\alpha$ by connecting the MRI formulation of accretion to the $\alpha$-disk model (Eq.~\eqref{eq:local turbulent parameter}), and compute the effective turbulent parameter $\bar{\alpha}$ (Eq.~\eqref{eq:alpha bar}). 
\end{itemize}

After each Step 6, we check that $\Sigma_{\rm{gas}}$ and $\bar{\alpha}$ are such that the right-hand side of Eq.~\eqref{eq:Accretion rate1} is radially constant equal to $\dot{M}_{\rm{acc}}$ given by Eq.~\eqref{eq:Accretion rate2}, and so at any disk radii. If not, we keep on iterating from Step 1 to Step 6 with the updated $\Sigma_{\rm{gas}}$ and $\bar{\alpha}$ from the previous iteration. The equilibrium solution is reached once the three following convergence conditions are met for the last $15$ consecutive iterations: (1) $\dot{M}_{\rm{acc}}$ is constant from iteration i-1 to iteration i within a $5\%$ error; (2) the right-hand side of Eq.~\eqref{eq:Accretion rate1} is radially constant equal to $\dot{M}_{\rm{acc}}$ within a $5\%$ error; (3) $\Sigma_{\rm{gas}}$ is constant from iteration i-1 to iteration i within a $5\%$ error. 

Our code converges toward a solution for the following reason. For a given fixed total disk gas mass, gas temperature profile, and stellar parameters, $\dot{M}_{\rm{acc}}$ varies from iteration i-1 to iteration i only if $\bar{\alpha}$ varies (see Eq.~\eqref{eq:Accretion rate2}). Once $\bar{\alpha}$ is constant from iteration i-1 to iteration i, $\dot{M}_{\rm{acc}}$ becomes constant as well, and so does $\Sigma_{\rm{gas}}$ (Eq.~\eqref{eq:gas surface density}). In this study, we constrain and choose the r.m.s. magnetic field strength $B$ such that the MRI-activity is maximally efficient permitted by the non-ideal MHD effects considered (see Sect.~\ref{sect:B field}). Consequently, $\bar{\alpha}$ converges through iterations for the steady-state accretion solution to satisfy this condition.

We initially assume that the protoplanetary disk is fully turbulent, and arbitrarily set $\bar{\alpha}$ to be radially constant equal to $\bar{\alpha}_{\rm{init}} = 0.1$ (in order to obtain the first $\dot{M}_{\rm{acc}}$ and $\Sigma_{\rm{gas}}$, required to initiate the iteration process). We checked that this choice does note affect the equilibrium solutions found.

\section{Results - Fiducial Model} \label{sect:results-fiducial model}

Applying our MRI-driven disk accretion model to describe the steady-state regime, we can now investigate the outer region structure ($r \gtrsim 1\,$au) of steadily viscously accreting protoplanetary disks. We first present a detailed description of the equilibrium solution obtained using the method described in Sect.~\ref{sect:methodology}, for the fiducial model ($M_{\star} = 1 \, M_{\odot}$, $L_{\star} = 2 \, L_{\odot}$, $M_{\rm{disk}} = 0.05 \,M_{\star}$, $a_{\rm{dust}} = 1 \,\mu$m, $f_{\rm{dg}} = 10^{-2} $, $\alpha_{\rm{hydro}} = 10^{-4}$): the different accretion layers (Sect.~\ref{sect:behavior of the different accretion layers}); the disk structure (Sect.~\ref{sect:disk structure}); and the disk turbulence level (Sect.~\ref{sect:turbulence}). We then present the equilibrium solutions arising from variations in some of our fiducial parameters in Sect.~\ref{sect:results-parameter study}.

\subsection{MRI-active Layer, Dead and Zombie Zone} \label{sect:behavior of the different accretion layers}

When Ohmic resistivity and ambipolar diffusion are the only non-ideal MHD effects considered, the MRI-active layer is sandwiched between two inactives regions: the dead and the zombie zone.

\begin{figure*}
\centering
\includegraphics[width=0.80\textwidth]{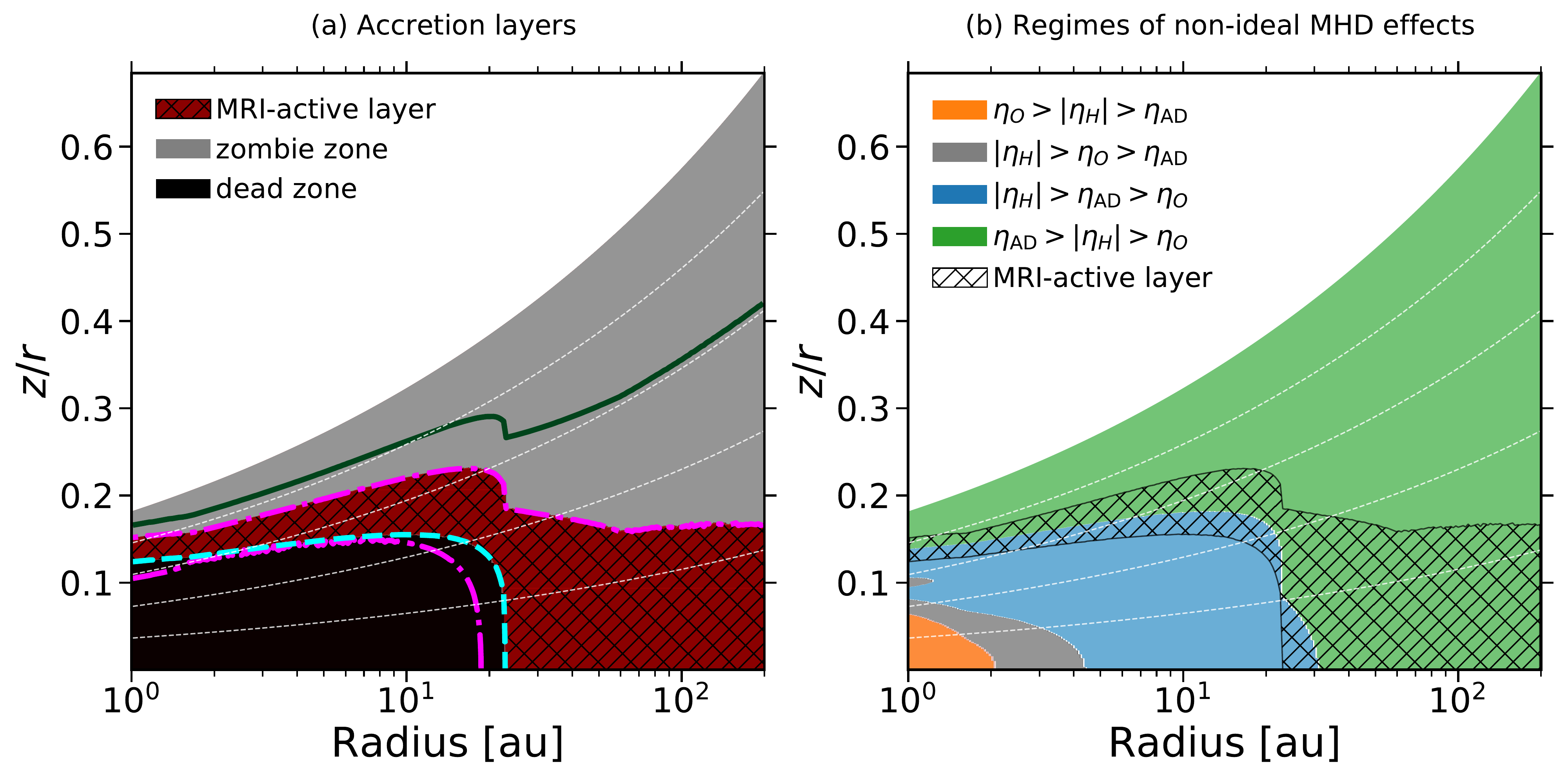}
 \caption{\textit{Panel (a)}: Steady-state accretion layers as a function of location in the disk. The black colored area corresponds to the dead zone. The gray colored area corresponds to the zombie zone. The red and black hatched area corresponds to the MRI-active layer. The region within the magenta dash-dotted lines defines where the ambipolar condition ($\beta > \beta_{\rm{min}}$) is satisfied in the disk; while the region above the cyan dashed line defines where the Ohmic condition ($\Lambda > 1$) is fulfilled. The region above the green solid line corresponds to where $\beta > 1$. The white dashed lines correspond to the surfaces $z = 1 \, H_{\rm{gas}}$, $z = 2 \, H_{\rm{gas}}$, $z = 3 \, H_{\rm{gas}}$ and $z = 4 \, H_{\rm{gas}}$; from bottom to top, respectively. \textit{Panel (b)}: Regimes of the non-ideal MHD effects by showing the dominant magnetic diffusivities as a function of location in the disk. The orange colored region means that the Ohmic resistivity dominates ($\eta_O$ is the highest magnetic diffusivity), the gray and blue colored regions mean that the Hall effect dominates ($|\eta_H|$ is the highest magnetic diffusivity), and the green colored regions means that ambipolar diffusion dominates ($\eta_{\rm{AD}}$ is the highest magnetic diffusivity). The MRI-active layer is shown by the black hatched area. The white dashed lines are the same as in panel (a).}
 \label{fig:accretion_layers}
\end{figure*}

Figure~\ref{fig:accretion_layers}(a) shows the stratification of the disk with the three different accretion layers. First, we notice that the dead zone sits in the innermost regions where the gas density is the highest, whereas the zombie zone is located in the disk atmosphere where the gas density is low. Second, the MRI-active layer only develops in the upper layers for $r \lesssim 23 \,$au, whereas the MRI operates from the mid-plane for $r \gtrsim 23 \,$au. As we will see in Sect.~\ref{sect:ionization level}, the ionization level is high enough only in the upper layers sitting right above the dead zone so that the MRI can only develop in a thin layer for $r \lesssim 23 \,$au; until ambipolar diffusion prohibits it in the very upper layers. Conversely, for $r \gtrsim 23 \,$au, the ionization level is high enough right from the mid-plane so that the magnetic field can efficiently couple to the charged particles in the gas-phase to trigger the MRI. Third, in Fig.~\ref{fig:accretion_layers}(a), the region in between the magenta dash-dotted lines defines where the ambipolar condition ($\beta > \beta_{\rm{min}}$) is satisfied in the disk; whereas the region above the cyan dashed line defines where the Ohmic condition ($\Lambda > 1$) is fulfilled. We note that: (1) the upper envelope of the MRI-active layer is utterly set by the ambipolar condition; (2) its lower envelope is mainly determined by the Ohmic condition (although the ambipolar condition plays a role for $2 \, \rm{au} \lesssim r \lesssim 9 \,$au). The latter result is different from previous studies \citep[e.g.,][]{2013ApJ...764...65M}, where they find that both the upper and lower envelope of the MRI-active layer are completely set by the ambipolar condition around a Sun-like star. We can explain this difference as follows: First, they did not describe the steady-state accretion solution, resulting in a disk structure different from what is presented here. Second and most importantly, they used the total Alfv\'{e}n velocity in their Ohmic condition, whereas we used its vertical component. Since the vertical component is $5$ times less than the total quantity in our model ($B^{2} \approx 25 B^{2}_{z}$), our Ohmic Elsasser number is lower; resulting in a more stringent Ohmic condition in our study, and thus a more extended dead zone (both radially and vertically).

Figure~\ref{fig:accretion_layers}(b) shows which one of the non-ideal MHD effects dominate the magnetic diffusivities in the disk. The MRI-active layer is overplotted and shown by the black hatched area. We notice that ambipolar diffusion dominates in most regions of the protoplanetary disk (especially the zombie zone), followed by the Hall effect, and finally the Ohmic resistivity. As expected, ambipolar diffusion dominates the low-density regions such as the upper layers or the outermost regions (mid-plane included), Ohmic resistivity dominates the innermost and densest regions, and the Hall effect comes into play for intermediate regions. Furthermore, most of the MRI-active layer sits in the ambipolar-dominated region of the disk; implying that the MRI can be sustained there, but it is weakened compared to its ideal limit ($0.1 \lesssim \rm{Am} \lesssim 10$ in the MRI-active layer, as shown in Fig.~\ref{fig:Elsasser_numbers}(a)). Particularly, the upper envelope of the MRI-active layer sits well within the region where the ambipolar magnetic diffusivity dominates; confirming that ambipolar diffusion sets it as seen above. On the other hand, Ohmic resistivity does not dominate the magnetic diffusivities where the lower envelope for the MRI-active layer sits. This result seems different from what we saw above, where we found that mainly Ohmic resistivity is important to determine the lower envelope of the MRI-active layer. This shows that Ohmic resistivity is the most stringent non-ideal MHD effect to overcome for the MRI to operate, and does not need to dominate the magnetic diffusivities to have a strong impact on it.   

Finally, it is quite striking that the Hall effect dominates the magnetic diffusivities in most of the dead zone as well as the inner regions of the MRI-active layer. Particularly, Figure~\ref{fig:accretion_layers}(b) shows that the lower envelope of the MRI-active layer sits in the Hall-dominated region. We further discuss the implications in Appendix~\ref{sect:the hall effect}. 

\subsection{Disk Structure} \label{sect:disk structure}

\subsubsection{Gas} \label{sect:gas structure}

\begin{figure}
\includegraphics[width=0.45\textwidth]{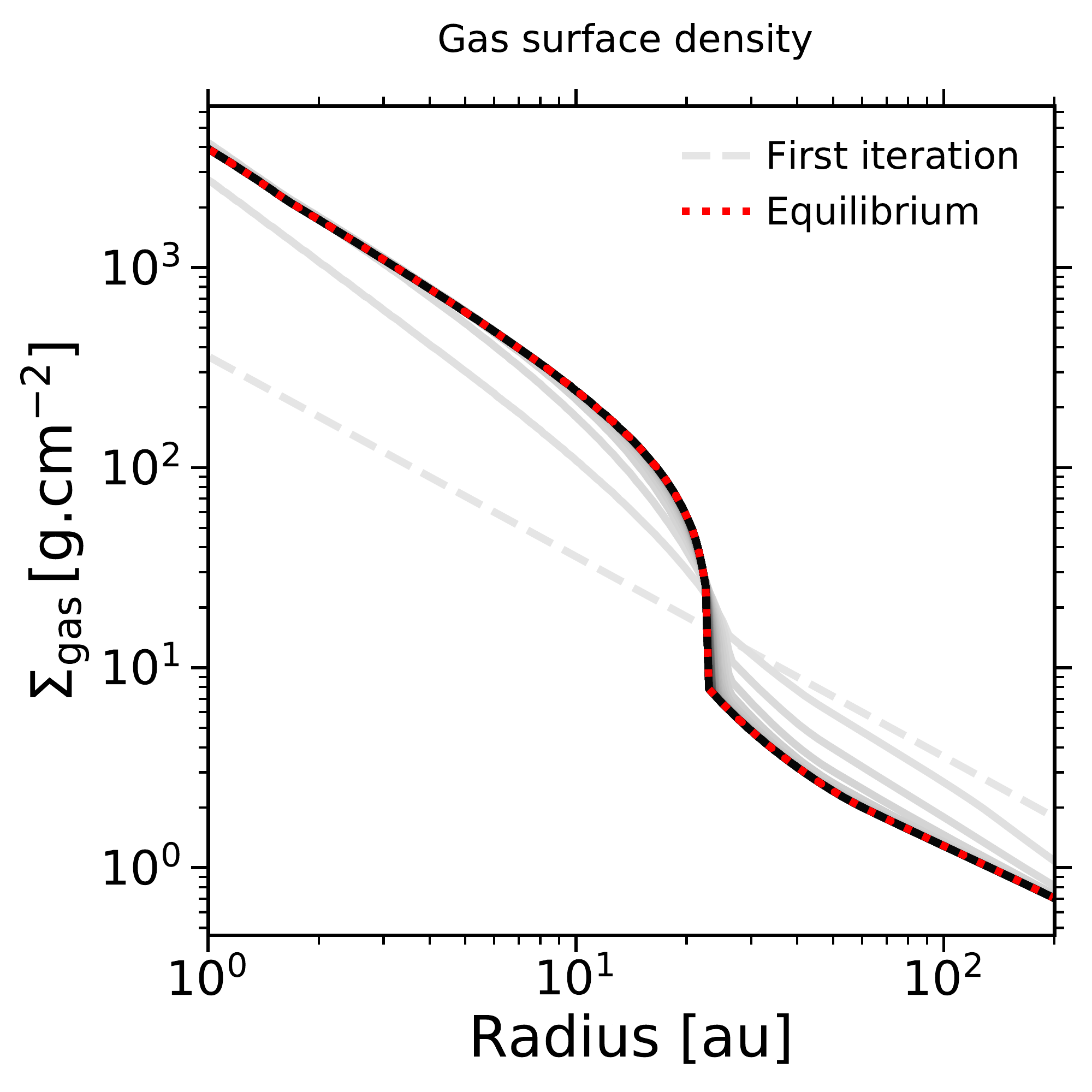}
 \caption{Steady-state gas surface density $\Sigma_{\rm{gas}}$ (Eq.~\eqref{eq:gas surface density}) as a function of radius, for the fiducial model. The gray dashed line shows the initial gas surface density by initially assuming that the effective turbulent parameter $\bar{\alpha}$ is radially constant equal to $\bar{\alpha}_{\rm{init}} = 0.1$. From light-colored solid lines to dark ones, we show how the gas surface density converges through iterations until the equilibrium solution is reached (red dotted line).}
 \label{fig:convergence_SIGMA_gas}
\end{figure}

The gas surface density is self-consistently derived -alongside the effective turbulence $\bar{\alpha}$- to ensure steady-state accretion (Eq.~ \eqref{eq:gas surface density}). Figure~\ref{fig:convergence_SIGMA_gas} shows the steady-state radial profile (red dotted line) obtained through iterations.

It can be divided into two regions: a high-density region for $r \lesssim 23 \,$au (mid-plane dead zone), and a low-density region for $r \gtrsim 23 \,$au (mid-plane MRI-active layer); where $r = 23 \,$au corresponds to the mid-plane dead zone outer edge. We note that the transition at the dead zone outer edge is sharp, and appears as a discontinuity. As we will see in Sect.~\ref{sect:turbulence}, the effective turbulent parameter $\bar{\alpha}$ sharply increases by jumping from a low turbulence state in the non-MRI regions to a high turbulence state in the MRI-active layer at  $r = 23 \,$au. Since the accretion rate must be kept radially constant to ensure steady-state accretion, the gas surface density compensates by sharply decreasing at that location. 

The profile displayed in Fig.~\ref{fig:convergence_SIGMA_gas} is actually expected. Indeed, if one initially assumes the disk to be out of equilibrium with its gas surface density following e.g., the self-similar solution from \citet{1974MNRAS.168..603L}, the accretion rate will be radially variable; where the outer regions (MRI-active layer) accrete more than the inner ones (dead zone) on average. Consequently, by letting viscously evolving the gas, one will find that the gas is depleted from the MRI-active layer to be accumulated into the dead zone. 

\subsubsection{Ionization level} \label{sect:ionization level}

\begin{figure*}
\centering
\includegraphics[width=\textwidth]{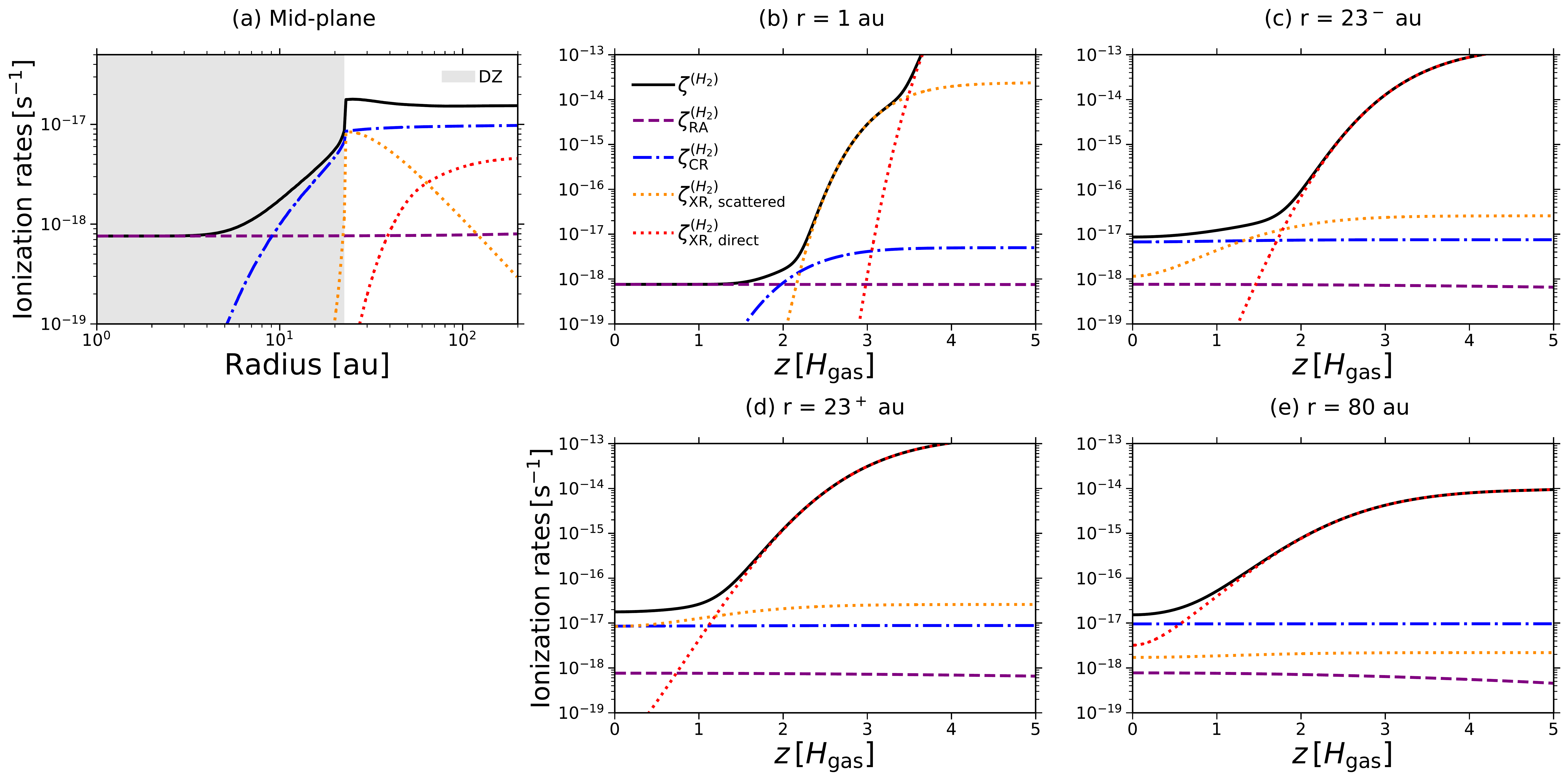}
 \caption{Steady-state ionization rates, for the fiducial model. The total ionization rate for H$_2$ ($\zeta^{(H_2)}$, black solid line) is the sum of the following contributions: the radionuclides ionization rate for H$_2$ ($\zeta^{(H_2)}_{\rm{RA}}$, purple dashed line) defined by Eq.~\eqref{eq:radionuclides ionization rate}; the galactic cosmic rays ionization rate for H$_2$ ($\zeta^{(H_2)}_{\rm{CR}}$, blue dash-dotted line) defined by Eq.~\eqref{eq:cosmic ray ionisation rate}; the scattered X-rays contribution of the X-rays ionization rate for H$_2$ ($\zeta^{(H_2)}_{\rm{XR,\,scattered}}$, dark orange dotted line) defined by Eq.~\eqref{eq:Scattered X-ray ionisation rate}, and its direct X-rays contribution ($\zeta^{(H_2)}_{\rm{XR,\,direct}}$, red dotted line) defined by Eq.~\eqref{eq:Direct X-ray ionisation rate}. \textit{Panel (a)}: Mid-plane ionization rates as a function of radius. The gray shaded area corresponds to the radial locations within the mid-plane dead zone. \textit{Panel (b)}: Ionization rates computed at $r = 1\,$au (a radial location well within the dead zone) as a function of height $z$. \textit{Panel (c)}: Same as panel (b) but for $r = 23^{-} \,$au, corresponding to the radial location right before the mid-plane dead zone outer edge ($r = 23 \,$au). \textit{Panel (d)}: Same as panel (b) but for $r = 23^{+} \,$au, corresponding to the radial location right after the mid-plane dead zone outer edge. \textit{Panel (e)}: Same as panel (b) but for $r = 80 \,$au, corresponding to a radial location well within the MRI-active layer. A 2D map of the total ionization rate for H$_2$ ($\zeta^{(H_2)}$) and its different contributions can be found in Fig.~\ref{fig:2D_map_disk_ionization}.}
 \label{fig:disk_ionization}
\end{figure*}

\begin{figure*}
\centering
\includegraphics[width=0.80\textwidth]{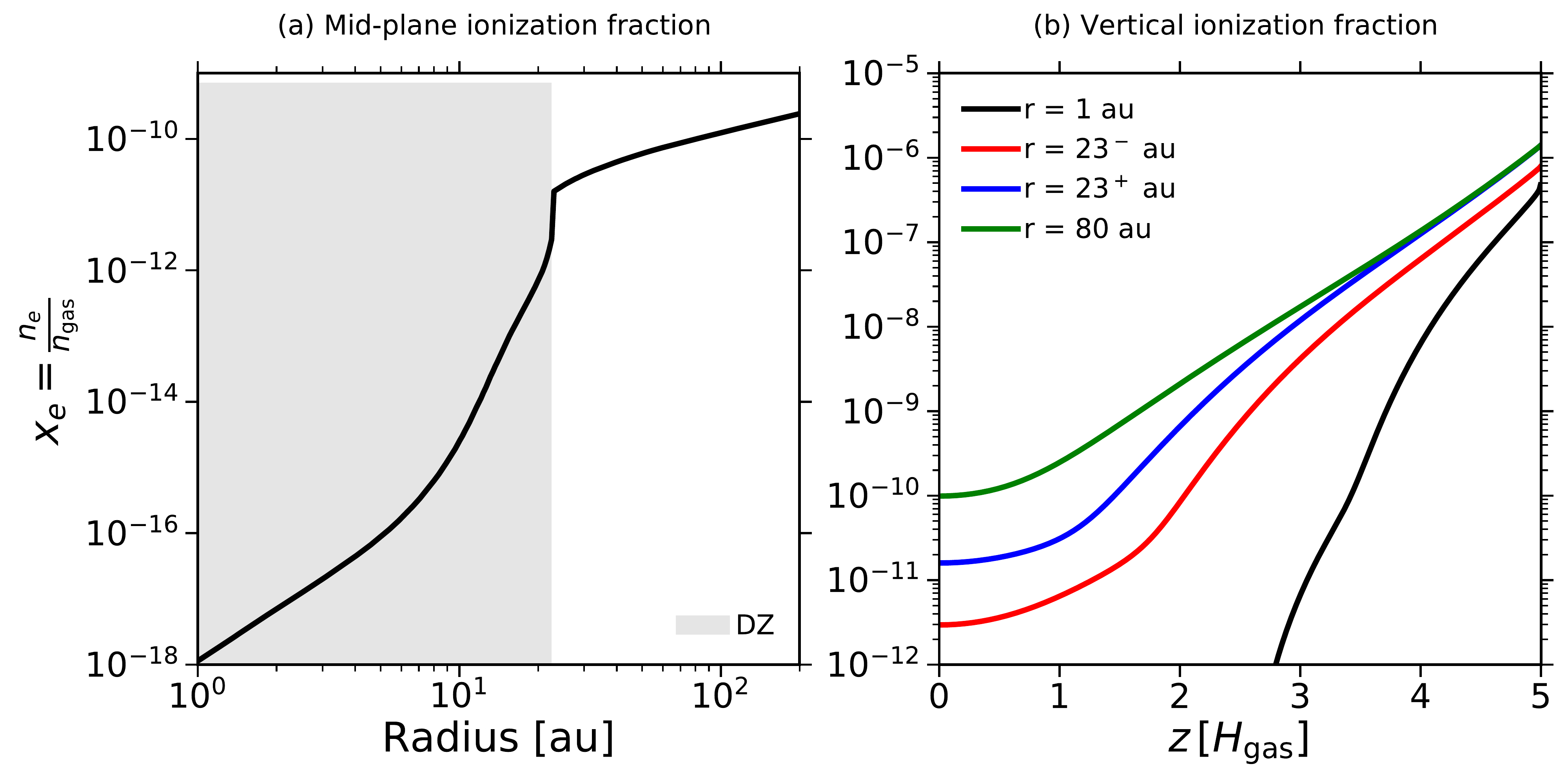}
 \caption{Steady-state ionization fraction $x_e$ ($n_e$ is defined by Eq.~\eqref{eq:free electrons number density}), for the fiducial model. \textit{Panel (a)}: Mid-plane ionization fraction as a function of radius. The gray shaded area corresponds to the radial locations within the mid-plane dead zone. \textit{Panel (b)}: Vertical profiles for the ionization fraction at the same radial locations as in Fig.~\ref{fig:disk_ionization}. The black solid line corresponds to $r = 1\,$au. The red solid line corresponds to $r = 23^{-}\,$au. The blue solid line corresponds to $r = 23^{+}\,$au. The green solid line corresponds to $r = 80\,$au.}
 \label{fig:disk_chemistry}
\end{figure*}

Since the gas content is mostly located within the dead zone, the ionization level is highly heterogeneous across the protoplanetary disk. Figures~\ref{fig:disk_ionization} and \ref{fig:disk_chemistry} show the steady-state total ionization rates for H$_2$ for the different non-thermal ionization sources considered (stellar X-rays, galactic cosmic rays, radionuclides) and the steady-state ionization fraction (free electron abundance), respectively.

Deep inside the dead zone ($r \lesssim 5 \,$au and $z \lesssim 1.5\, H_{\rm{gas}}$), the ionization solely comes from the decay of short/long-lived radionuclides (Figs.~\ref{fig:disk_ionization}(a) and \ref{fig:disk_ionization}(b)). In those innermost regions, the gas surface density is so high that the other non-thermal ionization sources cannot penetrate at all, resulting in a very low ionization fraction (Fig.~\ref{fig:disk_chemistry}(a)). For $9 \, \rm{au} \lesssim r \lesssim 23 \,$au and $z \lesssim 1.5\, H_{\rm{gas}}$, galactic cosmic rays can penetrate deep enough to dominate the ionization process, leading to an increase of the ionization fraction (Fig.~\ref{fig:disk_chemistry}(a), Fig.~\ref{fig:disk_chemistry}(b)). Right at the mid-plane dead zone outer edge ($r = 23 \,$au), the gas becomes tenuous enough for scattered stellar X-rays to come into play and contribute as much as the galactic cosmic rays (Fig.~\ref{fig:disk_ionization}(a)). As a result, the total ionization rate for H$_2$ gets a "boost" at the transition mid-plane dead zone/MRI-active layer (comparing the black solid curves in Figs.~\ref{fig:disk_ionization}(c) and \ref{fig:disk_ionization}(d)). We further notice that stellar X-rays (both the scattered and direct contributions) can overall penetrate deeper for $r \gtrsim 23 \,$au due to lower gas column densities. Consequently, the ionization fraction sharply increases at the transition dead zone/MRI-active layer (comparing the red and blue curves of Fig.~\ref{fig:disk_chemistry}(b)); resulting in enough charged particles in the gas-phase for the magnetic field to couple with, and triggering the MRI from the mid-plane (not only in the surface layers as it is the case for $r \lesssim 23 \,$au). Although the total ionization rate for H$_2$ is utterly dominated by stellar X-rays in the disk atmosphere (either through the scattered X-rays or direct X-rays contribution), Fig.~\ref{fig:disk_ionization}(a) shows that galactic comic rays always dominate at the mid-plane for $r \gtrsim 23\,$au. This behavior can be understood as follows: On one hand, the total stellar X-rays ionization rate (sum of the scattered and direct X-rays contributions) decreases over radius ($\propto r^{-2.2}$) unlike the ionization rate from galactic cosmic rays. Even though the gas is tenuous enough for the stellar scattered X-rays to penetrate deep enough, this emission has already lost most of its energy by traveling up to those regions. On the other hand, the stellar direct X-rays have a very small penetration depth -although they are the most energetic source of ionization considered here. The gas is thus not tenuous enough for them to efficiently ionize the mid-plane. Interestingly, we notice that the total ionization rate for H$_2$ roughly saturates at a value equal to $\zeta^{(H_2)} \approx 1.5 \times 10^{-17} \, \rm{s}^{-1}$ in the mid-plane MRI-active layer (Fig.~\ref{fig:disk_ionization}(a)): the galactic cosmic rays ionization rate saturates at its unattenuated value $ 10^{-17} \, \rm{s}^{-1}$, and the total stellar X-rays ionization rate roughly adds up $0.5 \times 10^{-17} \, \rm{s}^{-1}$). Consequently, the ionization fraction monotonously increases in the mid-plane MRI-active layer (Fig.~\ref{fig:disk_chemistry}(a)).

Finally, we use the fit from \citet{2009ApJ...701..737B} to compute the stellar X-rays ionization rate. Since the gas content is mostly located in the inner regions of the protoplanetary disk, the inner regions might absorb stellar direct X-rays before they could actually reach the outer regions. This self-shadowing event could thus prevent stellar X-rays to efficiently ionize the outer regions of the protoplanetary disk; resulting in an overall lower ionization level. We further discuss this point in Appendix~\ref{appendix:direct X-Ray ionisation rate}.

In summary, we find that (see Fig.~\ref{fig:2D_map_disk_ionization}): (1) the decay of short/long-lived radionuclides dominates the ionization process in the innermost regions of the dead zone; (2) galactic cosmic rays dominate the outermost regions of the dead zone, the whole mid-plane MRI-active layer, and most of the regions right about the mid-plane MRI-active layer; (3) stellar scattered X-rays dominate the lower layers of the disk atmosphere that sit right above the dead zone, and contribute as much as galactic cosmic rays right at the mid-plane dead zone outer edge; (4) stellar direct X-rays dominate the ionization process in the upper layers of the disk atmosphere, including the lower layers of the outermost regions; (5) the MRI is mainly driven by stellar X-rays, except close to the mid-plane where it is primarily driven by galactic cosmic rays.

\subsubsection{Magnetic Field Strength} \label{sect:magnetic field strength}

A lower limit on the magnetic field strength is the mean galactic field roughly equal to $ 10^{-5} \,$Gauss \citep[e.g.,][]{1996Natur.379...47B}, whereas an upper limit is the equipartition field (field such that the gas thermal pressure and the magnetic pressure are equal) which makes the shortest unstable MRI mode no longer fit within the disk gas scale height (hence, no MRI activity possible at all). In this study, we constrain the required magnetic field strength $B$ such that the MRI activity is maximally efficient permitted by Ohmic resistivity and ambipolar diffusion, at any locations in the disk.

\begin{figure*}
\centering
\includegraphics[width=0.80\textwidth]{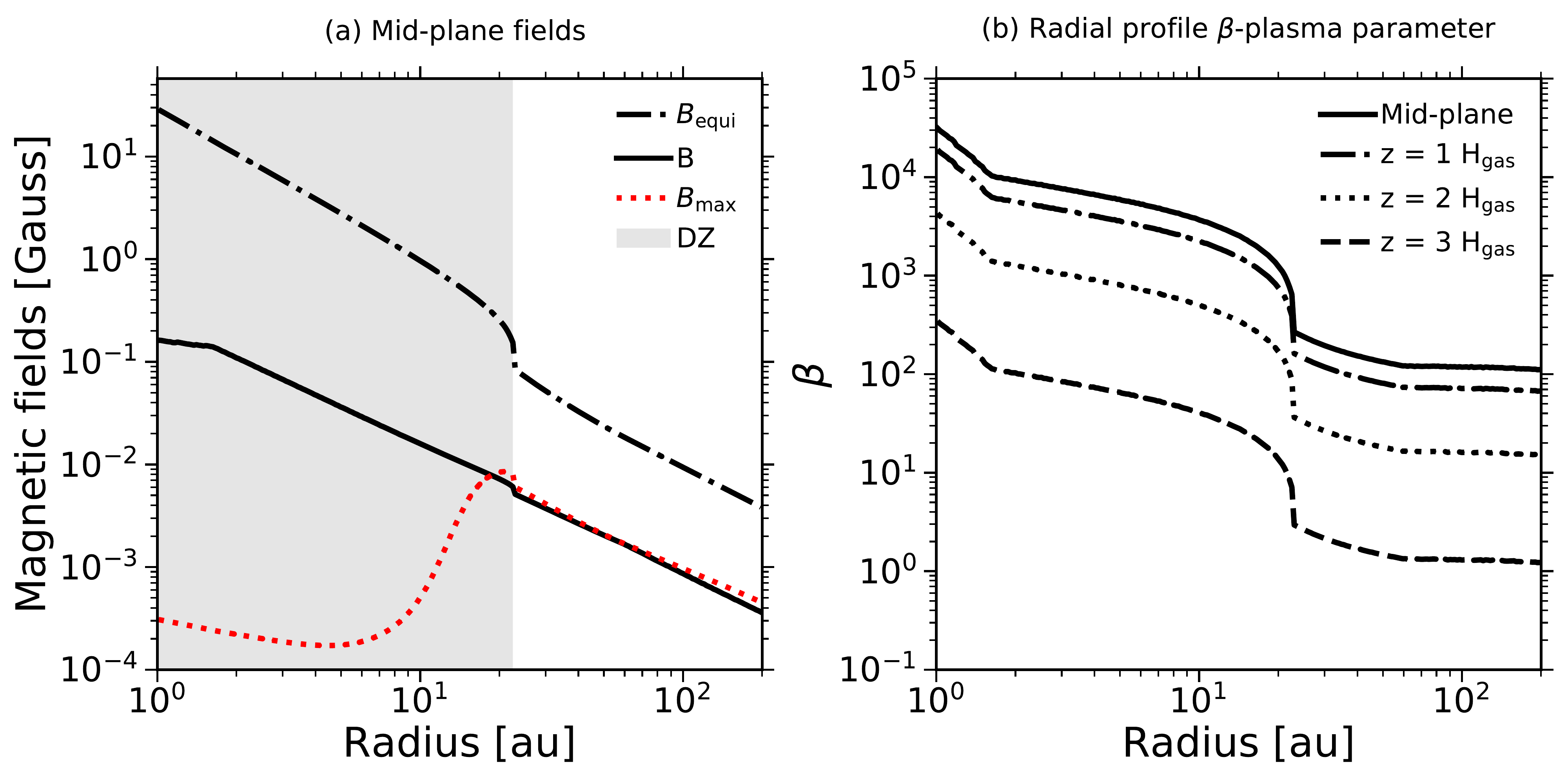}
 \caption{Steady-state quantities describing the disk magnetization, for the fiducial model. \textit{Panel (a)}: Mid-plane magnetic fields strength as a function of radius. The black solid line corresponds to $B$, the optimal r.m.s. magnetic field strength that is required for the MRI to be maximally efficient permitted by Ohmic resistivity and ambipolar diffusion (Sect.~\ref{sect:B field}). The red dotted line corresponds to $B_{\rm{max}} = \sqrt{\frac{8 \pi \rho_{\rm{gas}} c_s^{2}}{\beta_{\rm{min}}}}$, the threshold for the magnetic field strength from and above which the MRI-driven turbulence with such a field strength is prohibited by ambipolar diffusion, given the ambipolar condition ($\beta > \beta_{\rm{min}}$) of our active MRI criteria (Sect.~\ref{sect:criteria for active MRI}). The black dash-dotted line corresponds to $B_{\rm{equi}} = \sqrt{8 \pi \rho_{\rm{gas}} c_s^{2}}$, the equipartition field. The gray shaded area corresponds to the radial locations within the mid-plane dead zone. \textit{Panel (b)}: Radial profiles for the $\beta$-plasma parameter defined from the optimal r.m.s. magnetic field strength $B$ (Sect.~\ref{sect:criteria for active MRI}); for the mid-plane (solid line), $z = 1 \, H_{\rm{gas}}$ (dash-dotted line), $z = 2 \, H_{\rm{gas}}$ (dotted line) and $z = 3 \, H_{\rm{gas}}$ (dashed line).}
 \label{fig:magnetic_fields}
\end{figure*}

Figure~\ref{fig:magnetic_fields}(a) shows the various mid-plane fields of interest as a function of distance from the star. As expected, the highest value for $B$ is well below the equipartition field $B_{\rm{equi}} = \sqrt{8 \pi \rho_{\rm{gas}} c_s^{2}}$, while its lowest value is well above the mean galactic field. 
We notice that the optimal r.m.s. magnetic field $B$ displays a small discontinuity located at the mid-plane dead zone outer edge ($r = 23 \,$au). Although non-physical, this mathematical discontinuity expresses the fact that $B$ jumps from a solution in the dead zone where the turbulence is weak to another in the MRI-active layer where the turbulence is stronger. In the mid-plane MRI layer ($r \gtrsim 23 \,$au), $B$ is very close (but strictly lower) to $B_{\rm{max}} = \sqrt{\frac{8 \pi \rho_{\rm{gas}} c_s^{2}}{\beta_{\rm{min}}}}$, which corresponds to the maximal field strength from and above which the MRI-driven turbulence with such a field strength is prohibited by ambipolar diffusion (see Eqs.~\eqref{eq:magnetic field criterion for MRI} and \eqref{eq:beta min}). This is expected since we have constructed the field strength such that the MRI operates at its maximal efficiency. We further notice that the optimal r.m.s. magnetic field $B$ is higher than $B_{\rm{max}}$ for $r \lesssim 18 \,$au. It thus implies that the ambipolar condition is also not met in most of the mid-plane dead zone where the MRI is suppressed by Ohmic resistivity, as we saw in Sect.~\ref{sect:behavior of the different accretion layers} 

Figure~\ref{fig:magnetic_fields}(b) shows the $\beta$-plasma parameter radial profiles at various heights in the disk. Since the optimal r.m.s. magnetic field $B$ is assumed to be vertically constant, $\beta$-plasma decreases exponentially with increasing height; hence, the radial profiles at different disk gas scale height are re-scaled versions of the mid-plane profile (solid line). In the mid-plane dead zone ($r \lesssim 23 \,$au), the mid-plane $\beta$-plasma parameter has a mean value of roughly $8 \times 10^3$. As expected, the mid-plane dead zone is thus weakly magnetized, since the MRI cannot operate. In the mid-plane MRI-active layer ($r \gtrsim 23 \,$au), the mid-plane $\beta$-plasma parameter has a mean value of roughly $142$, implying that these regions are moderately magnetized.  We further notice that the mid-plane $\beta$ saturates in the MRI-active layer from $r \gtrsim 60 \,$au. This behavior can be understood as follows: As seen in the last section, the mid-plane ionization fraction increases with radius (Fig.~\ref{fig:disk_chemistry}(a)); resulting in the magnetic field to more efficiently couple to the gas which leads to a possible stronger MRI-driven local turbulence. However, ambipolar diffusion is the dominant non-ideal MHD effect in the regions where the gas is scarce, as we saw in Sect.~\ref{sect:behavior of the different accretion layers}. In such regions, a stronger local MRI-driven turbulence implies a stronger field strength ($\alpha \propto \beta^{-1}$), resulting in that stronger local turbulence to be prohibited by ambipolar diffusion. Consequently, the mid-plane $\beta$-plasma parameter saturating in the MRI-active layer from $r \gtrsim 60 \,$au expresses that the positive feedback on the MRI from a higher gas ionization degree is compensated by the negative feedback from ambipolar diffusion that prohibits turbulence with stronger magnetic fields.

\subsection{Turbulence} \label{sect:turbulence}

\begin{figure*}
\centering
\includegraphics[width=\textwidth]{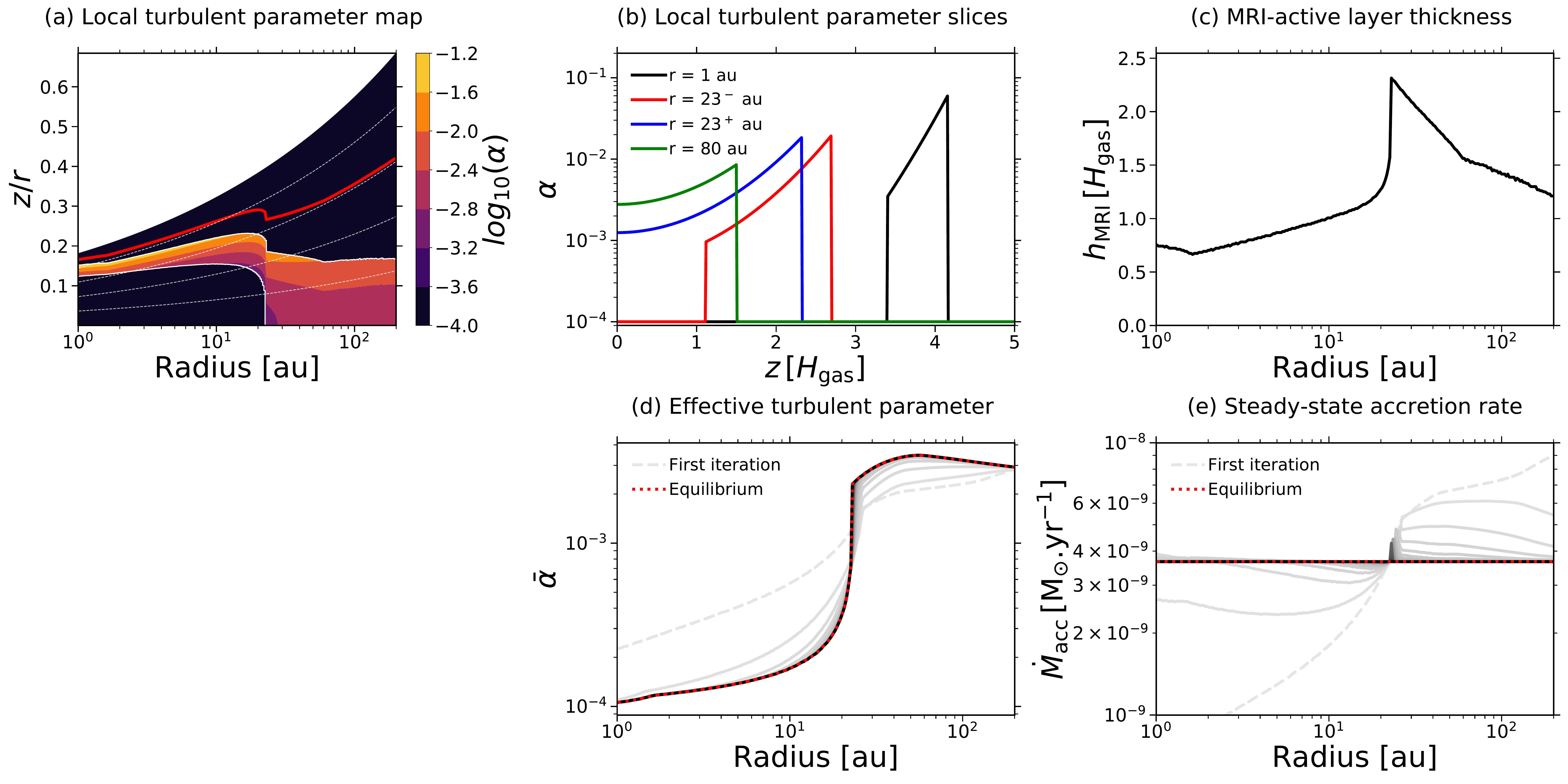}
 \caption{Steady-state quantities describing the turbulence state in the disk, for the fiducial model. \textit{Panel (a)}: Local turbulence parameter $\alpha$ (Eq.~\eqref{eq:local turbulent parameter}) as a function of location in the disk. The region above the red solid line corresponds to where $\beta > 1$. The white dashed lines correspond to the surfaces $z = 1 \, H_{\rm{gas}}$, $z = 2 \, H_{\rm{gas}}$, $z = 3 \, H_{\rm{gas}}$ and $z = 4 \, H_{\rm{gas}}$; from bottom to top, respectively. \textit{Panel (b)}: Vertical profiles for the local turbulent parameter $\alpha$ at various radial locations, with color codes as in Fig.~\ref{fig:disk_chemistry}. \textit{Panel (c)}: MRI-active layer thickness $h_{\rm{MRI}}$ as a function of radius, expressed in terms of the disk gas scale height $H_{\rm{gas}}$. \textit{Panel (d)}: Pressure-weighted vertically-integrated turbulent parameter $\bar{\alpha}$ (Eq.~\eqref{eq:alpha bar}) as a function of radius. The gray dashed line shows the profile after the first iteration. From light-colored solid lines to dark ones, we show how the pressure-weighted vertically-integrated turbulent parameter converges through iterations until the equilibrium solution is reached (red dotted line). \textit{Panel (e)}: Same as in panel (d), but for the accretion rate $\dot{M}_{\rm{acc}}$ (Eq.~\eqref{eq:Accretion rate2}).}
 \label{fig:turbulence}
\end{figure*}

In this study, the origin of turbulence is the radial transport of angular momentum redistributed within the protoplanetary disk by standard viscous torques: MRI magnetic torques in the active regions, and hydrodynamic torques in the non-active regions. Figure~\ref{fig:turbulence} shows various quantities of interest describing the properties of the steady-state disk turbulence. 

As expected, the disk is significantly more turbulent in the MRI-active layer than it is in the dead or zombie zone. From Figs.~\ref{fig:turbulence}(a) and \ref{fig:turbulence}(b), we notice that the local turbulent parameter $\alpha$ is higher at large radii and heights, in the MRI-active layer. Indeed, the ionization level is higher in those regions because the gas surface densities are lower, leading to more penetration by the non-thermal ionization sources (stellar X-rays in the disk atmosphere, and galactic cosmic rays at the disk mid-plane) as well as slower recombination. At the mid-plane dead zone outer edge ($r = 23 \,$au), the turbulence jumps from a low regime in the mid-plane dead zone to a high regime in the mid-plane MRI-active layer. More generally, this very steep transition happens at the boundaries between active and non-active regions; namely, at the upper and lower envelopes of the MRI-active layer. We obtain such a behavior due to our assumption on the MRI: either in a saturation level allowed by the non-ideal MHD effects, or completely shut off (see criteria in Sect.~\ref{sect:criteria for active MRI}). Due to this sharp change, Fig.~\ref{fig:turbulence}(d) shows that the effective turbulent parameter $\bar{\alpha}$ displays a small discontinuity at the mid-plane dead zone outer edge ($r = 23 \,$au); resulting in the discontinuity seen in the steady-state gas surface density (Fig.~\ref{fig:convergence_SIGMA_gas}). We will further discuss such a sharp transition in the gas surface density profile in Sect.~\ref{sect:discussion gas structure}. 

Since there is a thin turbulent MRI-active layer that sits right above the dead zone for $r \lesssim 23 \,$au, the effective turbulent parameter $\bar{\alpha}$ is higher than the minimum value $\alpha_{\rm{hydro}} = 10^{-4}$ in the dead zone, and increases toward the dead zone outer edge (see red dotted line in Fig.~\ref{fig:turbulence}(d)). If not located too high above the dead zone, active regions can indeed make the dead zone effectively more turbulent by launching turbulent waves \citep[e.g.,][]{2003ApJ...585..908F, 2009ApJ...704.1239O, 2010A&A...515A..70D, 2011ApJ...742...65O}. Although such waves are not described by our disk model, the effective turbulent parameter $\bar{\alpha}$ encodes the influence of the upper layers on the turbulence, since this quantity is the pressure-weighted vertical average of the local turbulence. In the innermost regions of the dead zone, $\bar{\alpha} \approx \alpha_{\rm{hydro}}$ because the MRI-active layer represents a very small fraction of the gas column densities. Since the vertical extent of the dead zone diminishes for larger radii, the active regions right above the dead zone increasingly represent a higher fraction of the gas column densities toward the dead zone outer edge; meaning that the contribution from the MRI-active layer on the effective turbulent parameter $\bar{\alpha}$ increasingly becomes larger. Overall though, we find that the mean value of $\bar{\alpha}$ is roughly $1.7 \times 10^{-4}$ in the dead zone. Therefore, the general behavior of the turbulence in these regions is a low regime, as expected.

Conversely, the red dotted line in Fig.~\ref{fig:turbulence}(d) shows that the mean value of $\bar{\alpha}$ in the outer region of the MRI-active layer ($r \gtrsim 23 \,$au) is roughly $3.1 \times 10^{-3}$, which is $19$ times higher than the mean value stated above for the dead zone. Interestingly, we notice that the effective turbulent parameter $\bar{\alpha}$ decreases for $r \gtrsim 60 \,$au. This seems to be at odds with the fact that the local turbulent parameter $\alpha$ saturates and is roughly constant for a given height at those radii (because $\alpha \propto \beta^{-1}$ and $\beta$ is roughly constant for a given height at those radii, as seen in Fig.~\ref{fig:magnetic_fields}(b)). To better understand this behavior, the MRI-active layer thickness $h_{\rm{MRI}}$ is shown in Fig.~\ref{fig:turbulence}(c). We notice that it diminishes comparatively to $H_{\rm{gas}}$ for $r \gtrsim 23 \,$au. Indeed, $h_{\rm{MRI}}$ increases toward larger radii, but slowly since local turbulence with stronger than permitted magnetic fields would be considered if the active layer was thicker at a given $r$, which is prohibited by ambipolar diffusion. As a result, the quantity $h_{\rm{MRI}}/H_{\rm{gas}}$ slowly decreasing for $r \gtrsim 23\,$au implying that the MRI-active layer increasingly represents a lower fraction of the gas column densities for increasing radii from $r \gtrsim 23\,$au. For a local turbulent parameter $\alpha$ roughly constant for $r \gtrsim 60\,$au and a reduction in how much the MRI-active layer contributes to the gas column densities, the effective turbulent parameter $\bar{\alpha}$ decreases for $r \gtrsim 60\,$au. 

In summary, we find that the effective turbulent parameter $\bar{\alpha}$ is strong when both the local MRI-driven turbulence $\alpha$ and the MRI-active layer thickness $h_{\rm{MRI}}$ (comparatively to $H_{\rm{gas}}$) are significant: the former tells us how much turbulence can be locally driven by the MRI, whereas the latter describes how much the MRI-active layer contributes to the gas column density at a given radius. 

We emphasize that the accretion rate $\dot{M}_{\rm{acc}}$ presented in Fig.~\ref{fig:turbulence}(e) represents the highest value possibly reachable given the set of criteria for active MRI employed, since we adopt the required magnetic field strength such that the MRI activity is maximally efficient (see Sect.~\ref{sect:B field}). In Appendix~\ref{sect:effect magnetic field strength}, we explore what would be the derived accretion rates for either weaker or stronger magnetic field strengths, and show that our choice for the magnetic field strength leads to the highest value for the accretion rate as expected. 

\section{Results - Parameter Study} \label{sect:results-parameter study}

We perform a series of simulations exploring the effect of various model parameters on the fiducial model described in the last section: total disk gas mass $M_{\rm{disk}}$ (Sect.~\ref{sect:effect total disk gas mass}), representative grain size $a_{\rm{dust}}$ (Sect.~\ref{sect:effect grain size}), vertically-integrated dust-to-gas mass ratio $f_{\rm{dg}}$ (Sect.~\ref{sect:effect dust-to-gas mass ratio}), and stellar properties -mass $M_{\star}$ and luminosity $L_{\star}$- (Sect.~\ref{sect:effect stellar properties}). In each series, we derive and show key quantities reached by our steady-state accretion disk model as a function of distance from the central star. Those quantities are: (a) the pressure-weighted vertically-integrated turbulent parameter $\bar{\alpha}$; (b) the gas surface density $\Sigma_{\rm{gas}}$; (c) the mid-plane total ionization rate $\zeta$; (d) the mid-plane optimal r.m.s. magnetic field strength $B$ that maximizes the MRI activity; (e) the accretion rate $\dot{M}_{\rm{acc}}$; (f) the mid-plane location of the dead zone outer edge $R_{\rm{DZ}}$ (corresponding to the dead zone maximal radial extent). A detailed description of the results are given in the following sections. In Appendix~\ref{appendix:Follow-up parameter study}, we additionally investigate the effect of magnetic field strength, hydrodynamic turbulent parameter $\alpha_{\rm{hydro}}$, gas temperature model $T(r)$, as well as the combined effect of stellar properties and total disk gas mass.

\subsection{The Effect of Disk Mass} \label{sect:effect total disk gas mass}

\begin{figure*}
\centering
\includegraphics[width=\textwidth]{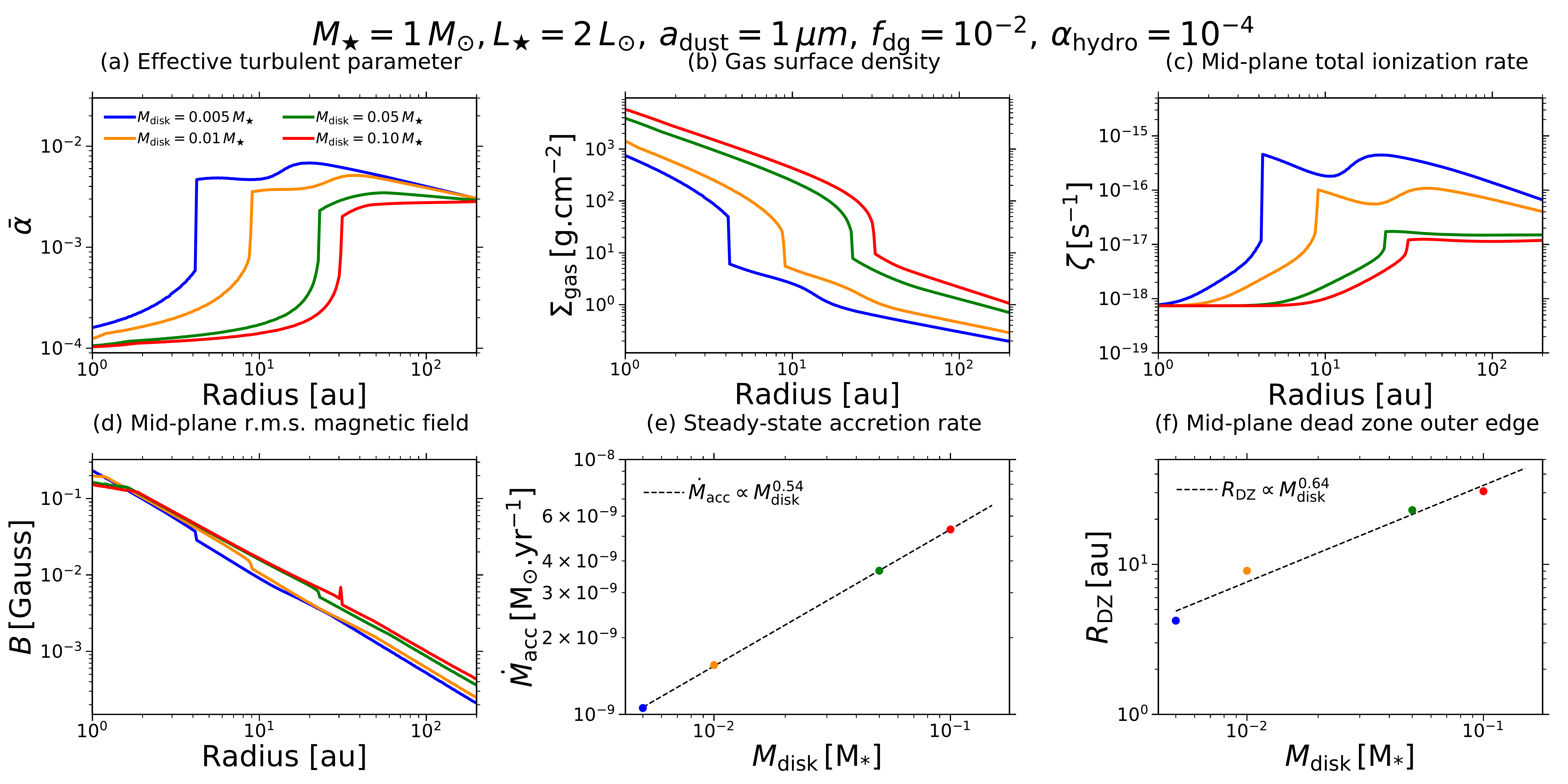}
 \caption{Effect of total disk gas mass on the equilibrium solution by varying the parameter $M_{\rm{disk}}$ from $0.005 \,M_\star$ (gas-poor disk) to $0.10 \,M_\star$ (gas-rich disk close to the GI limit): the case $M_{\rm{disk}} = 0.005 \,M_\star$ is shown in blue; $M_{\rm{disk}} = 0.01 \,M_\star$ in dark orange; $M_{\rm{disk}} = 0.05 \,M_\star$ in green (fiducial model); and $M_{\rm{disk}} = 0.10 \,M_\star$ in red. The panels show the various steady-state profiles of some key quantities, for the model parameters $M_{\star} = 1 \, M_{\odot}$, $L_{\star} = 2 \, L_{\odot}$, $a_{\rm{dust}} = 1\, \mu$m, $f_{\rm{dg}} = 10^{-2}$ and $\alpha_{\rm{hydro}} = 10^{-4}$. \textit{Panel (a)}: Pressure-weighted vertically-integrated turbulent parameter $\bar{\alpha}$ (Eq.~\eqref{eq:alpha bar}). \textit{Panel (b)}: Gas surface density $\Sigma_{\rm{gas}}$ (Eq.~\eqref{eq:gas surface density}). \textit{Panel (c)}: Mid-plane total ionization rate $\zeta$ (Eq.~\eqref{eq:total ionization rate}). \textit{Panel (d)}: Mid-plane optimal r.m.s. magnetic field strength $B$ that is required for the MRI to be maximally efficient permitted by Ohmic resistivity and ambipolar diffusion. \textit{Panel (e)}: Accretion rate $\dot{M}_{\rm{acc}}$ (Eq.~\eqref{eq:Accretion rate2}). \textit{Panel (f)}: Mid-plane radial dead zone outer edge location $R_{\rm{DZ}}$, defined as the mid-plane radial location where both the Ohmic and ambipolar condition are met for the first time (Sect.~\ref{sect:criteria for active MRI}). The black dashed lines in panels (e) and (f) correspond to a power-law fit of $\dot{M}_{\rm{acc}}$ and $R_{\rm{DZ}}$ as a function of $M_{\rm{disk}}$, respectively.}
 \label{fig:effect_M_disk}
\end{figure*}

The amount of gas enclosed within the protoplanetary disk sets many of the disk structure properties (e.g., ionization level). It is then expected that the variation in total disk gas mass will significantly impact the equilibrium solution. To quantify this further, we run a set of simulations where $M_{\rm{disk}}$ varies from a gas-poor disk motivated by the ALMA observations in the continuum \citep[see e.g.,][assuming a gas-to-dust ratio of $100$]{2016ApJ...831..125P} to a gas-rich disk close to the gravitational instability (GI) limit. Our disk model does not treat gravito-turbulence, hence the GI limit sets the upper limit for the total disk gas mass. We determine such an upper limit by choosing a total disk gas mass value so that $Q \gtrsim 2$ is satisfied at every radii $r$, where $Q$ is the Toomre parameter \citep[][]{1964ApJ...139.1217T}. In practice, we choose $M_{\rm{disk}}$ in the list $\{0.005, 0.01, 0.05, 0.10\} \,M_\star$ with $M_{\star} = 1\, M_{\odot}$. The results can be seen in Fig.~\ref{fig:effect_M_disk}. 

As expected for a steady-state viscously accreting protoplanetary disk, the accretion rate and total disk gas mass follow a tight correlation (Fig.~\ref{fig:effect_M_disk}(e)). Particularly, we find that $\dot{M}_{\rm{acc}} \propto M^{0.54}_{\rm{disk}}$. In our model, $c_s^2 \propto T \propto r^{-1/2}$. Consequently, Eq.~\eqref{eq:Accretion rate2} gives:
\begin{equation*}
    \dot{M}_{\rm acc} \propto \frac{L_\star^{1/4}M_{\rm disk}}{\sqrt{M_\star}\int_{r_{\rm min}}^{r_{\rm min}}\bar{\alpha}^{-1}dr}.
\end{equation*}

\noindent For fixed $L_\star$ and $M_\star$, the quantity $\dot{M}_{\rm acc}/M_{\rm disk}$ is entirely determined by the integral $\int_{r_{\rm min}}^{r_{\rm min}}\bar{\alpha}^{-1}dr$. We find that $\left[\int_{r_{\rm min}}^{r_{\rm min}}\bar{\alpha}^{-1}dr \right] \propto M^{0.46}_{\rm{disk}}$, hence the relation between $\dot{M}_{\rm{acc}}$ and $R_{\rm{DZ}}$. Physically, this relation can be simply understood: a higher total disk gas mass implies more gas available to be accreted onto the central star, hence a higher accretion rate.

Additionally, we find a similar correlation between the dead zone maximal radial extent and the total disk gas mass $R_{\rm{DZ}} \propto M^{0.64}_{\rm{disk}}$. Figure~\ref{fig:effect_M_disk}(f) shows that the dead zone maximal radial extent (corresponding to the mid-plane dead zone outer edge) can be as small as $\approx 4 \,$au for low total disk gas masses, and as large as $\approx 31 \,$au for high total disk gas masses. For a massive disk, the gas surface density is higher at every radii; leading to a lower ionization state compared to a low total disk gas mass (Fig.~\ref{fig:effect_M_disk}(c)). Since the ionization fraction is lower, there are less charged particles in the gas-phase for the magnetic field to couple with. Consequently, the mid-plane dead zone outer edge is located further away because the location where the ionization level is enough to trigger the MRI is reached further away. With the same argument on the overall lower gas ionization degree, we can explain why the mean value of the overall effective turbulent parameter $\bar{\alpha}$ is lower for disks with a higher total disk gas mass. 

Interestingly, we notice that $\bar{\alpha}$ displays a different shape for the two less massive disks. Instead of only steeply increasing right at the dead zone outer edge, it also displays a second shallow but significant increase in the MRI-active layer. We can explain the shape displayed for the two less massive disks by looking at their mid-plane total ionization rate (Fig.~\ref{fig:effect_M_disk}(c)):  For such low total disk gas masses, the ionization process (mid-plane included) is dominated by stellar X-rays. The scattered X-rays contribution dominates right at the dead zone outer edge (first steep increase in $\zeta$) but quickly decreases for larger radii, while the direct X-rays contribution increases from the dead zone outer edge but cannot penetrate deep enough yet. As a result, the mid-plane total ionization rate immediately decreases beyond the dead zone outer edge, until the direct X-rays contribution can finally penetrate deep enough to compensate for the decrease of the scattered X-rays contribution and provide another boost to the mid-plane ionization (second shallow increase in $\zeta$). 

Furthermore, we find that that the optimal r.m.s. magnetic field strength $B$ is weaker for a lower total disk gas mass once the mid-plane MRI-active layer is reached due to ambipolar diffusion (Fig.~\ref{fig:effect_M_disk}(d)). Since a lower total disk gas mass implies a higher gas ionization degree at the mid-plane, the ambipolar Elsasser number Am is slightly higher once the mid-plane MRI-active layer is reached (for example, we find that Am can be up to three times higher for $M_{\rm{disk}} = 0.005\,M_{\star}$ compared to $M_{\rm{disk}} = 0.05\,M_{\star}$). This implies that $\beta_{\rm{min}}(\rm{Am})$ is slightly lower for a lower total disk gas mass in such regions (it can be up to four times lower for $M_{\rm{disk}} = 0.005\,M_{\star}$ compared to $M_{\rm{disk}} = 0.05\,M_{\star}$). However, the gas surface densities are much lower across the disk for a lower total disk gas mass, leading to an overall lower gas volume density $\rho_{\rm{gas}}$ (it is at best four times lower for $M_{\rm{disk}} = 0.005\,M_{\star}$ compared to $M_{\rm{disk}} = 0.05\,M_{\star}$). Since the threshold for the magnetic field strength from and above which the MRI-driven turbulence with such a field strength is prohibited by ambipolar diffusion ($B_{\rm{max}}$) is such that $B_{\rm{max}} \propto \sqrt{\frac{\rho_{\rm{gas}}}{\beta_{\rm{min}}(\rm{Am})}}$, and $\rho_{\rm{gas}}$ decreases faster than $\beta_{\rm{min}}(\rm{Am})$ decreases for a lower total disk gas mass, we conclude that $B_{\rm{max}}$ is lower, hence a lower $B$.            

Finally, a striking result is that the effective turbulence level roughly becomes independent of the total disk gas mass considered at $r \approx 200\,$au (all the $\bar{\alpha}$ converge toward the same value $\approx 3 \times 10^{-3}$). This shows how ambipolar diffusion operates as a regulator to compensate for the positive feedback on the MRI from a higher gas ionization degree for a lower total disk gas mass, as explained in the previous paragraph.

\subsection{The Effect of Grain size} \label{sect:effect grain size}

\begin{figure*}
\centering
\includegraphics[width=\textwidth]{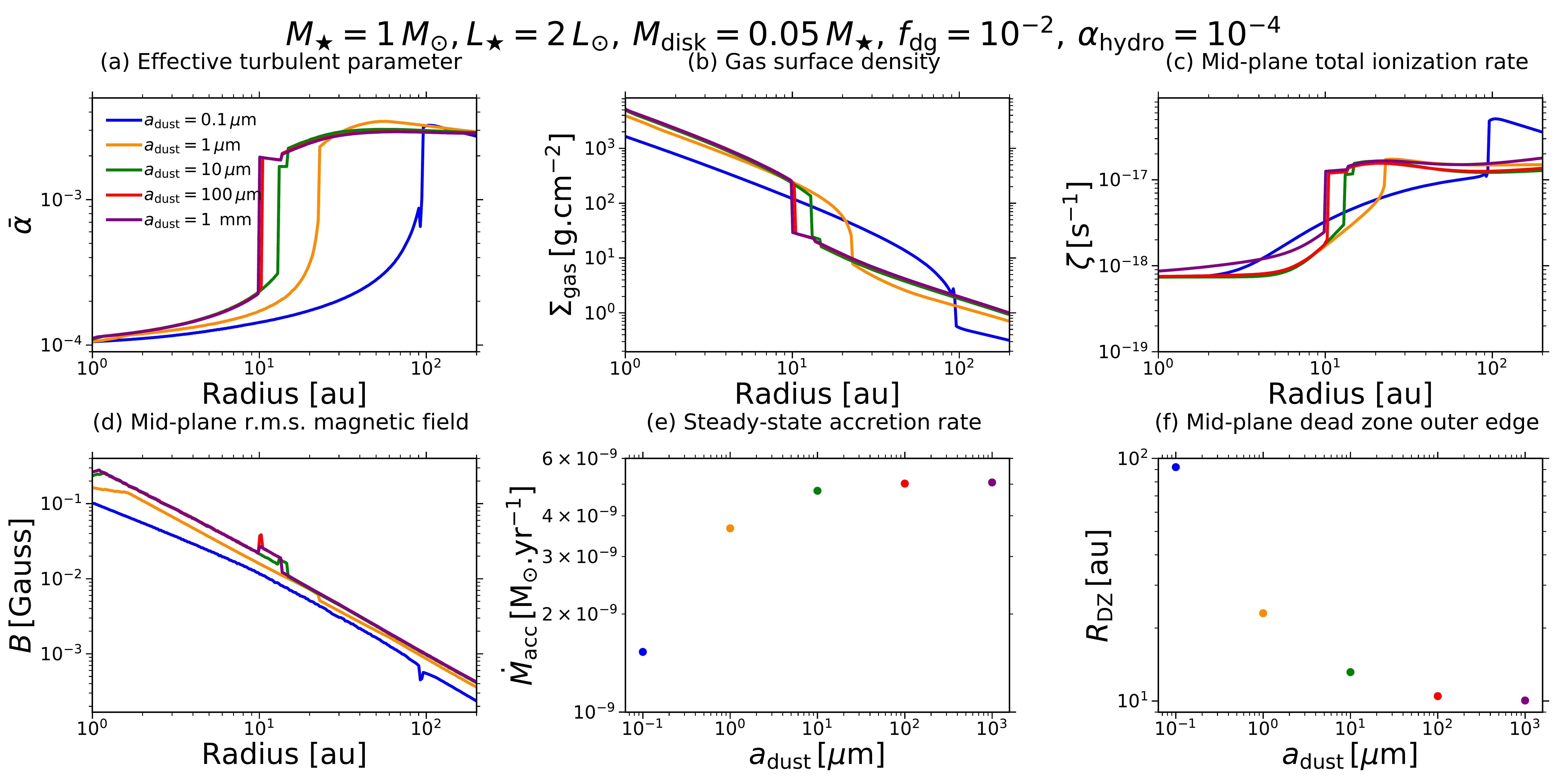}
 \caption{Effect of grain size on the equilibrium solution by varying the parameter $a_{\rm{dust}}$ from $0.1 \,\mu$m to $1 \,$mm: the case $a_{\rm{dust}} = 0.1 \,\mu$m is shown in blue; $a_{\rm{dust}} = 1 \,\mu$m in dark orange (fiducial model); $a_{\rm{dust}} = 10 \,\mu$m in green; $a_{\rm{dust}} = 100 \,\mu$m in red; and $a_{\rm{dust}} = 1 \,$mm in purple. The panels show the various steady-state profiles of some key quantities, for the model parameters $M_{\star} = 1 \, M_{\odot}$, $L_{\star} = 2 \, L_{\odot}$, $M_{\rm{disk}} = 0.05\, M_{\star}$, $f_{\rm{dg}} = 10^{-2}$ and $\alpha_{\rm{hydro}} = 10^{-4}$. Panels (a)-(f) show the same key quantities as in Fig.~\ref{fig:effect_M_disk}.}
 \label{fig:effect_grain_size}
\end{figure*}

The representative grain size at a given location in the disk is subject to significant change over time due to processes such as coagulation, fragmentation or radial drift. Depending on whether the representative grain size is skewed toward small or large dust particles, the equilibrium solution is expected to be very different. In our disk model, we do not yet account for dust evolution or multiple dust species. Nevertheless, we can still roughly quantify the importance of grain size on the equilibrium solution by running a set of simulations where we uniformly vary the dust particles size from sub-micron to millimeter. In practice, each dust particle has the same radius $a_{\rm{dust}}$, with $a_{\rm{dust}}$ in the list $\{10^{-5},10^{-4},10^{-3},10^{-2},10^{-1}\} \,$cm. The results can be seen in Fig.~\ref{fig:effect_grain_size}.

A smaller grain size implies a lower accretion rate and a more extended dead zone (Figs.~\ref{fig:effect_grain_size}(e) and \ref{fig:effect_grain_size}(f)). Smaller dust particles have indeed higher cross-section areas allowing them to efficiently capture free electrons or ions (overall lower ionization level, Fig.~\ref{fig:effect_grain_size}(c)). Since there are less charged particles in the gas-phase for the magnetic field to couple with, the MRI cannot be triggered easily and can even be shut off in most of the disk by strong Ohmic resistivity and ambipolar diffusion. Particularly, for sub-micron grains and at the mid-plane, the MRI can develop in the outermost regions of the disk only ($r \gtrsim 92 \,$au, Fig.~\ref{fig:effect_grain_size}(a)). For $r \lesssim 92 \,$au, the MRI can only operate in the surface layers, since the effect of grains on the ionization chemistry is diminished and stellar X-rays can more efficiently ionize the gas. Nonetheless, it is not enough to compensate for the fact that most of the disk is MRI-dead; leading to a mean value of a few $10^{-4}$ for $\bar{\alpha}$. Overall the accretion rate is thus much lower for sub-micron grains.

For larger grain sizes, the overall ionization level in the disk becomes higher for two reasons: the cross-section is significantly reduced resulting in the gas-phase recombination to more easily dominate the recombination process; and dust settling becomes important. Regarding the latter, larger dust particles easily decouple from the gas in the upper layers of the disk atmosphere, where the turbulence level is low due to ambipolar diffusion (see Sect.~\ref{sect:turbulence}). Consequently, settling can overcome vertical stirring, resulting in the dust to be concentrated in regions closer to the mid-plane. Dust settling implies a higher local dust-to-gas ratio closer to the mid-plane, hence a higher mid-plane radionuclides ionization rate since the ionization from radionuclides is mainly due to $^{26}$Al locked into grains. The effect of dust settling can be seen in Fig.~\ref{fig:effect_grain_size}(c), where the total mid-plane ionization rate is the highest for $a_{\rm{dust}} = 1\,$mm, for $r \lesssim 4\,$au.

Another salient result is that the steady-state accretion solution becomes independent of the grain size choice for $a_{\rm{dust}} \gtrsim 100 \, \mu$m. We can thus infer that there exists a threshold for the grain size above which the grains have little impact on the ionization chemistry, hence on the ionization level, and ultimately on the MRI.

Finally, once the mid-plane MRI-active layer is reached, we notice that the corresponding mean value of $\bar{\alpha}$ is roughly independent of the grain size, in the range $2 \times 10^{-3} - 3 \times 10^{-3}$ (Fig.~\ref{fig:effect_grain_size}(a)). We saw previously that the disk becomes more MRI-dead when smaller grain sizes are considered, implying that the gas surface densities in the MRI-active layer are lower compared to the ones for larger grain sizes (required to preserve the total disk gas mass constant). Since the gas surface densities are lower in the MRI-active layer for smaller grain sizes (Fig.~\ref{fig:effect_grain_size}(b)), we would expect more charged particles available in the gas-phase (due to stronger ionization) in such regions compared to the case of larger grain sizes. However, these charged particles are more efficiently captured when the dust grains are small, implying that the effective ionization level is actually not higher than the ones for larger grain sizes. Once the MRI-active layer is reached, its mean effective turbulence level is thus roughly constant regardless of the grain size.

\subsection{The Effect of Dust-to-gas mass ratio} \label{sect:effect dust-to-gas mass ratio}

Whether small dust particles are depleted or accumulated is expected to be crucial for the equilibrium solution, since the higher the concentration of those is, the harder it is for the MRI to operate. We run two sets of simulations where we: (1) consider an overall change in the vertically-averaged dust-to-gas mass ratio from $10^{-5}$ ($99.9\%$ depletion of dust particles relative to standard ISM) to $10^{-1}$ (Sect.~\ref{sect:without dust trapping}); (2) impose local dust enhancements either at the expected dead zone outer edge location for the fiducial model ($r = 23 \,$au); or at two locations in the disk, respectively, $r = 5 \,$au which is within the mid-plane dead zone and $r = 60 \,$au which is within the mid-plane MRI-active layer (Sect.~\ref{sect:with dust trapping}).

\subsubsection{Overall change} \label{sect:without dust trapping}

\begin{figure*}
\centering
\includegraphics[width=\textwidth]{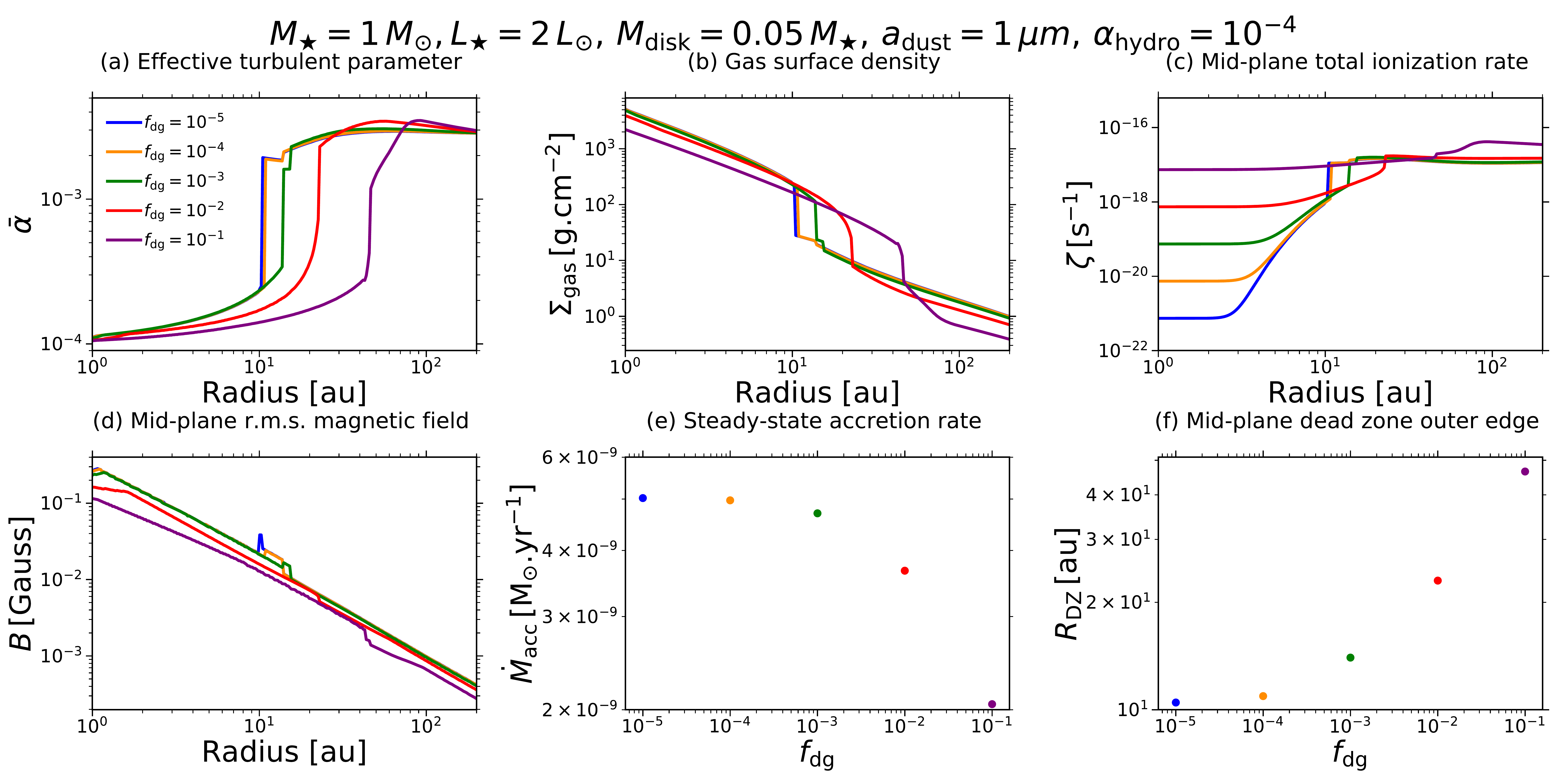}
 \caption{Effect of overall change in the vertically-integrated dust-to-gas mass ratio on the equilibrium solution by varying the radially constant parameter $f_{\rm{dg}}$ from $10^{-5}$ ($99.9$ dust depletion relative to standard ISM) to $10^{-1}$ (dust-rich disk): the case $f_{\rm{dg}} = 10^{-5}$ is shown in blue; $f_{\rm{dg}} = 10^{-4}$ in dark orange; $f_{\rm{dg}} = 10^{-3}$ in green; $f_{\rm{dg}} = 10^{-2}$ in red (fiducial model); and $f_{\rm{dg}} = 10^{-1}$ in purple. The panels show the various steady-state profiles of some key quantities, for the model parameters $M_{\star} = 1 \, M_{\odot}$, $L_{\star} = 2 \, L_{\odot}$, $M_{\rm{disk}} = 0.05\, M_{\star}$, $a_{\rm{dust}} = 1\, \mu$m and $\alpha_{\rm{hydro}} = 10^{-4}$. Panels (a)-(f) show the same key quantities as in Fig.~\ref{fig:effect_M_disk}.}
 \label{fig:effect_f_dg_no_trapping}
\end{figure*}

For this set of simulations, the vertically-averaged dust-to-gas mass ratio $f_{\rm{dg}}$ is chosen radially-constant in the list $\{10^{-5},10^{-4},10^{-3},10^{-2},10^{-1}\}$. The results can be seen in Fig.~\ref{fig:effect_f_dg_no_trapping}.

Figures~\ref{fig:effect_f_dg_no_trapping}(e) and \ref{fig:effect_f_dg_no_trapping}(f) show that protoplanetary disks with an overall higher depletion of micron-sized dust particles have a higher accretion rate and a more compact dead zone. For a disk poor in micron-sized grains, Ohmic resistivity and ambipolar diffusion are less stringent on the MRI-activity because grains have less impact on the recombination process, happening then mostly in the gas-phase; although the overall ionization level in the innermost regions is lower compared to less depleted disks due to less radionuclides available (Fig.~\ref{fig:effect_f_dg_no_trapping}(c)).

For a disk rich in micron-sized grains, the ionization from the decay of short/long-lived radionuclides is stronger. However, it is not enough to compensate for the more stringent non-ideal MHD effects on the MRI-driven turbulence, since a higher number of grains can more efficiently sweep up free electrons and ions from the gas-phase. Particularly, the case $f_{\rm{dg}} = 10^{-1}$ leads to the lowest accretion rate ($\approx 2 \times 10^{-9} \, M_{\odot}.\rm{yr}^{-1}$) and the largest dead zone maximal radial extent ($\approx 47\,$au).   

We notice that the steady-state accretion solution becomes independent of the vertically-averaged dust-to-gas mass ratio choice for $f_{\rm{dg}} \lesssim 10^{-4}$. Similarly to what we saw for the effect of grain size, there exists a threshold below which the dust depletion is so high that the grains barely have an impact on the ionization chemistry. Consequently, an even higher depletion would not change the ionization level, and ultimately the MRI. 

Finally, the mean effective turbulence level is roughly independent of the vertically-averaged dust-to-gas mass ratio once the MRI-active layer is reached (Fig.~\ref{fig:effect_f_dg_no_trapping}(a)). This can be explained by a similar argument used in the previous section: Although the gas surface densities in the MRI-active layer are lower for higher $f_{\rm{dg}}$ (which is expected to result in more charged particles available in the gas-phase due to stronger ionization), these charged particles are more efficiently captured by the dust for higher $f_{\rm{dg}}$. It implies that the effective ionization level is actually not higher than the ones for lower $f_{\rm{dg}}$; hence the mean effective turbulence level in the MRI-active layer being roughly constant regardless of $f_{\rm{dg}}$. 

\subsubsection{Local enhancements} \label{sect:with dust trapping}

\begin{figure*}
\centering
\includegraphics[width=\textwidth]{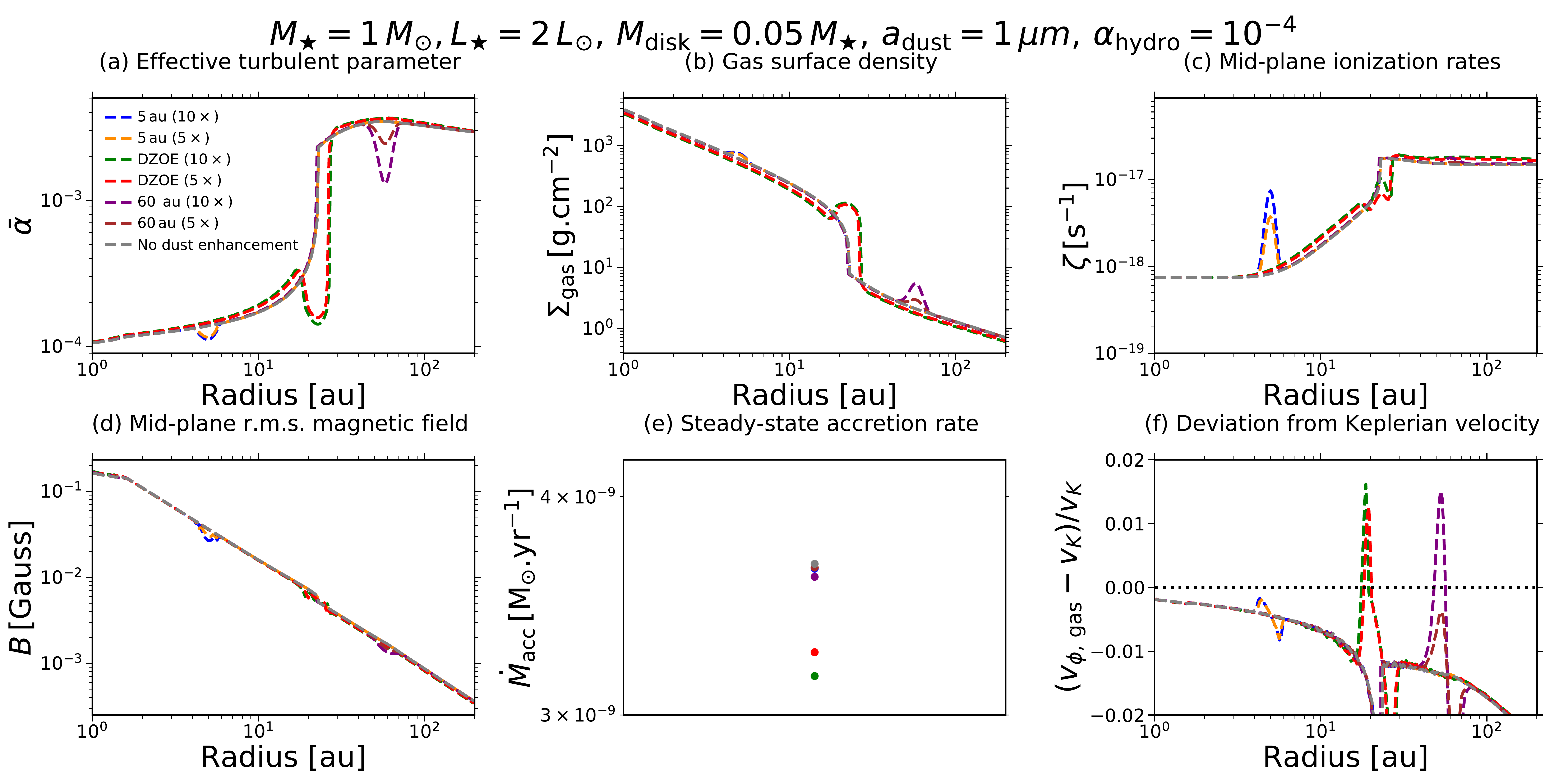}
 \caption{Effect of local enhancements in the vertically-integrated dust-to-gas mass ratio on the equilibrium solution by imposing $f_{\rm{dg}}$ to be radially variable with a local Gaussian perturbation at various locations (see the dust-to-gas mass ratio profiles in Fig.~\ref{fig:f_dg}). The unperturbed value of $f_{\rm{dg}}$ is the same for all the simulations, equal to $10^{-2}$. The "$5\,$au ($10\times$)" case (blue) corresponds to dust enhancement at $r = 5\,$au; where $f_{\rm{dg}}$ varies from its unperturbed value to $10^{-1}$. The "$5\,$au ($5\times$)" case (dark orange) is similar to the previous case, but $f_{\rm{dg}}$ goes up to $5 \times 10^{-2}$ instead. The "DZOE ($10\times$)" case (green) corresponds to dust enhancement at the expected dead zone outer edge location for the fiducial model ($r = 23 \,$ au); where $f_{\rm{dg}}$ varies from its unperturbed value to $10^{-1}$. The "DZOE ($5\times$)" case (red) is similar to the previous case, but $f_{\rm{dg}}$ goes up to $5 \times 10^{-2}$ instead. The "$60\,$au ($10\times$)" case (purple) corresponds to dust enhancement at $r = 60\,$au; where $f_{\rm{dg}}$varies from its unperturbed value to $10^{-1}$. The "$60\,$au ($5\times$)" case (brown) is similar to the previous case, but $f_{\rm{dg}}$ goes up to $5 \times 10^{-2}$ instead. The "No dust enhancement" case (gray) corresponds to the fiducial model. The panels show the various steady-state profiles of some key quantities, for the model parameters $M_{\star} = 1 \, M_{\odot}$, $L_{\star} = 2 \, L_{\odot}$, $M_{\rm{disk}} = 0.05\, M_{\star}$, $a_{\rm{dust}} = 1\, \mu$m and $\alpha_{\rm{hydro}} = 10^{-4}$. Panels (a)-(e) show the same key quantities as in Fig.~\ref{fig:effect_M_disk}. \textit{Panel (f)}: Deviation from Keplerian velocity where $v_{\phi,\,\rm{gas}}$ is the azimuthal gas velocity, and $v_{K} = r \Omega_{K}$ is the Keplerian velocity. The black dotted line corresponds to $v_{\phi,\,\rm{gas}} =  v_{K}$.}
 \label{fig:effect_f_dg_trapping}
\end{figure*}

In the first scenario, a fixed local dust enhancement (either moderate or strong) is imposed at the expected dead zone outer edge location for the fiducial model ($r = 23 \,$au). By doing so, we want to investigate how the equilibrium solution changes if dust particles have accumulated at the dead zone outer edge before the steady-state accretion regime is reached. In the second scenario, we impose a fixed local dust enhancement (either moderate or strong) at either $r = 5 \,$au or $r = 60\,$au. Here our goal is to study how local dust enhancements due to e.g., spontaneous traffic-jams can impact on the steady-state accretion solution if they occur before the steady-state accretion regime is reached. Such traffic-jams can form at locations where the gas pressure is enhanced (without the gas pressure gradient to necessarily flip of sign). The results can be seen in Fig.~\ref{fig:effect_f_dg_trapping}. 

For both scenarios, a moderate local dust enhancement refers to a local change in the vertically-averaged dust-to-gas mass ratio from $10^{-2}$ to $5 \times 10^{-2}$ (referred as "$5 \times$ enhancement"), whereas a strong one refers to a local change from  $10^{-2}$ to $10^{-1}$ (referred as "$10 \times$ enhancement"). Local dust enhancements are implemented by adding up Gaussian perturbations to a background vertically-integrated dust-to-gas mass ratio:
\begin{equation}
    B_{i}(r) \equiv A \exp{\left[- \frac{\left(r - r_{p,i}\right)^2}{2 \omega_{i}^2} \right]},
    \label{eq:dust particles traps}
\end{equation}

\noindent where $A$ is the dust enhancement level (either $A = 4$ for a moderate $5 \times$ enhancement, or $A = 9$ for a strong $10 \times$ enhancement), and $r_{p,i}$, $\omega_{i}$ are the center and the width of the Gaussian perturbation, respectively. $r_{p,i} = 23 \,$au for the first scenario, whereas $r_{p,i} = 5 \,$au or $r_{p,i} = 60\,$au for the second scenario. In all cases, $\omega_{i} \equiv H_{\rm{gas}}$ where $H_{\rm{gas}}$ is the disk gas scale height defined in Eq.~\eqref{eq:gas scale height}. The effective vertically-averaged dust-to-gas mass ratio is thus defined as 
\begin{equation}
    f_{\rm{dg}}(r) = f_{\rm{dg,\, bkg}} \left[1 + B_{i}(r)\right],
    \label{eq:perturbed dust-to-gas ratio}
\end{equation}

\noindent where we choose the background value $f_{\rm{dg,\, bkg}} = 10^{-2}$ to be the standard ISM value. The different profiles for the vertically-averaged dust-to-gas mass ratio can be found in Fig.~\ref{fig:f_dg}.

Figure~\ref{fig:effect_f_dg_trapping} shows that imposing local dust enhancements before the steady-state accretion regime is reached can lead to the formation of steady-state pressure maxima close to the locations where dust has locally accumulated. 

A local enhancement of dust implies a local increase of: (1) $^{26}$Al locked into grains; and (2) the number of grains that can efficiently sweep up free electrons and ions. The former induces a local increase of the radionuclides ionization rate ("bumps" in Fig.~\ref{fig:effect_f_dg_trapping}(c)), hence a local increase of the ionization fraction in the gas-phase. The latter leads to a weaker MRI-driven turbulence where dust has accumulated. Figure~\ref{fig:effect_f_dg_trapping}(a) shows that a lower value for $\bar{\alpha}$ (i.e., a dip) is found at the locations where the dust has accumulated. It thus means that the local increase of the ionization fraction due to the decay of short/long-lived radionuclides is not enough to balance the decrease of the MRI activity due to stronger non-ideal MHD effects. To maintain steady-state accretion ($\dot{M}_{\rm{acc}} = \rm{cst}$), a dip in $\bar{\alpha}$ must be compensated by a bump in $\Sigma_{\rm{gas}}$, hence in the gas pressure. If the perturbation (bump) in the gas pressure profile is significant enough, it can become a local pressure maximum: a location in the disk where the azimuthal gas velocity $v_{\phi,\,\rm{gas}}$ is super-Keplerian; namely, $v_{\phi,\,\rm{gas}} > v_K$, with $v_K = r \Omega_K$ the Keplerian velocity. Here we define the azimutal gas velocity as $v_{\phi,\,\rm{gas}} \approx v_{K} \left(1 - \frac{\eta_{\rm{gas}}}{2}\right)$, where $\eta_{\rm{gas}}$ is the gas pressure support parameter \citep[see e.g.,][]{2017ASSL..445...91P}   

The nature of such gas pressure perturbations (bumps) imposed by local dust enhancements depends on where and how much the dust has accumulated compared to the background value. Indeed, the effective turbulent parameter $\bar{\alpha}$ is much higher than the lower limit $\alpha_{\rm{hydro}} = 10^{-4}$ in the MRI-active layer; whereas $\bar{\alpha}$ is closer to $\alpha_{\rm{hydro}}$ in the dead zone (see the fiducial model in Sect.~\ref{sect:turbulence}). Additionally, the more dust particles accumulate at a given location in the disk, the deeper the dip in $\bar{\alpha}$ is, regardless of the location. Consequently, we expect local dust enhancements in the MRI-active layer to be able to generate deeper dips in $\bar{\alpha}$ compared to the ones in the dead zone, hence possibly producing stronger perturbations in the gas pressure profile; all in all depending on the enhancement level. Figure~\ref{fig:effect_f_dg_trapping}(f) shows that the perturbation in the gas pressure profile at $r = 5\,$au (within the dead zone) -caused by the local dust enhancement at that location- does not correspond to a pressure maximum; and so regardless of how much the dust has been locally enhanced ($5 \times$ or $10 \times$). Here we emphasize that this could change if the turbulence driven by hydrodynamic instabilities would be lower (e.g., $\alpha_{\rm{hydro}} = 10^{-5}$), since a lower $\alpha_{\rm{hydro}}$ value implies a possible deeper dip in $\bar{\alpha}$. Conversely, at $r = 60\,$au (within the MRI-active layer), we clearly see that whether the perturbation corresponds to a pressure maximum or not depends on how much dust has been locally enhanced. Interestingly, dust accumulation at the dead zone outer edge forms a pressure maximum there for both a $5 \times$ and $10 \times$ enhancement. This suggests that a spontaneous steady-state pressure maximum can be generated at the dead zone outer edge, and so even from a moderate amount of dust particles accumulation. 

Such spontaneous steady-state pressure maxima are formed over a viscous evolution timescale $t_{\rm{visc,\,bump}} = \frac{\Delta r^{2}}{3\bar{\nu}(r = r_{\rm{bump}})}$; where $r_{\rm{bump}}$ is the location of the pressure maximum, $\bar{\nu} = \frac{\bar{\alpha} c_{s}^{2}}{\Omega_K}$ is the effective kinematic viscosity at every disk radii $r$, and $\Delta r = H_{\rm{gas}}(r_{\rm{bump}})$ is the effective width of the bump (corresponding to the width of the local dust accumulation producing such a pressure maximum). Table~\ref{tab:Table2} summarizes the viscous formation timescales for the potential steady-state pressure maxima formed by the various local dust enhancement scenarios. We can conclude that when steady-state pressure maxima can form, they do so within the typical disk lifetime \citep[$5-10\,$Myrs, see e.g.,][]{2001ApJ...553L.153H}. 

In summary, we find that: (1) no steady-state pressure maxima are expected to be formed in the dead zone (except if the turbulence driven by hydrodynamic instabilities is lower than $\alpha_{\rm{hydro}} = 10^{-4}$; (2) steady-state pressure maxima are expected to be formed in the MRI-active layer only if the local dust enhancements level is strong ($10 \times$); (3) a steady-state pressure maximum can be formed at the dead zone outer edge, and so even for a moderate local dust enhancement ($5 \times$ is enough); (4) steady-state pressure maxima are formed within the disk lifetime, hence promoting further dust trapping and inducing a self-increase of their trapping strength.
\begin{table}
    \begin{center}
    \begin{tabular}{ |c||c|c|c|}
    \hline
    \hline
     $\mathbf{r_{p,i}\,}$\textbf{[au]} & $\mathbf{5}$ & $\mathbf{23}$ & $\mathbf{60}$ \\
    \hline
    Pressure maximum? ($5 \times$ enhancement) & \xmark & \cmark & \xmark  \\
    
    Formation timescale [Kyrs] & - & 37 & -\\ 
    \hline
    Pressure maximum? ($10 \times$ enhancement) & \xmark & \cmark & \cmark  \\
    
    Formation timescale [Kyrs] & - & 41 & 18 \\ 
    \hline
    \hline
\end{tabular}
\caption{Summary of the viscous formation timescales for the potential steady-state pressure maxima formed by local dust enhancement at various locations in the protoplanetary disk. We note that $r = 23\,$au corresponds to the mid-plane dead zone outer edge location for the fiducial model described in Sect.~\ref{sect:results-fiducial model}.}
\label{tab:Table2}
\end{center}
\end{table}

\subsection{The Effect of Stellar Properties} \label{sect:effect stellar properties}

\begin{figure*}
\centering
\includegraphics[width=\textwidth]{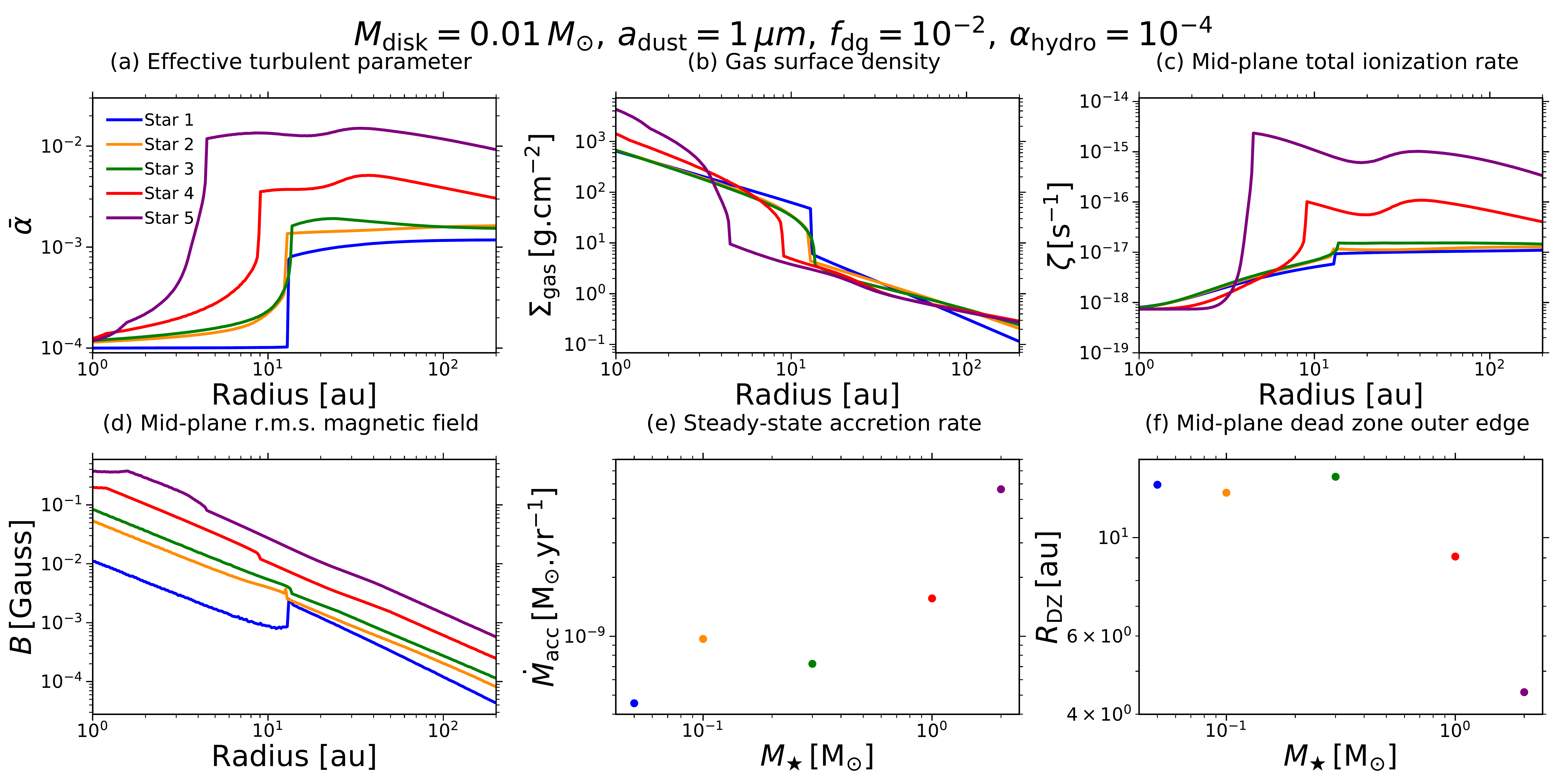}
 \caption{Effect of stellar properties on the equilibrium solution by varying the parameters $M_{\star}$ and $L_{\star}$ altogether, all the way from accreting brown dwarfs to herbig stars. The total disk gas mass does not depend on stellar mass and is such that $M_{\rm{disk}} = 0.01 \, M_{\odot}$. The "Star 1" case (blue) corresponds to a star of mass $M_{\star} = 0.05 \, M_{\odot}$ and bolometric luminosity $L_{\star} = 2 \times 10^{-3} \, L_{\odot}$. The "Star 2" case (dark orange) corresponds to a star of mass $M_{\star} = 0.1 \, M_{\odot}$ and bolometric luminosity $L_{\star} = 8 \times 10^{-2} \, L_{\odot}$. The "Star 3" case (green) corresponds to a star of mass $M_{\star} = 0.3 \, M_{\odot}$ and bolometric luminosity $L_{\star} = 0.2\, L_{\odot}$. The "Star 4" case (red) corresponds to the fiducial model where the stellar mass is $M_{\star} = 1 \, M_{\odot}$ and the bolometric luminosity is $L_{\star} = 2 \, L_{\odot}$. The "Star 5" case (purple) corresponds to a star of mass $M_{\star} = 2 \, M_{\odot}$ and bolometric luminosity $L_{\star} = 20 \, L_{\odot}$. The panels show the various steady-state profiles of some key quantities, for the model parameters $M_{\rm{disk}} = 0.05\, M_{\star}$, $a_{\rm{dust}} = 1 \, \mu$m, $f_{\rm{dg}} = 10^{-2}$ and $\alpha_{\rm{hydro}} = 10^{-4}$. Panels (a)-(f) show the same key quantities as in Fig.~\ref{fig:effect_M_disk}.}
 \label{fig:effect_M_star}
\end{figure*}

Finally, we explore how the stellar properties impact on the equilibrium solution by running five simulations, covering the possible range from very low-mass stars to Herbig-like stars. For each star of interest, both the stellar mass and bolometric luminosity change and are taken from Table~\ref{tab:Table3}. Moreover, we avoid the effect of total disk gas mass on the equilibrium solution (see Sect.~\ref{sect:effect total disk gas mass}) by setting the total disk gas mass independent of stellar mass equal to $0.01 \, M_{\odot}$. The results can be seen in Fig.~\ref{fig:effect_M_star}.
\begin{table}
    \begin{center}
    \begin{tabular}{ |p{2.8cm}||p{1.2cm}|p{1.2cm}|}
    \hline
    \hline
    \small{\textbf{STELLAR MODEL}} & \small{$\mathbf{M_{\star} \, [M_{\odot}]}$} & \small{$\mathbf{L_{\star} \, [L_{\odot}]}$}\\
    \hline
    Star 1 & $0.05$ & $2 \times 10^{-3}$ \\
    \hline
    Star 2 & $0.1$ & $8 \times 10^{-2}$ \\
    \hline
    Star 3 & $0.3$ & $0.2$ \\
    \hline
    Star 4 & $1$ & $2$ \\
    \hline
    Star 5 & $2$ & $20$ \\
    \hline
    \hline
\end{tabular}
\caption{Summary of the stellar parameters (mass $M_{\star}$ and luminosity $L_{\star}$) used to investigate the effect of stellar properties on the equilibrium solution. Star 4 model corresponds to the Sun-like star used for the fiducial model described in Sect.~\ref{sect:results-fiducial model}.}
\label{tab:Table3}
\end{center}
\end{table}

Going from star 1 to star 5 models, both the stellar mass and bolometric luminosity increase (the latter way faster than the former). A higher bolometric luminosity leads to (slightly) higher gas temperature (see Eq.~\eqref{eq:Temperature}), and higher stellar X-ray luminosity ($L_{\rm{XR}} \propto L_{\star}$). Since the effect of gas temperature is weak and barely impacts on the MRI activity (see Sect.~\ref{sect:effect temperature model}), a higher bolometric luminosity implies a positive feedback on the MRI because it produces a higher stellar X-ray luminosity, generating more charged particles in the gas-phase.

Figures~\ref{fig:effect_M_star}(e) and \ref{fig:effect_M_star}(f) show that star 4 and 5 models have a higher accretion rate and a more compact dead zone compared to star 1-3 models (the highest accretion rate and most compact dead zone are obtained for star 5 model). One way to understand this result is by looking at Fig.~\ref{fig:effect_M_star}(c). We notice that the mid-plane ionization rate is much higher for star 4 and 5 models compared to star 1-3 models, with the distinctive shape seen in Sect.~\ref{sect:effect total disk gas mass} for $M_{\rm{disk}} = 0.005\,M_{\star}$ or $M_{\rm{disk}} = 0.01,M_{\star}$. Such a shape indicates that stellar X-rays utterly dominate the ionization process in most of the disk, included at the mid-plane. Since the ionization level is much higher for star 4 and 5 models, the non-ideal effects are less stringent (see e.g., Fig.~\ref{fig:effect_M_star}(d) where the optimal r.m.s. magnetic field strength is higher for star 4 and 5 models). As a result, the MRI can be more easily triggered (more compact dead zones) and the MRI-driven turbulence is stronger (higher accretion rates) for high-mass and more luminous stars. 

Conversely, we notice a weaker dependence of $R_{\rm{DZ}}$ and $\dot{M}_{\rm{acc}}$ on $M_{\star}$ for star 1-3 models. With a sub-dominant ionization from stellar X-rays at the mid-plane, the significant increase of $L_{\star}$ (hence $L_{\rm{XR}}$) from star 1 to 3 models does not have as much impact as seen for star 4 and 5 models, since galactic cosmic rays dominate the ionization process at the mid-plane. As a result, $R_{\rm{DZ}}$ and $\dot{M}_{\rm{acc}}$ primarily depends on $M_{\star}$, and not on $M_{\star}$ and $L_{\star}$, as it is the case for star 4 and 5 models ($R_{\rm DZ}$ depends on $L_{\star}$ only weakly through $T$ for low-mass and less luminous stars). Since $M_{\star}$ weakly increases from star 1 to 3 models, we can explain the weaker dependence of $R_{\rm{DZ}}$ and $\dot{M}_{\rm{acc}}$ on $M_{\star}$ for low-mass and less luminous stars.  

Finally, we note that the aspect ratio can be as large as $0.4$ in the disk outer regions for star 1 and 2 models. We thus caution the reader that the profiles shown for those stars may not be valid for $r \gtrsim 40 \,$au, where the thin disk approximation reaches its limit (the aspect ratio is $< 1$, but not $\ll 1$). Nevertheless, we do expect the global picture for the effect of stellar properties to still hold as described. 

Here we investigated the effect of stellar properties alone by setting the total disk gas mass to the fixed stellar-independent value $M_{\rm{disk}} = 0.01 \, M_{\odot}$. However, this choice leads to very peculiar total disk gas masses relatively to their stellar mass for star 1 model ($M_{\rm{disk}} = 0.2 \, M_{\star}$) and star 5 model ($M_{\rm{disk}} = 0.005 \, M_{\star}$). Although $M_{\rm{disk}} = 0.2 \, M_{\star}$ for star 1 model, we note that the corresponding Toomre parameter $Q$ is still higher than $2$ at all radii; implying that the disk around this star is not self-gravitating. In Appendix~\ref{sect:combined effect of stellar properties and total disk gas mass}, we present a set of simulations with a stellar-dependent total disk gas mass ($M_{\rm{disk}} = 0.05 \, M_{\star}$), which represents a more realistic choice.

\section{Discussion} \label{sect:discussion}

\subsection{Gas Structure} \label{sect:discussion gas structure}

For steady-state accretion, we expect the mass flow through the disk to be uniform only if the disk is more massive at radii where the effective turbulent parameter $\bar{\alpha}$ is lower (see Eq.~\eqref{eq:Accretion rate1}); namely, where the dead zone sits. For the fiducial model, we find that the dead zone indeed contains most of the gas content enclosed in the domain considered ($67 \%$). This has been previously seen by studies such as \citet{2008ApJ...689..532T}. Here we confirm their result with a more consistent steady-state model that accounts for detailed considerations of the MRI with non-ideal MHD effects (Ohmic resistivity and ambipolar diffusion), instead of a parametric version. 

\citet{2008ApJ...689..532T} also shows that no steady-state pressure maximum at the dead zone outer edge is expected in a 1+1D approach (see their Fig.~3). We extend their study by showing that it holds only if dust accumulation has been inefficient at the dead zone outer edge before the disk reaches its steady-state accretion regime. The formation of a pressure bump at that location is indeed expected to be a transient phenomenon, since the gas pile-up would be smeared out by density waves or Reynolds stress driven by the turbulent layers \citep[see e.g.][]{2016A&A...596A..81P}. However, if a sufficient amount of dust particles accumulate at the dead zone outer edge before the steady-state accretion regime is reached, the gas structure will display a steady-state pressure bump at that location which can be formed within the disk lifetime (see Sect.~\ref{sect:with dust trapping}), hence promoting further dust trapping.

A striking feature of our steady-state gas surface density profiles is the discontinuity at the dead zone outer edge ($r = 23 \,$au for the fiducial model). As mentioned in Sect.~\ref{sect:results-fiducial model}, this discontinuity arises from our on/off criteria for active MRI, inducing a steep change in the local turbulent parameter $\alpha$ at each transition from low to high turbulence regime. To confirm whether this is a real discontinuity in the mathematical sense or a numerical artifact due to a lack of resolution, we re-run the fiducial model with a refined radial grid around the mid-plane dead zone outer edge ($r = 23\,$au). With the much higher resolution that provides the refined radial grid ($259$ cells to describe the region $r \in \left[21, 25\right]\,$au, instead of $9$ cells for the fiducial radial grid), we still find that $\Sigma_{\rm{gas}}$ is discontinuous at that location. In reality though, such a steep transition in the gas surface density would not happen: as proposed by \citet{2010MNRAS.402.2436Y}, the gas would rearrange itself triggered by e.g., Rayleigh adjustment in a timescale shorter (close to the dynamical timescale) than the disk takes to reach its steady-state accretion regime (viscous evolution timescale). Consequently, we expect our steady-state accretion solution to be valid everywhere but at the transition dead zone/MRI-active layer; where the transition should be a smoother version of our solution. This discrepancy does not change the qualitative results drawn in this study.

\begin{figure}
\includegraphics[width=0.45\textwidth]{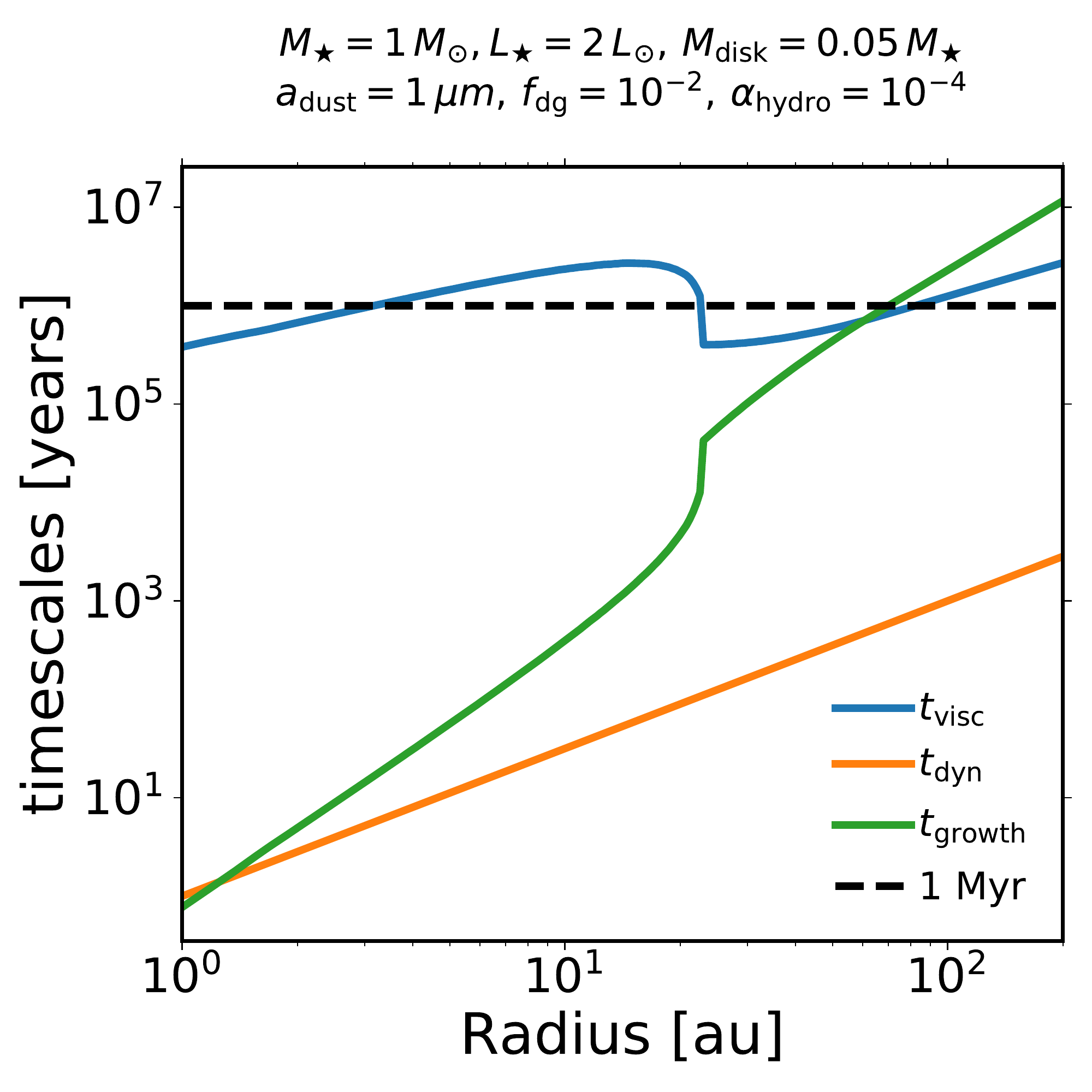}
 \caption{Various local timescales as a function of radius, for the fiducial model described in Sect.~\ref{sect:results-fiducial model}. The blue solid line ($t_{\rm{visc}}$) corresponds to the viscous evolution timescale. The orange solid line ($t_{\rm{dyn}}$) corresponds to the dynamical timescale. The green solid line ($t_{\rm{growth}}$) corresponds to the growth timescale for the case of micron-sized particles. The dashed black line corresponds to $1\,$Myr.} 
 \label{fig:timescales}
\end{figure}

As shown by \citet{2018ApJ...861..144M} (see their Appendix~B), the total inward accretion rate $\dot{M}_{\rm{acc}}$ can always be decomposed, at each radius, as the sum of the accretion rates within the individual vertical layers of the disk (MRI-active layer, dead and zombie zone). Since we impose steady-state accretion, $\dot{M}_{\rm{acc}}$ is radially constant in our equilibrium solutions. Doing so though, means that the accretion rates within the individual layers are unavoidably radially variable, as there are no separate conditions to demand invariance. For example, we find for the fiducial model that the accretion rate within the dead zone is negative for $r \gtrsim 10\,$au to compensate the accretion within the MRI-active layer that exceeds the total value; hence ensuring the total inward accretion rate $\dot{M}_{\rm{acc}}$ to be radially constant (see Fig.~\ref{fig:accretion_rates_individual_layers}). Consequently, we could expect the gas to build-up at some locations or excavate at others (e.g., $r = 10\,$au); hence driving the disk away from our equilibrium solution. To confirm that our solution is valid in a dynamical time-averaged sense, the local viscous evolution timescale over which the gas build-up/excavation changes the vertical density structure must be much longer than the local dynamical timescale needed for the disk to relax back to a hydrostatic equilibrium vertical profile. In Fig.~\ref{fig:timescales}, we show the local viscous evolution timescale ($t_{\rm{visc}} = \frac{r^{2}}{3\bar{\nu}}$, where $\bar{\nu} = \frac{\bar{\alpha} c_{s}^{2}}{\Omega_K}$) as well as the local dynamical timescale ($t_{\rm{dyn}} = \frac{2 \pi}{\Omega_K}$). Since we find $t_{\rm{dyn}} \ll t_{\rm{visc}}$ everywhere, we can thus conclude that the gas volume density perturbations -induced by the variable accretion rates in the individual layers- are vertically smoothed out much more rapidly than they can grow and drive away the disk from our steady-state accretion solution. We confirm that it is the case for all equilibrium solutions presented in this work. 

If the protoplanetary disk is at some stage of its evolution out of equilibrium, the gas surface density will change so that the disk evolves toward a steady-state (our solutions). The local viscous evolution timescale $t_{\rm{visc}}$ tells us in how much time the steady-state accretion is reached, at a given radius, assuming that the disk is purely viscously evolving. For the fiducial model, we find that $t_{\rm{visc}}$ is in the range $0.4 - 2.7 \,$Myrs; with a mean value of $1.5 \,$Myrs in the dead zone and $1\,$Myr in the MRI-active layer (see Fig.~\ref{fig:timescales}, where the longest viscous evolution timescale is reached in the dead zone at $r \approx 15 \,$au). The steady-state accretion regime can thus be established, at most radii, within the disk lifetime; and particularly before the disk dispersal phase by e.g., internal photoevaporation that can disperse the gas content after $1 - 4$\,Myrs of evolution \citep[see e.g.,][]{2019MNRAS.487.3702O, 2021ApJ...909..109K}. Consequently, the protoplanetary disk could accumulate a spatially extended long-lived inner disk gas reservoir (the dead zone) accreting a few $10^{-9}\, M_{\odot}.\rm{yr}^{-1}$, before photoevaporation starts carving a hole; possibly resulting in transition disks with large gaps and high accretion rates once photoevaporation starts being at play, as shown by G\'{a}rate et al. (accepted).

\subsection{Accretion Rates for different Stellar Masses} \label{sect:discussion accretion rates}

In steady-state accretion models such as ours, there is a tight correlation between total disk gas masses and accretion rates (see Fig.~\ref{fig:effect_M_disk}(e)). In principle, for given stellar and dust properties, we could estimate the total disk gas mass from an observed accretion rate (assuming that the disk has reached a steady-state MRI-driven accretion regime). 

\begin{figure}
\includegraphics[width=0.45\textwidth]{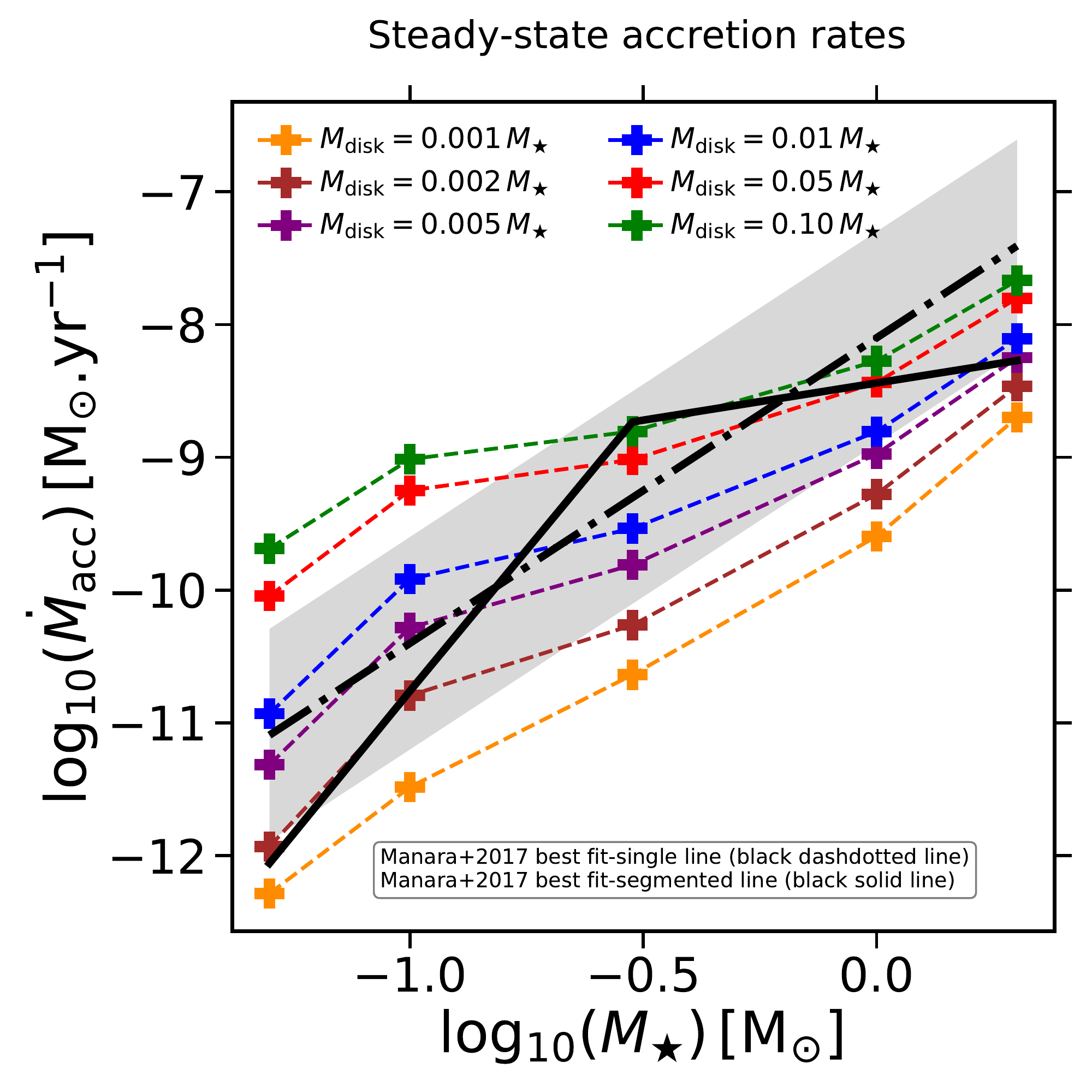}
 \caption{Accretion rate as a function of stellar mass. The black dash-dotted and solid lines correspond to the single-line and segmented-line best fit models, respectively, from \citet{2017A&A...604A.127M}. The $1\sigma$ dispersion around the single-line best fit model is shown by the gray area. The "+" symbols represent the accretion rates obtained for various total disk gas masses and for each star taken from Table~\ref{tab:Table3}, using our steady-state accretion model. Here we assume that $a_{\rm{dust}} = 1\, \mu$m, $f_{\rm{dg}} = 10^{-2}$ and $\alpha_{\rm{hydro}} = 10^{-4}$.} 
 \label{fig:M_accretion_VS_M_star}
\end{figure}

We perform this exercise with the results reported in \citet{2017A&A...604A.127M}. In their work, the authors present a study of a sample of 94 young stars with disks in the Chamaeleon I star-forming regions \citep[$1-6\,$Myr, see][their Table~1 and references therein]{2021arXiv210406838V}. They analyze the spectra of those objects obtained with the ESO VLT/X-shooter to derive, among others, the $\dot{M}_{\rm{acc}}-M_{\star}$ relation. \citet{2017A&A...604A.127M} show that this relation could be better described with a model more complex than a "single-line" fit (although the single-line fit is not statistically excluded): a "segmented-line" fit with a break at $M_{\star} \approx 0.3\, M_{\odot}$, delimiting a linear relation at higher $M_{\star}$ and steeper relation at lower $M_{\star}$. We present the single-line as well as the segmented-line best fits in Fig.~\ref{fig:M_accretion_VS_M_star}, with the 1$\sigma$ dispersion around the former (gray area). From our ends, we run six simulations corresponding to a total disk gas mass $M_{\rm{disk}} \in \{0.001,0.002,0.005,0.01,0.05,0.10\} \,M_{\star}$, for each stars taken from Table~\ref{tab:Table3}. The results are shown in Fig.~\ref{fig:M_accretion_VS_M_star}.

To quantitatively match the accretion rates of very low-mass and less luminous stars (star 1 and 2 models) obtained from \citet{2017A&A...604A.127M}'s $\dot{M}_{\rm{acc}}-M_{\star}$ relation, we find that their total disk gas mass must be lower compared to the more massive and luminous counterparts (star 3-5 models). For star 1 and 2 models, Fig.~\ref{fig:M_accretion_VS_M_star} shows that the two best fits from \citet{2017A&A...604A.127M} are consistent with a total disk gas mass that is roughly in the range $0.002 \,M_{\star} - 0.005 \,M_{\star}$. Conversely, for star 3 and 4 models, the two best fits are consistent with a total disk gas mass that is roughly in the range $0.01 \,M_{\star} - 0.05 \,M_{\star}$; whereas they are consistent with a total disk gas mass that is roughly in the range $0.01 \,M_{\star} - 0.10 \,M_{\star}$ for star 5 model. Our results support the ideas that the steeper relation seen in the accretion rates for very low-mass and less luminous stars may be caused by: (1) a faster evolution, since we need a total disk gas mass values that are nowadays low; or (2) the fact that very low-mass and less luminous stars initially form with a low total disk gas mass (relatively to their stellar mass). Previous studies suggest that the faster evolution of disk around very low-mass stars could be explained by faster accretion, since these disks could potentially be entirely MRI active \citep[e.g.,][]{doi:10.1146/annurev-astro-081915-023347}. However, as pointed out by \citet{2013ApJ...764...65M}, this idea seems unlikely because ambipolar diffusion becomes stronger at the low densities expected for the disk around those stars; hence preventing the MRI to easily operate everywhere. Without any other mechanisms to explain faster evolution in the context of MRI-driven accretion, the idea of very low-mass and less luminous stars being initially formed with a lower total disk gas mass compared to high-mass and more luminous stars seems more probable. 

We note that our accretion rates presented in Fig.~\ref{fig:M_accretion_VS_M_star} depend on the choice made for $r_{\rm{max}}$, the outer boundary of our radial grid ($r_{\rm{max}} = 200\,$au in this study). For fixed model parameters and given a field strength, the steady-state accretion solution is unique. By solely varying $r_{\rm{max}}$, we change how the gas mass is distributed in space: a lower $r_{\rm{max}}$ implies higher gas surface densities (since the same amount of gas is now distributed across a more confined domain), whereas a higher $r_{\rm{max}}$ leads to lower ones. We find that both the accretion rate and the dead zone maximal radial extent increase with decreasing $r_{\rm{max}}$ (not shown here). The former dependence can be explained by the fact that a lower $r_{\rm{max}}$ means less distance for the gas to travel before being accreted onto the central star. The latter dependence is justified by the fact that a lower $r_{\rm{max}}$ implies an overall lower ionization level in the disk. Consequently, all the accretion rates presented in Fig.~\ref{fig:M_accretion_VS_M_star} are expected to increase if $r_{\rm{max}} < 200\,$au, and decrease if $r_{\rm{max}} > 200\,$au. Additionally, we find that the effect of $r_{\rm{max}}$ is more prominent for lower total disk gas masses (not shown here). Nonetheless, the main conclusions of this section do not change. Indeed, the gas radius for a disk around a very low-mass and less luminous star is more likely to be lower than our fiducial $200\,$au \citep[see e.g.,][]{Kurtovic_2021}, and also lower than the gas radius for a disk around a high-mass and more luminous star. As a result, the accretion rate of a very low-mass and less luminous stars is expected to increase compared to what is displayed in Fig.~\ref{fig:M_accretion_VS_M_star}, implying that even lower total disk gas masses will be needed to explain the steeper relation seen in the accretion rates.

\subsection{The Need for Gas, Dust and Stellar Evolution} \label{sect:discussion the need for gas/dust evolution}

The detailed parameter study conducted in Sect.~\ref{sect:results-parameter study} shows that the following parameters play a crucial role in shaping the steady-state accretion solution: (1) the total disk gas mass; (2) the representative grain size; (3) the vertically-averaged dust-to-gas mass ratio; and (4) the stellar X-ray luminosity. In general, those parameters may change during the evolution of the protoplanetary disk. We will thus need to relax the steady-state accretion assumption in future studies. Doing so means that we will need to implement some missing ingredients.

First, we will need to consider gas evolution. During the evolution of class II protoplanetary disks, the gas content is removed from the disk due to accretion onto the star and winds (e.g., magnetocentrifugal or photoevaporative). Additionally, there is no source term as in class 0/I disks where the envelope in-fall can feed in gas. In general, the total disk gas mass thus inevitably decreases over time. In Sect.~\ref{sect:effect total disk gas mass}, we saw that the equilibrium solution strongly depends on the total disk gas mass with which the disk reaches its steady-state accretion regime: a lower total disk gas mass implies a lower accretion rate, and a more compact dead zone. Consequently, for a given initial total disk gas mass, we expect the gas surface density to start evolving toward the equilibrium solution corresponding to this total disk gas mass. Since the total disk gas mass decreases over time, the gas surface density will re-adjust by evolving toward a new steady-state solution corresponding to that new total disk gas mass. Gas evolution is crucial here, since it will tell us how the gas structure changes over time; hence, from what steady-state solution to another the disk will tend assuming pure viscous evolution.

Second, we will need to implement dust evolution including growth processes. Here we have assumed a fiducial vertically-averaged dust-to-gas mass ratio of $10^{-2}$ and small dust grains of constant size $1\,\mu$m. Such values may be expected in the early stages of dust evolution in the disk. However, dust particles grow, fragment, drift, stir and settle during the evolution of protoplanetary disks. In the case of our fiducial model, Fig.~\ref{fig:timescales} shows that micron-sized grains grow in a timescale ($t_{\rm{growth}}$) that is shorter than the timescale the gas needs to evolve on large scales (viscous evolution timescale $t_{\rm{visc}}$), for $r \lesssim 60\,$au. Here $t_{\rm{growth}}$ is computed using the mid-plane value of Eq.~(30) from \citet{2016SSRv..205...41B}; where we take the relative velocity $\Delta v$ to be the Brownian relative velocity defined as $\Delta v = \sqrt{8k_{\rm B}T/\pi m_{\rm {red}}}$, with $m_{\rm {red}}$ the reduced mass ($m_{\rm {red}} = m_{\rm dust}/2$ for a collision of equal micron-sized grains. On one hand, it implies that our steady-state accretion solution may describe well the gas for $r \gtrsim 60\,$au because such a regime can be reached before micron-sized grains can grow substantially. On the other hand, dust growth becomes important for $r \lesssim 60\,$au. In Sect.~\ref{sect:effect grain size} and Sect.~\ref{sect:effect dust-to-gas mass ratio}, we found that both a larger grain size and a higher dust depletion imply a higher accretion rate, a higher $\bar{\alpha}$ overall, and a more compact dead zone. Consequently, we would expect the following cycle to happen, where the disk turbulence state oscillates between a low and high regime. For the initial conditions $M_{\rm{disk}} = 0.05 \, M_{\star}$, $a_{\rm{dust}} = 1 \, \mu$m and $f_{\rm{dg}} = 10^{-2}$, the gas will start evolving toward the corresponding equilibrium solution. Since $t_{\rm{growth}} \leq t_{\rm{visc}}$ for $r \lesssim 60\,$au, the dust can efficiently grow by coagulation to representative grain sizes skewed toward larger sizes in such regions, and the vertically-averaged dust-to-gas mass ratio of micron-sized grains can substantially drop before this steady-state accretion solution can actually be reached. As a result, the gas will evolve toward a new equilibrium solution corresponding to a disk with a higher $\bar{\alpha}$ overall. A higher $\bar{\alpha}$ in the disk implies a more efficient fragmentation that can replenish micron-sized grains. The representative grain sizes can thus be skewed toward micron-sized particles again, and the vertically-averaged dust-to-gas mass ratio of micron-sized grains can re-increase; resulting in the gas evolving toward a steady-state solution close to the initial one, corresponding to a less turbulent disk. Due to a low disk turbulence level, the dust can effectively grow and less frequently fragment again, hence looping the cycle. A similar cycle has been reported in \citet{2012ApJ...753L...8O} where growth, settling and fragmentation were considered alongside Ohmic resistivity (see their Fig.~3). The authors found that this cycle is actually a "damped oscillation", i.e. the disk turbulence state and grain size distribution converge and stop oscillating rather than indefinitely oscillating. The equilibrium solution corresponding to $M_{\rm{disk}} = 0.05 \, M_{\star}$, $a_{\rm{dust}} = 1 \, \mu$m and $f_{\rm{dg}} = 10^{-2}$ is thus physically motivated if the cycle mentioned above can be completed in a timescale shorter than the viscous evolution timescale (otherwise the total disk gas mass decreases) and the drift timescale (otherwise dust particles are removed from the disk, and the overall dust content drops); and if the grain size distribution at the end of the cycle is comparable to the one at the beginning of the cycle (i.e., the grain size distribution is skewed toward micron-sized grains). Additionally, we note that such a cycle is also likely to happen in the steady-state pressure maxima generated by local dust enhancements (see Sect.~\ref{sect:with dust trapping}), possibly generating pressure maxima that could "come and go" over time. In order to provide a more consistent picture for this cycle, we need a time-dependent framework with dust evolution included.

Finally, stellar evolution will need to be accounted for. As we saw in Sect.~\ref{sect:effect stellar properties}, the stellar X-ray luminosity $L_{XR}$ plays a fundamental role because it quantifies how much the gas can be ionized by stellar X-rays: the higher the stellar X-ray luminosity is, the more turbulent the disk is overall. Depending on the stellar type, this quantity is more or less constant over time \citep[see e.g,][their Fig.~5]{2021ApJ...909..109K}. As a result, a time-dependent framework needs to consider the time-dependence of $L_{XR}$.

\subsection{Model Limitations and Justification for Model Assumptions} \label{sect:assumptions}

In our 1+1D global MRI-driven disk accretion model, we adopt a number of simplifications:

\begin{enumerate}[itemsep=2mm]
    \item We focus our study on the MRI-driven accretion and assume maximal efficiency for the MRI activity. For this assumption to hold, the disk must be threaded by a net vertical magnetic field, otherwise the MRI is very inefficient \citep{2013ApJ...764...66S}. In the presence of a net vertical field though, it seems inevitable for the disk to launch a magnetized wind that efficiently carries angular momentum away and drives accretion at rates consistent with observations \citep[see e.g.,][]{1982MNRAS.199..883B,2013ApJ...769...76B,2021A&A...650A..35L}. To be consistent with the assumption of maximal efficiency for the MRI activity, wind-driven accretion thus needs to be included in our model in the future. Furthermore, the highest accretion rate reached by our various steady-state MRI-driven accretion models for a disk around a solar-type star corresponds to the canonical value $\dot{M}_{\rm{acc}} \approx 10^{-8} \, M_{\odot}.\rm{yr}^{-1}$ (for $M_{\rm{disk}} = 0.13\,M_\star$, $a_{\rm{dust}} = 10 \, \mu$m, $f_{\rm{dg}} = 10^{-3}$, and $\alpha_{\rm{hydro}} = 10^{-4})$. Nonetheless, it seems that this value can only be reached for a disk close to be self-gravitating. For typical class II disks, MRI/wind hybrid models are thus necessary to capture the turbulence level in the disk (wind and MRI-driven accretion can co-exist as shown by \citet{2021arXiv210610167C}).
    
    \item We assume that the MRI activity reaches a maximally efficient state permitted by Ohmic resistivity and ambipolar diffusion. Such a condition is fulfilled by constraining and adopting the magnetic field strength so that it corresponds to the highest field strength that would still allow the MRI to operate. In reality, there are no clear reasons why the field would follow such a configuration. Actually, there is ample evidence from MHD simulations that the saturation level of the MRI-driven accretion stress is controlled by the strength of the net vertical magnetic field, \citep[e.g.,][]{1995ApJ...440..742H,2010ApJ...718.1289S,2011ApJ...742...65O,2013ApJ...775...73S,2013ApJ...764...66S}. In principle, the radial distribution of the net vertical magnetic flux is determined by the diffusion and advection of the large-scale field lines, although its details are yet to be fully understood \citep[e.g.,][]{1994MNRAS.267..235L,2014MNRAS.441..852G,2014ApJ...785..127O,2014ApJ...797..132T,2017ApJ...836...46B,2018ApJ...857...34Z}. As such, a more consistent picture could be obtained by combining our calculations of the non-ideal MHD terms (magnetic diffusivities) -derived from a careful modeling of the gas ionization degree- to 3D MHD simulations where the magnetic field strength would be solved self-consistently (beyond the scope of this paper).
     
    \item We consider stellar X-rays, galactic cosmic rays, and the decay of short/long-lived radionuclides as the main ionization sources. In this approach, two important ionization sources have been neglected: 
    \begin{itemize}[label=\textbullet]
        \item FUV. It can fully ionize atomic carbon and sulfur. \citet{2011ApJ...735....8P} show that FUV can penetrate a vertical gas column density $\Sigma_{\rm{FUV}} \in [0.01-0.1] \,$ g.cm$^{-2}$, and produce much higher ionization than other sources. Additionally, \cite{2013ApJ...775...73S} found that the MRI is the most effective in the FUV layer in the outer disk. Depending at which disk height the penetration depth $\Sigma_{\rm{FUV}}$ is reached, FUV could thus be important and needs to be included in future works. 
         \item Stellar energetic particles (SPs). They become the dominant H$_2$ ionization source in the warm molecular layer of the disk above the CO ice line, and can increase the column densities of HCO$^{+}$ by a factor in the range $3-10$ for disk radii $r \lesssim 200\,$au, with a higher impact in the disk regions with low cosmic rays ionization \citep[see][]{2017A&A...603A..96R}. The inclusion of the SPs could change our equilibrium solution because it could boost the ionization level and thus impact the number density of HCO$^{+}$. This effect should be included in future studies.
    \end{itemize}
    
    \item For the disk dust structure, grains are assumed to be a mono-disperse population of size $a_{\rm{dust}}$. In reality, protoplanetary disks have a distribution of grain sizes determined by the balance between collisional coagulation, fragmentation and radial drift. One of the main goals for a future work is to self-consistently implement a size distribution directly from dust evolution models. For simplicity, we also do not consider the sublimation of water ice, and thus the change in dust properties across the protoplanetary disk. Furthermore, we assume that the vertical dust density follows a Gaussian distribution \citep[e.g.,][when attempting to interpret dust continuum observations]{2008A&A...489..633P}. Nevertheless, this assumption may not hold at high altitude in the protoplanetary disk, as seen in studies of dust settling driven by the MRI in the ideal MHD limit \citep[see][]{2009A&A...496..597F}.
    
    \item For the disk gas structure, we assume the gas to be vertically isothermal, and passively heated by stellar irradiation. Here we neglect viscous heating. This could be added in future studies by solving for the self-consistent disk vertical structure adopting a similar approach described in e.g., \citet{1998ApJ...500..411D, 1999ApJ...527..893D, 2021MNRAS.504..280J}. Nevertheless, we do not expect such an implementation to change our results because viscous heating is inefficient in the dead zone \citep{2011ApJ...732L..30H,2019ApJ...872...98M,2021arXiv210513101M} as well as in the regions where the MRI can operate (low gas density).
    
    \item Other than magnetic disk winds, we also neglect some processes that could drive the angular momentum transport: (1) the Hall shear instability (HSI) causing strong radial angular momentum transport if the magnetic field is aligned with the rotation axis of the protoplanetary disk \citep[see e.g.,][]{2008MNRAS.385.1494K, 2014A&A...566A..56L,2015ApJ...798...84B}; (2) gravito-turbulence that may arise in regions of the disk where the enclosed local disk gas mass is such that $Q \approx 2$, with a cooling timescale longer than the orbital timescale \citep[see e.g.,][]{2021A&A...650A..49B}. For the latter, the total disk gas mass of all our disk models has been chosen such that the Toomre parameter $Q$ is always higher than $2$ ($\approx 3$ in practice). The gravito-turbulence is thus expected not to dominate the disk turbulence state in our steady-state accretion solutions.
    
    \item We assume that HCO$^{+}$ is the only ion species in the gas-phase, reflecting the fact that CO is the most abundant molecule after H$_2$. However, this assumption may not hold everywhere in the disk. For temperatures below $25\,$K (CO freeze-out), the dominant ions are $\rm{NH}_{x}^{+}$, where $x=1,\,2,\,3$. In the disk inner part where $T \gtrsim 100\,$K, the metal ion Mg$^+$ can desorb from the solid grains and tend to recombine in longer timescales \citep{2013ApJ...765..114D}. Nonetheless, we do not expect such ions to significantly modify our equilibrium solutions for $25\,\rm{K}\lesssim T \lesssim 100\,$K.
\end{enumerate}

\section{Summary and Conclusions} \label{sect:summary and conclusions}

We present a 1+1D global framework to study the outer protoplanetary disk ($r \gtrsim 1\,$au) which accretes viscously, solely due to the MRI and hydrodynamic instabilities. The disk is heated by stellar irradiation, and ionized by stellar X-rays, galactic cosmic rays as well as radionuclides. A semi-analytical chemical model is employed to capture the charge state of the disk dust-gas mixture, hence carefully modeling the gas ionization degree. Additionally, the magnetic field strength is constrained and adopted to maximize the MRI activity. For given stellar and disk parameters, the effective viscosity parameter $\bar{\alpha}$ can thus be determined self-consistently under the framework of viscously-driven accretion, from detailed considerations of the MRI with Ohmic resistivity and ambipolar diffusion.

We employ our framework to investigate the structure of a steadily accreting protoplanetary disk. To obtain the steady-state MRI-driven accretion solution corresponding to a maximally efficient MRI activity, the gas surface density, gas ionization state, magnetic field as well as the viscosity and accretion rate are calculated self-consistently.

For the fiducial protoplanetary disk model considered, we find that:
\begin{enumerate}[label=(\roman*),itemsep=2mm]

    \item The accretion rate reached by this disk model is $\dot{M}_{\rm{acc}} \approx 3.7 \times 10^{-9} \, M_{\odot}.\rm{yr}^{-1}$, the mean effective turbulence level in the MRI-active layer is $\left<\bar{\alpha}\right>_{\rm{MRI}} \approx 3 \times 10^{-3}$, and the dead zone maximal radial extent is $R_{\rm{DZ}} = 23 \,$au. For studies using an ad hoc prescription for the Shakura-Sunyaev $\alpha$-parameter, a value of a few $10^{-3}$ is found to be more appropriate to describe the viscosity in the MRI-active layer; accounting for ambipolar diffusion.
        
    \item The steady-state accretion regime of such a disk is reached within the disk lifetime, particularly before the disk dispersal phase. The disk could thus accumulate a spatially extended long-lived inner disk gas reservoir, before internal photoevaporation starts carving a hole.
    
    \item A steady-state pressure maximum at the dead zone outer edge does not form.

    \item Ambipolar diffusion quenches the MRI activity over most of the disk. The upper envelope of the MRI-active layer is set by ambipolar diffusion, whereas Ohmic resistivity primarily sets its lower envelope. 
\end{enumerate}

Additionally, we perform an exhaustive parameter study to investigate what model parameters are crucial to MRI-driven accretion in protoplanetary disks, as well as the extent to which the accretion is efficient (namely, the location of the dead zone outer edge and the value of the accretion rate). We adopt the fiducial parameters, but vary one at a time the following model parameters: the total disk gas mass ($M_{\rm{disk}}$), the representative grain size ($a_{\rm{dust}}$), the vertically-averaged dust-to-gas mass ratio ($f_{\rm{dg}}$), the stellar properties (both $M_{\star}$ and $L_{\star}$), the hydrodynamic turbulent parameter ($\alpha_{\rm{hydro}}$), and the gas temperature model ($T$). We find that:
\begin{enumerate}[label=(\roman*),itemsep=2mm]
    \item The accretion rate and dead zone outer edge location as a function of the total disk gas mass are approximately power laws: $\dot{M}_{\rm{acc}} \propto M_{\rm{disk}}^{0.54}$ and $R_{DZ} \propto M_{\rm{disk}}^{0.64}$. A more massive disk implies a higher accretion rate and a more extended dead zone (both radially and vertically).
    
    \item A smaller grain size leads to a smaller accretion rate and a more extended dead zone. When sub-micron dust particles are considered ($a_{\rm{dust}} = 0.1\, \mu$m), the disk mid-plane is mainly MRI-dead and accretes much slower. Furthermore, the steady-state solution remains unchanged regardless of the grain size for $a_{\rm{dust}} \gtrsim 100\, \mu$m. This suggests that there exists a threshold for the grain size above which the grains have little impact on the ionization chemistry, hence on the MRI. 
    
    \item An overall higher depletion of micron-sized grains leads to a higher accretion rate and a more compact dead zone. Dust enhancement ($f_{\rm{dg}} = 10^{-1}$) compared to the standard ISM value strongly damps the MRI-driven accretion. Additionally, the steady-state accretion solution is independent of the vertically-integrated dust-to-gas mass ratio choice for $f_{\rm{dg}} \lesssim 10^{-4}$. As a result, there exists a threshold for the depletion of micron-sized grains below which the grains have little impact on the ionization chemistry, hence on the MRI.
    
    \item If a sufficient amount of dust particles have locally accumulated in a region where the MRI can operate before the disk reaches its steady-state accretion regime, the gas structure will adjust to form a steady-state pressure maximum close to the location of dust enhancement. These spontaneous steady-state pressure maxima can be formed within the disk lifetime, which might have direct consequences for planet formation by promoting further dust trapping.
    
    \item For a fixed stellar-independent total disk gas mass, a more massive and luminous star implies a higher accretion rate and a more compact dead zone. 
    
    \item The accretion rate and dead zone outer edge location as a function of the hydrodynamic turbulent parameter are approximately power laws: $\dot{M}_{\rm{acc}} \propto \alpha_{\rm{hydro}}^{0.31}$ and $R_{DZ} \propto \alpha_{\rm{hydro}}^{0.22}$. A higher $\alpha_{\rm{hydro}}$ leads to a higher accretion rate and a more extended dead zone. For $\alpha_{\rm{hydro}} = 10^{-3}$, the gas surface density profile is much smoother ($\Sigma_{\rm{gas}} \propto r^{-1.22}$) than the profiles obtained for lower $\alpha_{\rm{hydro}}$ values, and is actually close to the constant-$\alpha$ model solution ($\Sigma_{\rm{gas}} \propto r^{-1}$).  
    
    \item The choice for either a optically thin or thick radial gas temperature model has little impact on the accretion rate, dead zone size, and the overall equilibrium solution. 
\end{enumerate}

Finally, we investigate in more depths the effect of stellar properties by considering more realistic models where the total disk gas mass is stellar-dependent. We find that:
\begin{enumerate}[label=(\roman*),itemsep=2mm]

    \item For a fixed stellar-dependent total disk gas mass, a more massive and luminous star implies a higher accretion rate and a more extended dead zone. 

    \item The dead zone cannot become indefinitely large for high-mass and more luminous stars, since their high stellar X-ray luminosity regulates its radial extent. Similarly, the dead zone cannot become indefinitely small for low-mass and less luminous stars due to their too low stellar X-ray luminosity. 
    
    \item For the range of stars considered ($0.05 \,M_{\odot} - 2 \,M_{\odot}$), ambipolar diffusion is the primary factor determining the strength of the MRI-driven accretion.
    
    \item To explain the steeper relation seen in the observed accretion rates of very low-mass and less luminous stars \citep[][]{2017A&A...604A.127M} in the context of MRI-driven accretion, we find that very low-mass and less luminous stars need to be initially formed with a lower total disk gas mass compared to the high-mass and more luminous counterparts.
    
\end{enumerate}

In summary, we have shown that the MRI-driven accretion behavior crucially depends on the total disk gas mass, the representative grain size, the vertically-integrated dust to gas ratio, and the stellar X-ray luminosity. Since those quantities are expected to undergo substantial change throughout the evolution of the disk, we need to relax the steady-state assumption and account for gas, dust and stellar evolution. Furthermore, we have shown that dust accumulation at specific locations in the disk can lead to the spontaneous formation of steady-state pressure maxima. This strongly motivates the combination of dust evolution with careful modeling of the gas ionization degree in a time-dependent framework. Consequently, a self-consistent time-dependent coupling between gas, dust, stellar evolution models and our framework on million-year timescales is required in future studies.

\begin{acknowledgements}

We are thankful to the anonymous referee for the very detailed and constructive report. We thank Christelle Delage and Reema Joshi for providing great comments on the manuscript. We thank Marija Jankovic, Giovanni Rosotti, Dmitry Semenov, Carlo Manara and Nicol\'{a}s T. Kurtovic for fruitful discussions and very useful insights. This work made extensive use of the Astropy \citep{2013A&A...558A..33A}, Matplotlib \citep{Hunter:2007}, Numpy \citep{harris2020array}, Scipy \citep{2020SciPy-NMeth} software packages. T.N.D. and P.P. acknowledge support provided by the Alexander von Humboldt Foundation in the framework of the Sofja Kovalevskaja Award endowed by the Federal Ministry of Education and Research. S.O. is supported by JSPS KAKENHI Grant Numbers JP18H05438, JP19K03926, JP20H01948, and 20H00182.  M.F. acknowledges funding from the European Research Council (ERC) under the European Union’s Horizon 2020 research and innovation program (grant agreement No. 757957). N.D. acknowledges support provided by DFG under grant DU 414/20-1 (SPP 1992).

\end{acknowledgements}

\bibliographystyle{aa} 
\bibliography{Bibliography.bib}

\begin{appendix}

\section{Derivation of Eqs.~\eqref{eq:charge state distribution of dust particles}--\eqref{eq:free electrons number density}} \label{appendix:dZ2_ni_ne}

In this section, we derive the equations that describe the gas--dust charge reaction equilibrium, Eqs.~\eqref{eq:charge state distribution of dust particles}--\eqref{eq:free electrons number density}. We basically follow the derivation by \citet{2009ApJ...698.1122O}, but take the induced polarization force between ions/free electrons with charged grains into account. 
We begin with the equations describing the time evolution of $n_i$ and $n_e$ due to ionization and recombination,  
\begin{equation}
    \frac{dn_i}{dt} = \zeta n_{\rm gas} -s_i u_i\sigma_{\rm dust}n_{\rm dust}P_in_i - k n_i n_e,
    \label{eq:dnidt}
\end{equation}
\begin{equation}
    \frac{dn_e}{dt} = \zeta n_{\rm gas} -s_e u_e\sigma_{\rm dust}n_{\rm dust}P_en_e - k n_i n_e,
    \label{eq:dnedt}
\end{equation}
where $P_{i(e)}$ is the effective cross section averaged over grain charge $Z$ and normalized by $s_{i(e)} \sigma_{\rm dust}$, 
\begin{equation}
    P_{i(e)} = \langle \tilde{J}_{i(e)}(Z)\rangle,
\end{equation}
with $\tilde{J}_{i(e)}$ is
the normalized effective cross sections for grains of charge $Z$  and the brackets denote an average over $Z$. 
Accounting for the induced polarization forces and assuming $Z < 0$ (see the main text), we use Eqs.~(3.3) and (3.5) of \citet{1987ApJ...320..803D}:
\begin{equation}
    \tilde{J}_i = \left(1-\frac{Z}{\tau}\right)\left(1+\sqrt{\frac{2}{\tau-2Z}}\right),
\end{equation}
\begin{equation}
    \tilde{J}_e = \exp{\left(\frac{Z/\tau}{1 + \left(-Z\right)^{-\frac{1}{2}}}\right)} \left(1 + \sqrt{\frac{1}{4\tau - 3Z}}\:\right)^{2}.
\end{equation}
In comparison, \citet{2009ApJ...698.1122O} adopted $\tilde{J}_i = 1-Z/\tau$ and $\tilde{J}_e = \exp(Z/\tau)$ for $Z < 0$ by neglecting the polarization forces.

In ionization--recombination equilibrium with $dn_i/dt = 0$ and $dn_e/dt = 0$,   Eqs.~\eqref{eq:dnidt} and \eqref{eq:dnedt} reduce to two algebraic equations for $n_i$ and $n_e$. These equations can be easily solved to yield Eqs.~\eqref{eq:ions number density} and \eqref{eq:free electrons number density} \citep[see][for a similar calculation]{2009ApJ...698.1122O}.

To derive the equilibrium grain charge distribution, we follow the Appendix of \citet{2009ApJ...698.1122O} and write down the detailed balance equation,
\begin{equation}
    s_i u_i n_i \tilde{J}_i(Z) n_{\rm dust}(Z) = s_e u_e n_e \tilde{J}_e(Z+1) n_{\rm dust}(Z+1). 
    \label{eq:detailed}
\end{equation}
For $\tau \ga 1$, the grain charge distribution satisfies $\sqrt{\langle \Delta Z^2 \rangle} \gtrsim 1$ and hence can be viewed as a continuous function of $Z$ \citep[see the Appendix of][]{2009ApJ...698.1122O}. In this case,  one has $\tilde{J}_e(Z+1) \approx \tilde{J}_e(Z)$ and $n_{\rm dust}(Z+1) \approx n_{\rm dust}(Z) + dn_{\rm dust}/dZ$, and therefore 
Eq.~\eqref{eq:detailed} approximately reduces to 
\begin{equation}
    \frac{dn_{\rm dust}(Z)}{dZ}+ W(Z)n_{\rm dust}(Z) = 0,    
    \label{eq:dnddZ}
\end{equation}
where
\begin{equation}
    W(Z) \equiv 1 - \frac{s_i u_i n_i \tilde{J}_i(Z)}{s_e u_e n_e \tilde{J}_e(Z)}.
\end{equation}

To derive an approximate solution for \eqref{eq:dnddZ}, we define $Z_0$ such that $W(Z_0) = 0$, i.e., $s_i u_i n_i \tilde{J}_i(Z_0) = s_e u_e n_e \tilde{J}_e(Z_0)$, and $\delta Z \equiv Z - Z_0$. To the first order in $\delta Z$, one has
\begin{eqnarray}
    W(Z) &\approx& 
    -\frac{s_i u_i n_i}{s_e u_e n_e}\dfrac{d}{dZ_0}\left(\dfrac{ \tilde{J}_i(Z_0)}{ \tilde{J}_e(Z_0)}\right)\delta Z
    \nonumber \\
    &=& 
    -\frac{\tilde{J}_e(Z_0)}{\tilde{J}_i(Z_0)}\dfrac{d}{dZ_0}\left(\dfrac{ \tilde{J}_i(Z_0)}{ \tilde{J}_e(Z_0)}\right)\delta Z
    \nonumber \\
    &=& 
    -\left(\dfrac{d \ln \tilde{J}_i(Z_0)}{dZ_0}-\dfrac{d \ln \tilde{J}_e(Z_0)}{dZ_0}\right)\delta Z.
\label{eq:W_approx}
\end{eqnarray}
With Eq.~\eqref{eq:W_approx}, Eq. \eqref{eq:dnddZ} can be analytically solved, yielding $n_{\rm dust}(Z)$ given by Eq.~\eqref{eq:charge state distribution of dust particles} with $\langle Z \rangle = Z_0$ and 
\begin{eqnarray}
    \langle\Delta Z^2 \rangle
    &=&  -\left(\dfrac{d \ln \tilde{J}_i(\langle Z \rangle)}{d\langle Z \rangle}-\dfrac{d \ln \tilde{J}_e(\langle Z \rangle)}{d\langle Z \rangle}\right)^{-1}.
\end{eqnarray}
Finally, by approximating     
$\tilde{J}_{i(e)}(\langle Z \rangle) \approx  \langle\tilde{J}_{i(e)}( Z) \rangle = P_{i(e)}$ and using $\langle Z \rangle = -\Gamma \tau$, we obtain Eq.~\eqref{eq:charge dispersion}.
A complete expression of Eq.~\eqref{eq:charge dispersion} is 
\begin{eqnarray}
    \langle\Delta Z^2 \rangle 
    &=&
    {\tau} \left[\frac{ 2 \Gamma  \tau +3 \sqrt{\Gamma  \tau }}{2
   \left(\sqrt{\Gamma  \tau }+1\right)^2}+\frac{1}{\Gamma +1}
   \right.
   \nonumber \\
   && \quad -\frac{\sqrt{2} 
   }{(2 \Gamma +1) \left((2 \Gamma +1)   \sqrt{\frac{\tau}{2 \Gamma  +1
   }}+\sqrt{2}\right)}
   \nonumber \\
   && \left. \quad +\frac{3}{(3 \Gamma +4)^2 \sqrt{\frac{1}{3 \Gamma  +4  }}
   \left(\sqrt{\frac{1}{3 \Gamma   +4 }}+\sqrt{\tau}\right)}\right]^{-1}.
\end{eqnarray}

\section{Comparison between the semi-analytical chemical model and a chemical reaction network} \label{appendix:comparison between the semi-analytical chemical model and a chemical reaction network}

We aim to validate the semi-analytical chemical approach used in this study \citep[adapted from][]{2009ApJ...698.1122O} by comparing with a chemical reaction network. Particularly, we use the chemical reaction network by \citet{1980PASJ...32..405U}. We include the following charged species: free electrons, five ion species (H$^+$, H$_3^+$, HCO$^+$, He$^+$, and C$^+$) and charged grains. Among these, HCO$^+$ represents heavy molecular ions (e.g., H$_3$O$^+$) that recombine with free electrons dissociatively, and therefore at similar recombination rates. We neglect metal ions, assuming that all metal elements are locked into grains. The grains are allowed for a discrete charge distribution $Z \in [Z_{\rm min},Z_{\rm max}]$ with $Z_{\rm min} = \lfloor \langle Z \rangle - 4\langle \Delta Z^2 \rangle^{1/2} \rfloor$ and $Z_{\rm max} = \lceil \langle Z \rangle + 4\langle \Delta Z^2 \rangle^{1/2} \rceil$, where $\langle Z \rangle$ and $\sqrt{\langle \Delta Z^2 \rangle}$ are the mean grain charge and grain charge dispersion predicted from the semi-analytical model given by Eqs.~\eqref{eq:mean charge} and \eqref{eq:charge dispersion}, respectively. The radial and vertical profiles for the temperature, gas and dust densities as well as the total ionization rate are taken from the equilibrium solution obtained for the fiducial model described in Sect.~\ref{sect:results-fiducial model}.

\begin{figure*}
\centering
\includegraphics[width=\textwidth]{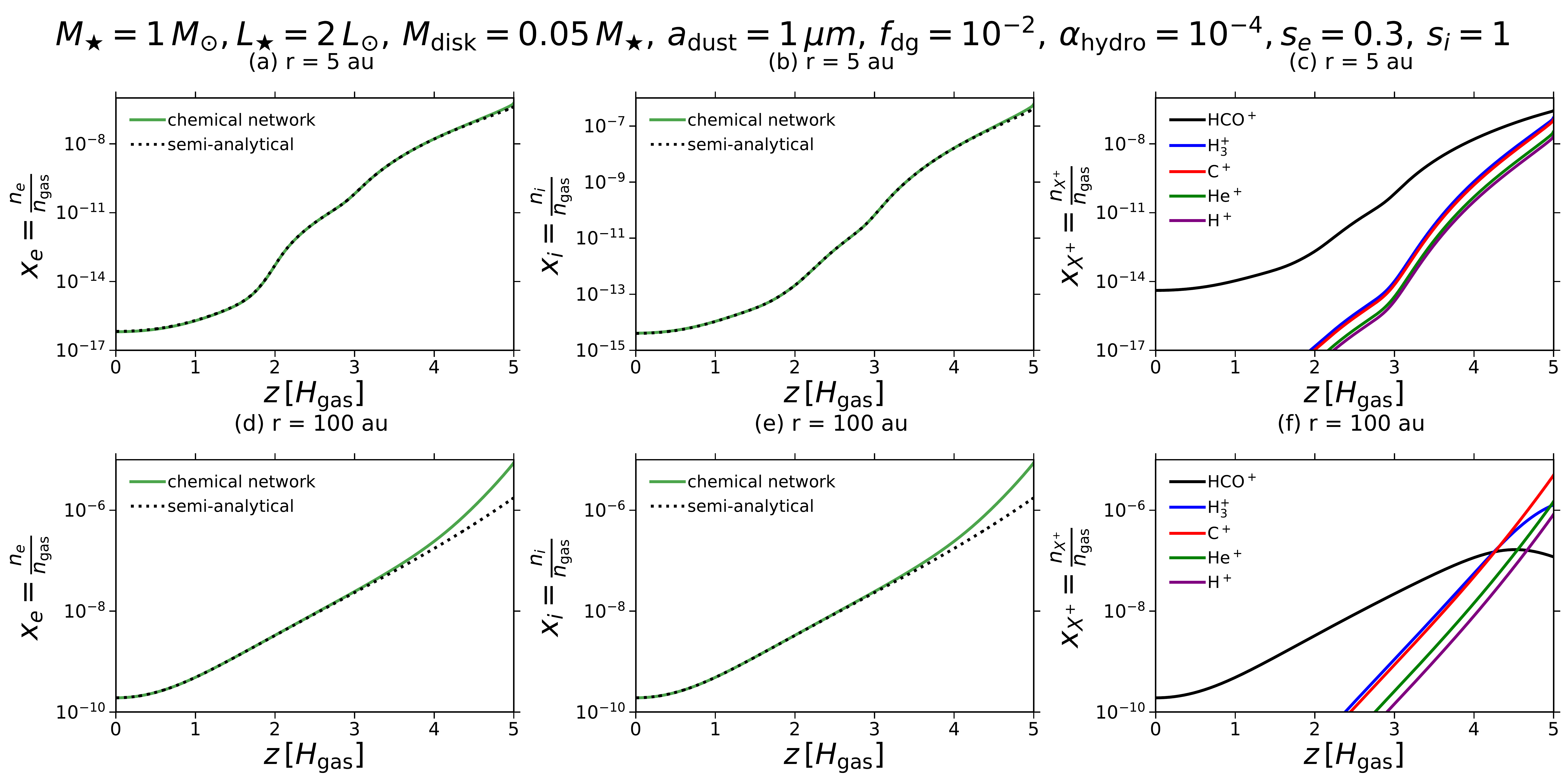}
 \caption{Comparison between the semi-analytical chemical model employed in this study and a chemical reaction network, for the fiducial model described in Sect.~\ref{sect:results-fiducial model}. The panels show various species abundances as a function of height $z$, at $r = 5\,$au and $r = 100 \,$au (upper and lower rows, respectively). \textit{Left panels}: free electron abundance from the chemical reaction network (green solid lines) and the semi-analytical model (black dotted lines). \textit{Central panels}: Total ion abundance from the chemical reaction network (green solid lines) and the semi-analytical model (black dotted lines). \textit{Right panels}: Abundances of individual ion species from the chemical reaction network. The black solid line corresponds to HCO$^{+}$, the blue solid line to H$_{3}^{+}$, the red solid line to C$^{+}$, the green solid line to He$^{+}$, and the purple solid line to H$^{+}$.}
 \label{fig:semi-analytical chemical model VS chemical network}
\end{figure*}

We calculate the abundances of various charged species in local ionization-recombination equilibrium for $0 \leq z/H_{\rm gas} \leq 5$, at $r = 5\,$au and $r =100\,$au. Figure~\ref{fig:semi-analytical chemical model VS chemical network} shows the results of the chemical network calculations. We plot the vertical profiles of the free electron abundance, total ion abundance, and the abundances of individual ion species. The results show that the dissociatively recombining  molecular ions HCO$^+$ are the dominant ions at all locations in the disk, except in the upper layers ($z \gtrsim 4\,H_{\rm gas}$) of the disk outer regions. In such high-altitude regions though, the gas density is so low that it does not significantly impact on our equilibrium solution (the solution is determined by $\bar{\alpha}$, which is a pressure-weighted vertically-averaged quantity). Consequently, the assumption that HCO$^{+}$ is the dominant ion species in most of the protoplanetary disk -made for the semi-analytical model employed in this study- is a good approximation. Figure~\ref{fig:semi-analytical chemical model VS chemical network} also displays the free electron and total ion abundances from the semi-analytical model. At $r = 5\,$au (resp., $r = 100\,$au), we find the semi-analytical calculations to be in good agreement with the chemical network for all $z \lesssim 5 \,H_{\rm gas}$ (resp., $z \lesssim 4 \,H_{\rm gas}$). At $r = 100\,$au and for $z \approx 4\,H_{\rm gas}$, the semi-analytical calculations slightly underestimate the free electron and total ion abundances, since the slowly recombining ion species H$_3^+$ and C$^+$ become as important as the fast recombining HCO$^+$. However, the relative error between the  chemical network and semi-analytical results is no larger than $40 \%$. This discrepancy becomes larger for $4 < z/H_{\rm gas} \lesssim 5$ because HCO$^{+}$ are no longer the dominant ions species. In those regions, the semi-analytical calculations can underestimate the free electron and total ion abundances up to a factor of $\approx 5$ (at $z = 5 \,H_{\rm gas}$). Nevertheless, the gas density is very low there, which means that our equilibrium solution is not sensitive to this discrepancy.    

\section{The stellar X-rays ionization rate} \label{appendix:direct X-Ray ionisation rate}

To compute the stellar X-rays ionization rate (Eq.~\eqref{eq:X-ray ionisation rate}), we have used the fit from \citet{2009ApJ...701..737B}, which is expressed as a function of vertical gas column densities. The underlying assumption is that the stellar X-rays penetrate the gas in the vertical direction, at any distance from the star. Although this approximation is good enough for galactic cosmic rays that originate from outside the protoplanetary disk, stellar X-rays are launched from around the central star. Depending on the exact location from where they originate (e.g., stellar surface or stellar jets), the use of vertical gas column densities to calculate the stellar X-rays ionization rate might become invalid.

\begin{figure}
\includegraphics[width=0.45\textwidth]{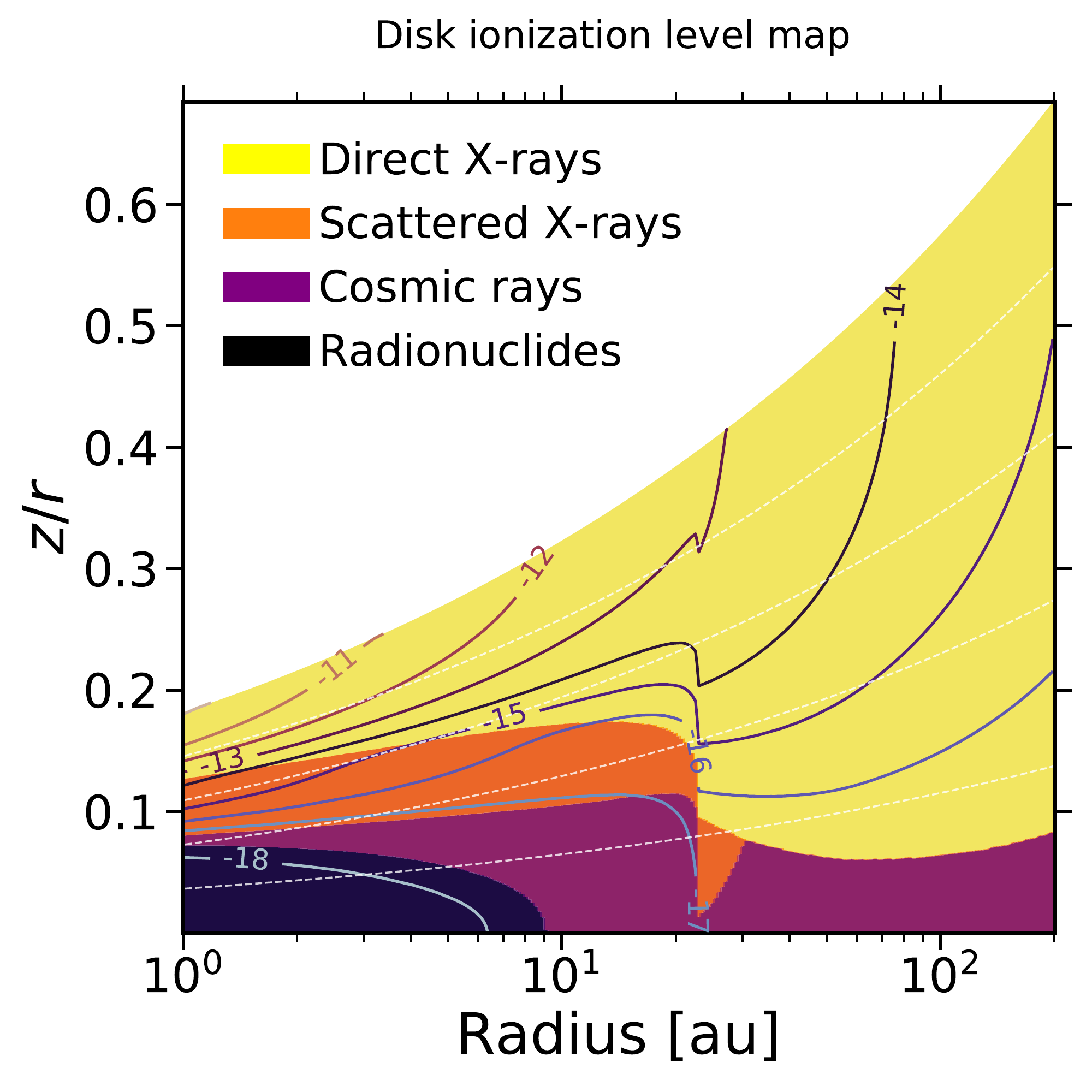}
 \caption{Ionization rate for H$_2$ $\zeta^{(H_2)}$ (defined in Sect.~\ref{sect:ionization sources}) and its different contributions as a function of location in the disk, for the fiducial model described in Sect.~\ref{sect:results-fiducial model}. For each color, the following non-thermal ionization source dominates $\zeta^{(H_2)}$: decay from short/long-lived radionuclides (black), galactic cosmic rays (purple) or stellar X-rays (orange and yellow for the scattered and direct contribution, respectively). The colored solid contour lines correspond to the surfaces of constant total ionization rate $\zeta^{(H_2)}$, ranging from $7.6 \times 10^{-19} \,$s$^{-1}$ to $1.8 \times 10^{-10} \,$s$^{-1}$. The white dashed lines correspond to the surfaces $z = 1 \, H_{\rm{gas}}$, $z = 2 \, H_{\rm{gas}}$, $z = 3 \, H_{\rm{gas}}$ and $z = 4 \, H_{\rm{gas}}$; from bottom to top, respectively.}
 \label{fig:2D_map_disk_ionization}
\end{figure}

For example, Fig.~\ref{fig:2D_map_disk_ionization} shows that stellar X-rays can efficiently penetrate and ionize the outer regions of the disk (via the direct X-rays contribution). In reality, this may not be the case if they have to penetrate the dense inner regions before reaching the outer ones: this extinction phenomenon corresponds to the self-shadowing event. It is expected to arise if stellar X-rays are launched from low-altitude regions such as the stellar surface. In this case, the direct contribution is quickly extinguished due to its small penetration depth, and the scattered contribution is substantially modified since they originate from the former. Conversely, if the stellar X-rays are launched from high-altitude regions above the disk mid-plane such as the stellar jets, they can travel without having to penetrate the dense inner regions; and thus reach and ionize the outer regions of the disk. 

To properly account for the self-shadowing event, one solution could be to implement ray tracing in our framework. However, doing so would greatly increase the computational complexity of the whole problem. Given our main goal (combining detailed considerations of the MRI with non-ideal MHD effects to gas and dust evolution models), we assume in this study that the stellar X-rays originate from high-altitude regions around the central star (e.g., stellar jets), so that the stellar X-rays ionization rate's fit given by \citet{2009ApJ...701..737B} holds. 

\section{Follow-up parameter study} \label{appendix:Follow-up parameter study}

In this appendix, we extend the parameter study conducted in Sect.~\ref{sect:results-parameter study} by performing a series of simulations which investigate: (1) the effect of magnetic field strength (Appendix~\ref{sect:effect magnetic field strength}); (2) the effect of hydrodynamic turbulent parameter $\alpha_{\rm{hydro}}$ (Appendix~\ref{sect:effect hydrodynamics turbulence}); (3) the effect of gas temperature model $T(r)$ (Appendix~\ref{sect:effect temperature model}); (4) the combined effect of stellar properties and total disk gas mass (Appendix~\ref{sect:combined effect of stellar properties and total disk gas mass}). We derive and show the same key quantities reached by our steady-state accretion disk model as in Sect.~\ref{sect:results-parameter study}.

\subsection{The Effect of Magnetic Field Strength}\label{sect:effect magnetic field strength}

In this present work, all the derived accretion rates $\dot{M}_{\rm{acc}}$ represent the highest value possibly reachable given the set of criteria for active MRI employed. Indeed, we adopt the required magnetic field strength such that the MRI activity is maximally efficient (see Sect.~\ref{sect:B field}). Here we explore -for the fiducial parameters ($M_{\star} = 1 \, M_{\odot}$, $L_{\star} = 2 \, L_{\odot}$, $M_{\rm{disk}} = 0.05\, M_{\star}$, $a_{\rm{dust}} = 1 \, \mu$m, $f_{\rm{dg}} = 10^{-2}$ and $\alpha_{\rm{hydro}} = 10^{-4}$)- what would be the derived accretion rates for either weaker or stronger magnetic field strengths, and show that our choice for the magnetic field strength leads to the highest value for the accretion rate as expected. 

\begin{figure*}
    \centering
    \includegraphics[width=\textwidth]{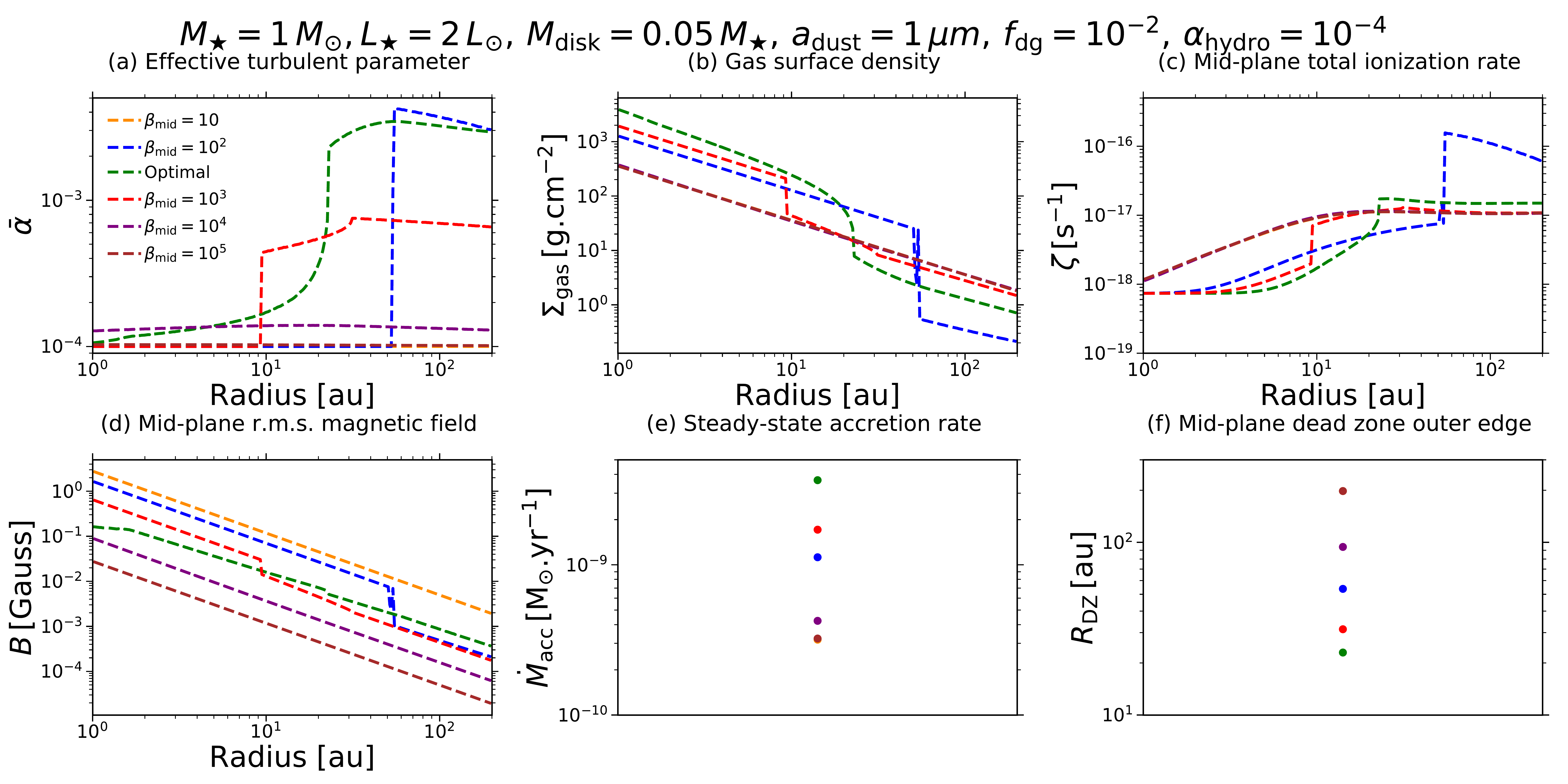}
     \caption{Effect of magnetic field strength on the equilibrium solution by exploring weaker and stronger scenarios compared to the optimal field. Various constant mid-plane $\beta$-plasma parameters $\beta_{\rm{mid}}$ have been considered such that $\beta_{\rm{mid}} \in \{10, 10^{2}, 10^{3}, 10^{4}, 10^{5}\}$. The "Optimal" case (green) corresponds to the field strength used for the fiducial model, i.e. the field strength that is required for the MRI to be maximally efficient permitted by Ohmic resistivity and ambipolar diffusion. The panels show the various steady-state profiles of some key quantities, for the model parameters $M_{\star} = 1 \, M_{\odot}$, $L_{\star} = 2 \, L_{\odot}$, $M_{\rm{disk}} = 0.05\, M_{\star}$, $a_{\rm{dust}} = 1 \, \mu$m, $f_{\rm{dg}} = 10^{-2}$ and $\alpha_{\rm{hydro}} = 10^{-4}$. Panels (a), (b), (c), (e) and (f) show the same key quantities as in Fig.~\ref{fig:effect_M_disk}. \textit{Panel (d)}: Mid-plane r.m.s. magnetic field strengths corresponding to the various $\beta_{\rm{mid}}$ as well as the optimal r.m.s. magnetic field strength (green).}
     \label{fig:impact_B_field}
\end{figure*}

To do so, we assume various constant mid-plane $\beta$-plasma parameters $\beta_{\rm{mid}}$ such that $\beta_{\rm{mid}} \in \{10, 10^{2}, 10^{3}, 10^{4}, 10^{5}\}$, calculate the corresponding r.m.s. mid-plane magnetic field strength $B = \sqrt{\frac{8 \pi \rho_{\rm{gas}} c_s^{2}}{\beta_{\rm{mid}}}}$, and assume that $B$ is vertically constant. The results can be seen in Fig.~\ref{fig:impact_B_field}.  

For $\beta_{\rm{mid}} = 10$, ambipolar diffusion utterly prohibits any MRI-driven turbulence at any locations in the disk because the magnetic field strength is too strong; leading to $\bar{\alpha} = \alpha_{\rm{hydro}} = 10^{-4}$ at every radii, hence to the lowest accretion rate. For $\beta_{\rm{mid}} = 10^{2}$, the MRI can only operate in regions beyond $54\,$au (there is no MRI-active layer sitting above the dead zone for $r\lesssim 54\,$au). Although the effective turbulence is higher in such regions for this scenario compared to the optimal one, the mid-plane dead zone outer edge is roughly located as twice as far, implying that the accretion rate is overall lower. For $\beta_{\rm{mid}} = 10^{3}$, the field strength is such that the MRI can only operate in the upper layers for $r \gtrsim 10\,$au, and from the mid-plane for $r \gtrsim 30\,$au. Since this field strength is overall lower than the one constrained for the optimal scenario in the MRI-active layer, the local MRI-driven turbulence is weaker (hence lower $\bar{\alpha}$). Finally, for $\beta_{\rm{mid}} = 10^{4}$ and $\beta_{\rm{mid}} = 10^{5}$, the MRI can only effectively operate in the surface layers (there is no longer a mid-plane MRI-active layer). On one hand, for $\beta_{\rm{mid}} = 10^4$, we find that the MRI could in theory develop at the mid-plane for $r \gtrsim 100 \,$au. In practice though, the induced mid-plane $\alpha_{\rm{MRI}} = 3.3 \times 10^{-5}$ (radially constant) is less than $\alpha_{\rm{hydro}} = 10^{-4}$. As a result, we consider such regions as dead by fiat (see Eq.~\eqref{eq:alpha MRI}). On the other hand, for $\beta_{\rm{mid}} = 10^5$, we find that the MRI cannot operate at the mid-plane because Eq.~\eqref{eq:ohmic Elsasser number criterion for MRI} is not fulfilled. Consequently, the mid-plane is weakly turbulent for both cases ($\bar{\alpha} \approx \alpha_{\rm{hydro}}$), since the MRI-active layer is located too high in the disk atmosphere to significantly contribute to the effective turbulence level.

\subsection{The Effect of Hydrodynamic Turbulent Parameter} \label{sect:effect hydrodynamics turbulence}

\begin{figure*}
\centering
\includegraphics[width=\textwidth]{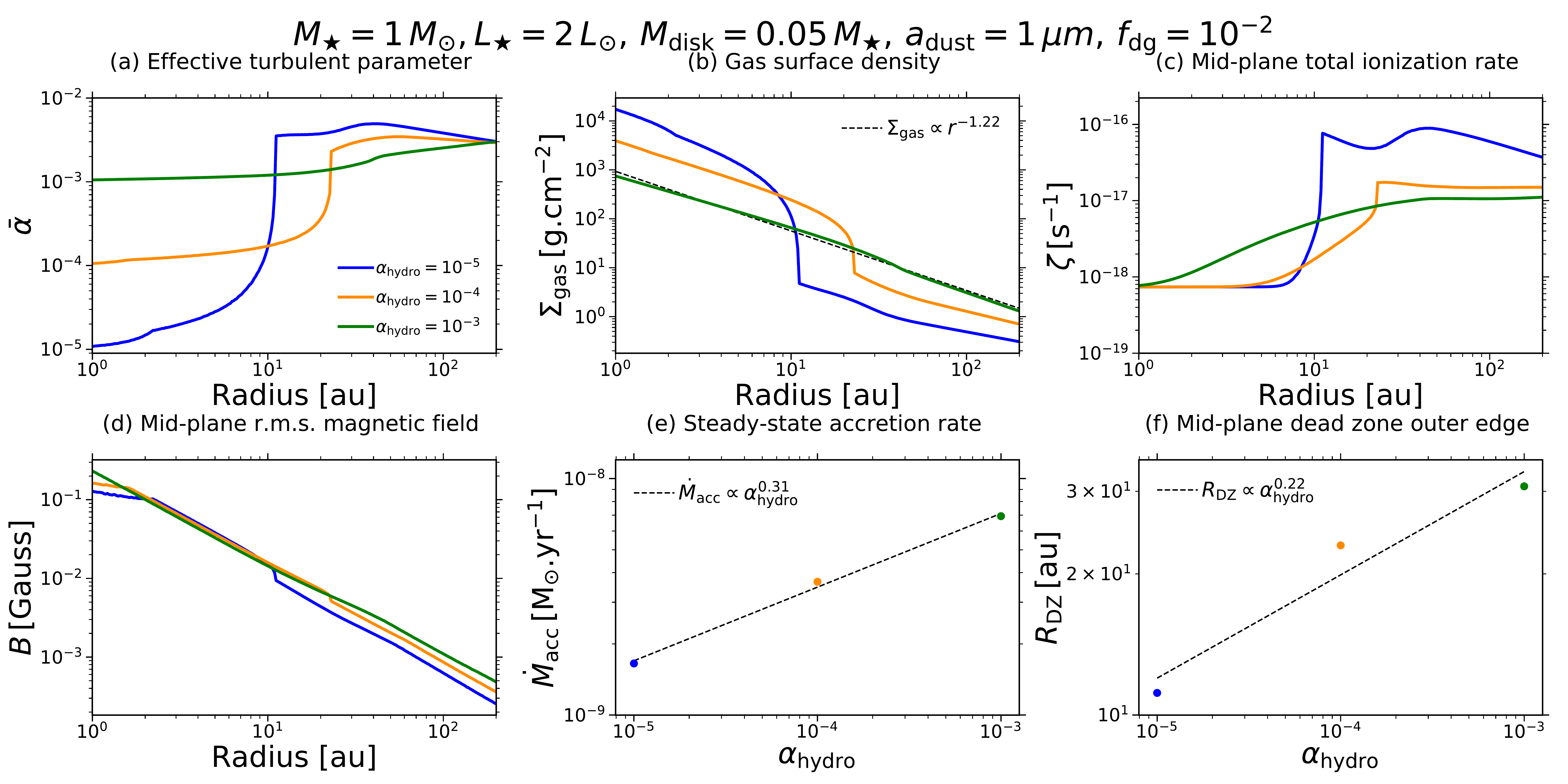}
 \caption{Effect of the hydrodynamic turbulent parameter on the equilibrium solution by varying $\alpha_{\rm{hydro}}$ from $10^{-5}$ to $10^{-3}$: the case $\alpha_{\rm{hydro}} = 10^{-5}$ is shown in blue; $\alpha_{\rm{hydro}} = 10^{-4 }$ in dark orange (fiducial model); and $\alpha_{\rm{hydro}} = 10^{-3}$ in green. The panels show the various steady-state profiles of some key quantities, for the model parameters $M_{\star} = 1 \, M_{\odot}$, $L_{\star} = 2 \, L_{\odot}$, $M_{\rm{disk}} = 0.05\, M_{\star}$, $a_{\rm{dust}} = 1 \, \mu$m and $f_{\rm{dg}} = 10^{-2}$. Panels (a)-(f) show the same key quantities as in Fig.~\ref{fig:effect_M_disk}. The black dashed line in panel (b) corresponds to a power-law fit of $\Sigma_{\rm{gas}}$ as a function of distance from the star $r$. The black dashed lines in panels (e) and (f) correspond to a power-law fit of $\dot{M}_{\rm{acc}}$ and $R_{\rm{DZ}}$ as a function of $\alpha_{\rm{hydro}}$, respectively.}
 \label{fig:effect_alpha_hydro}
\end{figure*}

Previous work on hydrodynamic instabilities suggest that $\alpha_{\rm{hydro}}$ could approximately range from $10^{-5}$ to $10^{-3}$ \citep[see e.g.,][]{10.1111/j.1365-8711.1998.01118.x,Klahr_2003, Nelson_2013, Raettig_2013, Lin_2015, Manger_2020, 2020ApJ...897..155F, 2021arXiv210601159B}. In our study, we choose the fiducial value $\alpha_{\rm{hydro}} = 10^{-4}$ motivated by the most recent studies on the VSI. We run a set of simulations to see how the variation in this parameter could potentially change the equilibrium solution. In practice, we take $\alpha_{\rm{hydro}}$ in the list $\{10^{-5},10^{-4},10^{-3}\}$. The results can be seen in Fig.~\ref{fig:effect_alpha_hydro}.

The stronger the hydrodynamic-driven turbulence is, the higher the overall effective turbulence level is; resulting in a shorter timescale for accretion to occur. Consequently, a larger $\alpha_{\rm{hydro}}$ value induces a larger accretion rate (Fig.~\ref{fig:effect_alpha_hydro}(e)). Overall though, we find a weak dependence of $\dot{M}_{\rm{acc}}$ on $\alpha_{\rm{hydro}}$ ($\dot{M}_{\rm{acc}} \propto \alpha^{0.31}_{\rm{hydro}}$). 

In this study, it is a requirement for any potential MRI-active regions to have a turbulence level that exceeds the turbulence set by hydroynamic stresses. For higher $\alpha_{\rm{hydro}}$, the MRI needs to generate stronger turbulence to fulfill this requirement. The dead zone maximal radial extent is thus larger (Fig.~\ref{fig:effect_alpha_hydro}(f)), since the gas is more ionized in the outer regions of the protoplanetary disk where this can happen. Similarly to what we saw in the previous paragraph, we find a weak dependence of $R_{\rm{DZ}}$ on $\alpha_{\rm{hydro}}$ ($R_{\rm{DZ}} \propto \alpha^{0.22}_{\rm{hydro}}$).  

Finally, the effective turbulence level is quenched to a few $10^{-3}$ by ambipolar diffusion (see fiducial model), implying that the profile for the effective turbulent parameter $\bar{\alpha}$ is the smoothest for $\alpha_{\rm{hydro}} = 10^{-3}$ (Fig.~\ref{fig:effect_alpha_hydro}(a)). Consequently, the corresponding steady-state gas surface density displays a much smoother profile $\Sigma_{\rm{gas}} \propto r^{-1.22}$ compared to the others obtained for lower $\alpha_{\rm{hydro}}$, close to the constant-$\alpha$ model solution $\Sigma_{\rm{gas}} \propto r^{-1}$ (Fig.~\ref{fig:effect_alpha_hydro}(b)). Knowing how the gas is distributed across the protoplanetary disk could thus provide hints on the overall turbulence set by hydrodynamic instabilities such as the VSI.

\subsection{The Effect of Temperature Model} \label{sect:effect temperature model}

\begin{figure*}
\centering
\includegraphics[width=\textwidth]{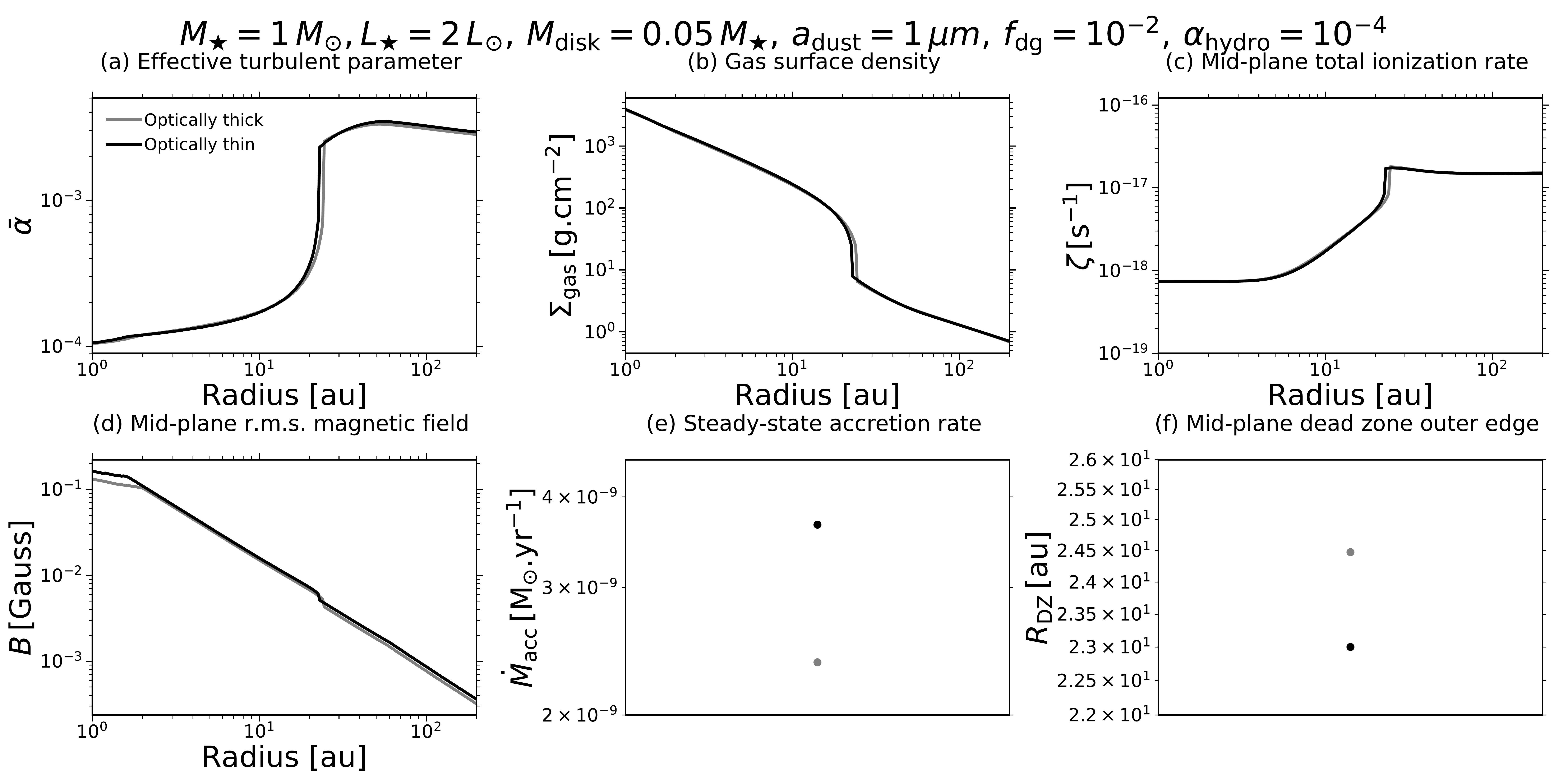}
 \caption{Effect of gas temperature on the equilibrium solution by varying the radial profile $T(r)$. The "Optically thick" case (gray) and the "Optically thin" case (black) correspond to gas temperature profiles that can be seen in Fig.~\ref{fig:Temperature_model}. The "Optically thin" case corresponds to the gas temperature profile used for the fiducial model described in Sect.~\ref{sect:results-fiducial model}. The panels show the various steady-state profiles of some key quantities, for the model parameters $M_{\star} = 1 \, M_{\odot}$, $L_{\star} = 2 \, L_{\odot}$, $M_{\rm{disk}} = 0.05\, M_{\star}$, $a_{\rm{dust}} = 1 \, \mu$m, $f_{\rm{dg}} = 10^{-2}$ and $\alpha_{\rm{hydro}} = 10^{-4}$. Panels (a)-(f) show the same key quantities as in Fig.~\ref{fig:effect_M_disk}.}
 \label{fig:effect_T}
\end{figure*}

There are a number of uncertainties in determining what the temperature really is in protoplanetary disks. To account for those, we run two simulations where the prescription of the gas temperature model assumed is different: optically thin vs. optically thick model. The results can be seen in Fig.~\ref{fig:effect_T}. 

\begin{figure}
\includegraphics[width=0.45\textwidth]{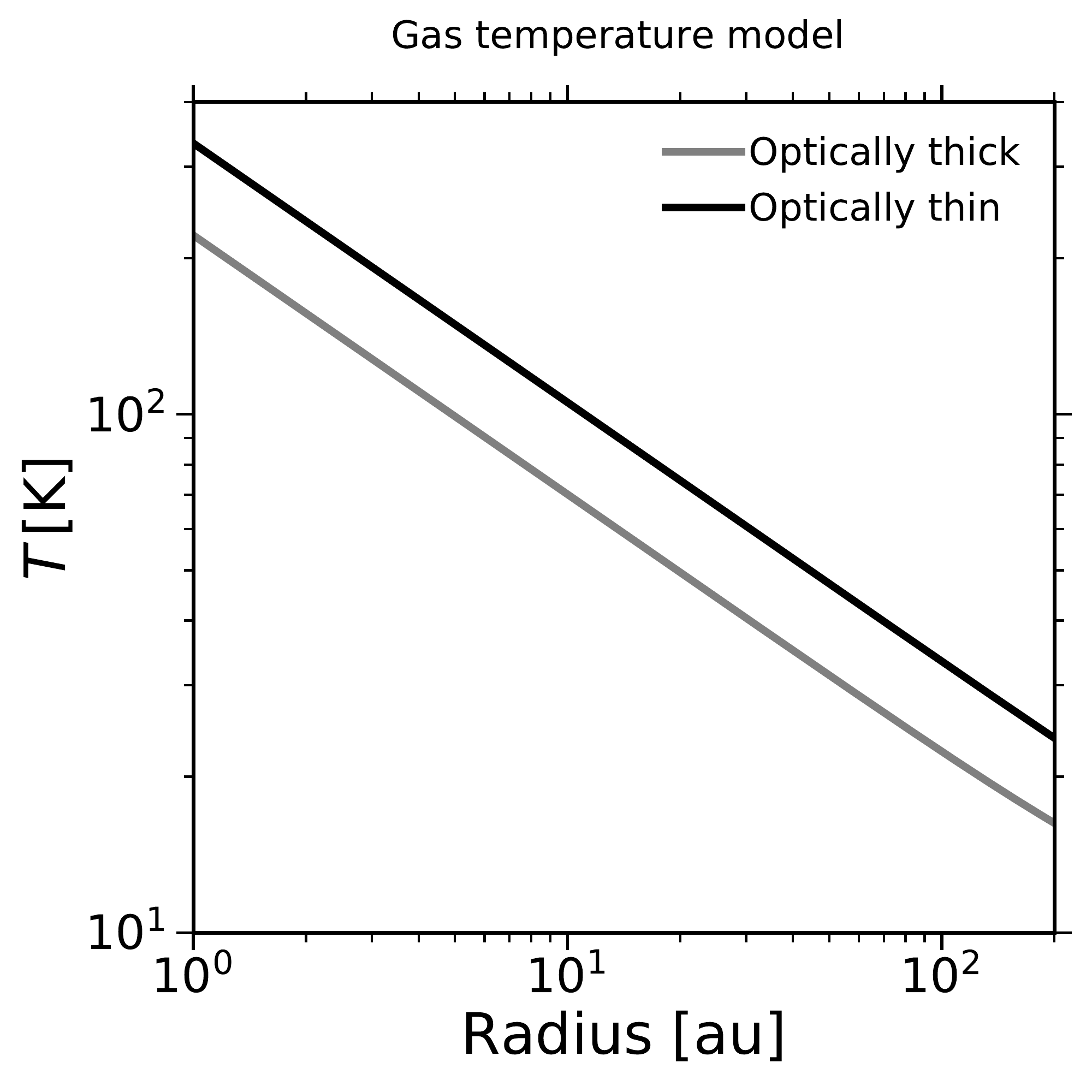}
 \caption{Gas temperature profiles for different model scenarios as a function of radius.}
 \label{fig:Temperature_model}
\end{figure}

The optically thin model refers to a radial gas temperature profile using Eq.~\eqref{eq:Temperature}, whereas the optically thick model refers to
\begin{equation*}
    T(r) = \left(\theta \frac{L_{\star}}{4 \pi \sigma_{\rm{SB}} r^{2}}+T_{\rm{bkg}}^{4}\right)^{\frac{1}{4}},
\end{equation*}

\noindent where $\theta = 0.05$ is the grazing angle under which the stellar light can illuminate the surface of the flared disk, $\sigma_{\rm{SB}}$ is the Stefan-Boltzmann constant, and $T_{\rm{bkg}} = 10 \,$K is the background gas temperature corresponding to the primordial temperature of the cloud prior to the collapse. This temperature profile is obtained by assuming that cooling from the two surfaces of the disk radiating as Planck functions can exactly compensate heating due to stellar irradiation. The two profiles for the gas temperature model can be found in Fig.~\ref{fig:Temperature_model}.

Overall, we notice that the choice made for the gas temperature model has very little impact on the equilibrium solution. This can be simply explained by noting that the optical thick model is only $1.5$ times colder than the thin model. Additionally, most of the disk parameters only weakly depend on the gas temperature profile $T$. For instance, the disk gas scale height scales as $H_{\rm{gas}} \propto \sqrt{T}$, the gas-phase recombination rate scales as $k \propto T^{-0.69}$, and the rate for the adsorption rate onto the grains scales as $\propto \sqrt{T}$. Consequently, the ionization chemistry is barely modified by the variation from optically thick to optically thin.

\subsection{Combined effect of stellar properties and total disk gas mass} \label{sect:combined effect of stellar properties and total disk gas mass}

\begin{figure*}
\centering
\includegraphics[width=\textwidth]{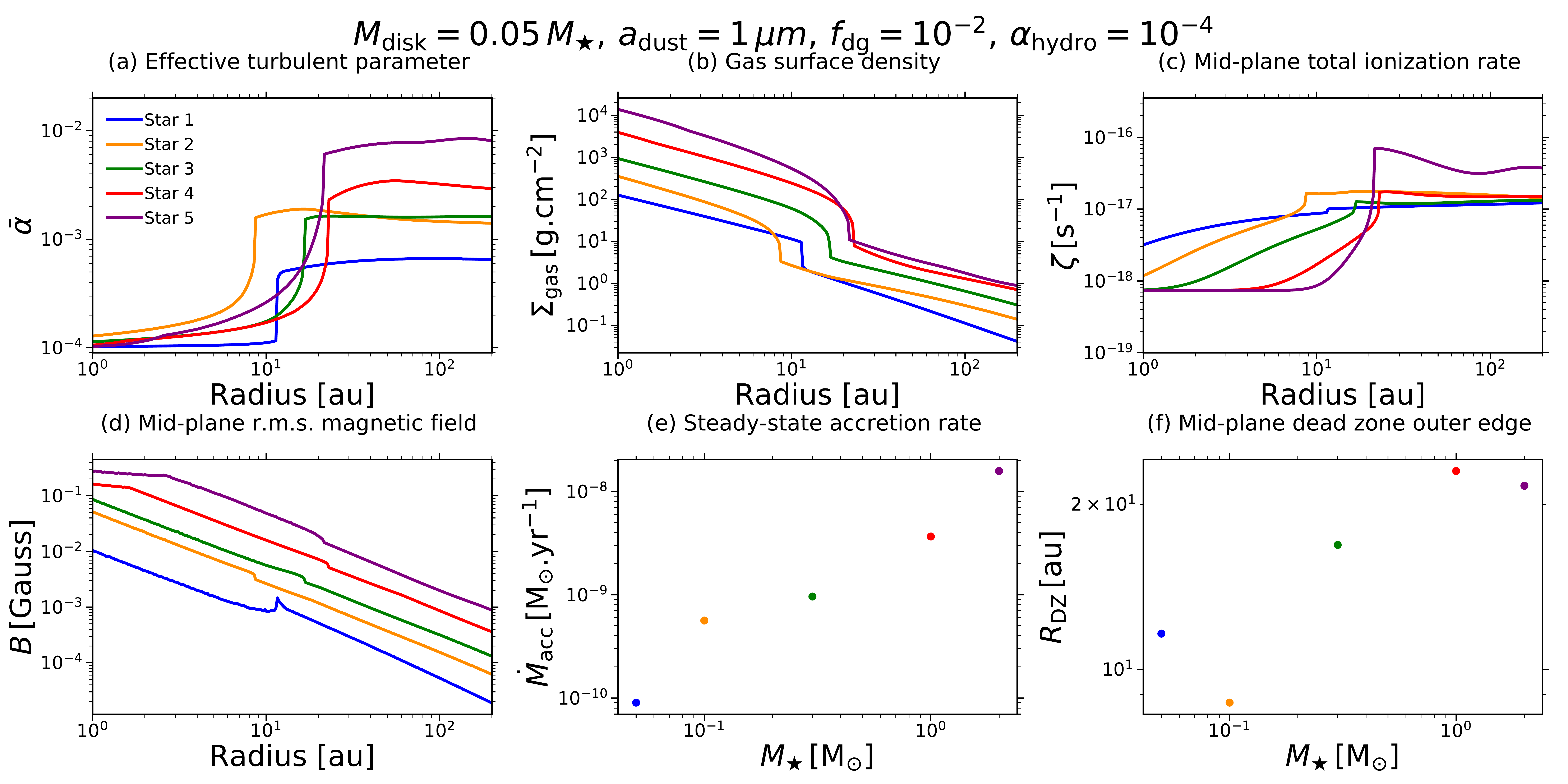}
 \caption{Combined effect of stellar properties and total disk gas mass on the equilibrium solution by varying the parameters $M_{\star}$ and $L_{\star}$ altogether, all the way from accreting brown dwarfs to herbig stars. The total disk gas mass depends on stellar mass and is such that $M_{\rm{disk}} = 0.05 \, M_{\star}$. The "Star 1" case (blue) corresponds to a star of mass $M_{\star} = 0.05 \, M_{\odot}$ and bolometric luminosity $L_{\star} = 2 \times 10^{-3} \, L_{\odot}$. The "Star 2" case (dark orange) corresponds to a star of mass $M_{\star} = 0.1 \, M_{\odot}$ and bolometric luminosity $L_{\star} = 8 \times 10^{-2} \, L_{\odot}$. The "Star 3" case (green) corresponds to a star of mass $M_{\star} = 0.3 \, M_{\odot}$ and bolometric luminosity $L_{\star} = 0.2\, L_{\odot}$. The "Star 4" case (red) corresponds to the fiducial model where the stellar mass is $M_{\star} = 1 \, M_{\odot}$ and the bolometric luminosity is $L_{\star} = 2 \, L_{\odot}$. The "Star 5" case (purple) corresponds to a star of mass $M_{\star} = 2 \, M_{\odot}$ and bolometric luminosity $L_{\star} = 20 \, L_{\odot}$. The panels show the various steady-state profiles of some key quantities, for the model parameters $M_{\rm{disk}} = 0.05\, M_{\star}$, $a_{\rm{dust}} = 1 \, \mu$m, $f_{\rm{dg}} = 10^{-2}$ and $\alpha_{\rm{hydro}} = 10^{-4}$. Panels (a)-(f) show the same key quantities as in Fig.~\ref{fig:effect_M_disk}.}
 \label{fig:effect_M_star_and_M_disk}
\end{figure*}

In Sect.~\ref{sect:effect stellar properties}, we investigated the effect of stellar properties alone by setting the total disk gas mass to $M_{\rm{disk}} = 0.01 \, M_{\odot}$ (stellar-independent choice). For each star model in Table~\ref{tab:Table3}, we now run a simulation with a more realistic choice for the total disk gas mass relatively to the stellar mass ($M_{\rm{disk}} = 0.05 \, M_{\star}$, stellar-dependent choice). The results can be seen in Fig.~\ref{fig:effect_M_star_and_M_disk}. 

The trend seen in sect.~\ref{sect:effect stellar properties} where high-mass and more luminous stars generate higher accretion rates still holds, which is expected because a higher total disk gas mass has a higher gas content that can be accreted onto the central star.

The dead zone maximal radial extent now displays a different trend compared to what we saw for the effect of stellar properties alone (Sect.~\ref{sect:effect stellar properties}). Indeed, a high-mass and more luminous star has a more massive disk, hence higher gas surface densities. The effect of total disk gas mass can thus compensate for the effect of stellar properties by significantly reducing the overall ionization level, resulting in a more extended dead zone. Nonetheless, Fig.~\ref{fig:effect_M_star_and_M_disk}(f) displays the interesting feature that the dead zone maximal radial extent ($R_{\rm{DZ}}$) saturates for star 4 model. This saturation can be attributed to stellar X-rays that dominates the mid-plane ionization process for stars more massive and luminous that star 4 model, as showed by Fig.~\ref{fig:effect_M_star_and_M_disk}(c). The high stellar X-ray luminosity ($L_{\rm{XR}}$) produced by such stars regulates the radial extent of the dead zone. Consequently, it seems that dead zones cannot become indefinitely large for high-mass and more luminous stars. To validate this idea, we run six additional simulations for each star model in Table~\ref{tab:Table3}, varying the total disk gas mass from $M_{\rm{disk}} = 0.001 \, M_{\star}$ to $M_{\rm{disk}} = 0.10 \, M_{\star}$. For all the total disk gas masses considered, we confirm that the dead zone maximal radial extent saturates for stars more massive and luminous than star 4 model. Similarly, we find that the dead zone maximal radial extent cannot become indefinitely small for stars less massive and luminous than star 2 model due their too low stellar X-ray luminosity (Fig.~\ref{fig:effect_M_star_and_M_disk}(f) shows that the dead zone maximal radial extent is higher for star 1 than star 2 model).

\section{Ambipolar diffusion and the Hall effect}

In this appendix, we show that the strong-coupling limit holds in our disk models, which is a requirement for the criterion employed to describe the effect of ambipolar diffusion on the MRI (Appendix~\ref{sect:strong-coupling limit}). Furthermore, we discuss the potential impact that the Hall effect could have on our steady-state accretion solutions (Appendix~\ref{sect:the hall effect}).

\subsection{Ambipolar diffusion in the Strong-coupling limit} \label{sect:strong-coupling limit}

In our study, we employ the \citet{2011ApJ...736..144B}'s criterion to account for the effect of ambipolar diffusion on the MRI (see Eqs.~\eqref{eq:magnetic field criterion for MRI} and \eqref{eq:beta min}). The underlying idea behind this criterion is that the disk is in the strong-coupling limit \citep[][]{1991pagd.book.....S}, i.e. the single fluid regime of neutrals. In this limit, local ionization equilibrium is a good enough approximation, and the magnetic diffusion coefficients can be directly evaluated from the total ionization rate and local thermodynamic quantities such as density and temperature. It requires that: (1) the ion inertia is negligible compared to the inertia of the neutrals, so that the ion density is entirely controlled by ionization equilibrium with the neutrals; (2) the ionization equilibrium is achieved on a timescale $t_{\rm{rcb}}$ shorter than the dynamical timescale $t_{\rm{dyn}} = \frac{2 \pi }{\Omega_K}$, implying that the ions are continuously created and destroyed on a timescale that is shorter than the timescale for the MRI to grow.

It is clear that the ionization fraction is extremely small in our protoplanetary disk models, and even so in the uppermost layers (see e.g., Fig.~\ref{fig:disk_chemistry}(b)). Hence, the inertia of charged particles is negligible compared to the inertia of neutrals, justifying the first requirement of the strong-coupling limit.

\begin{figure}
\includegraphics[width=0.50\textwidth]{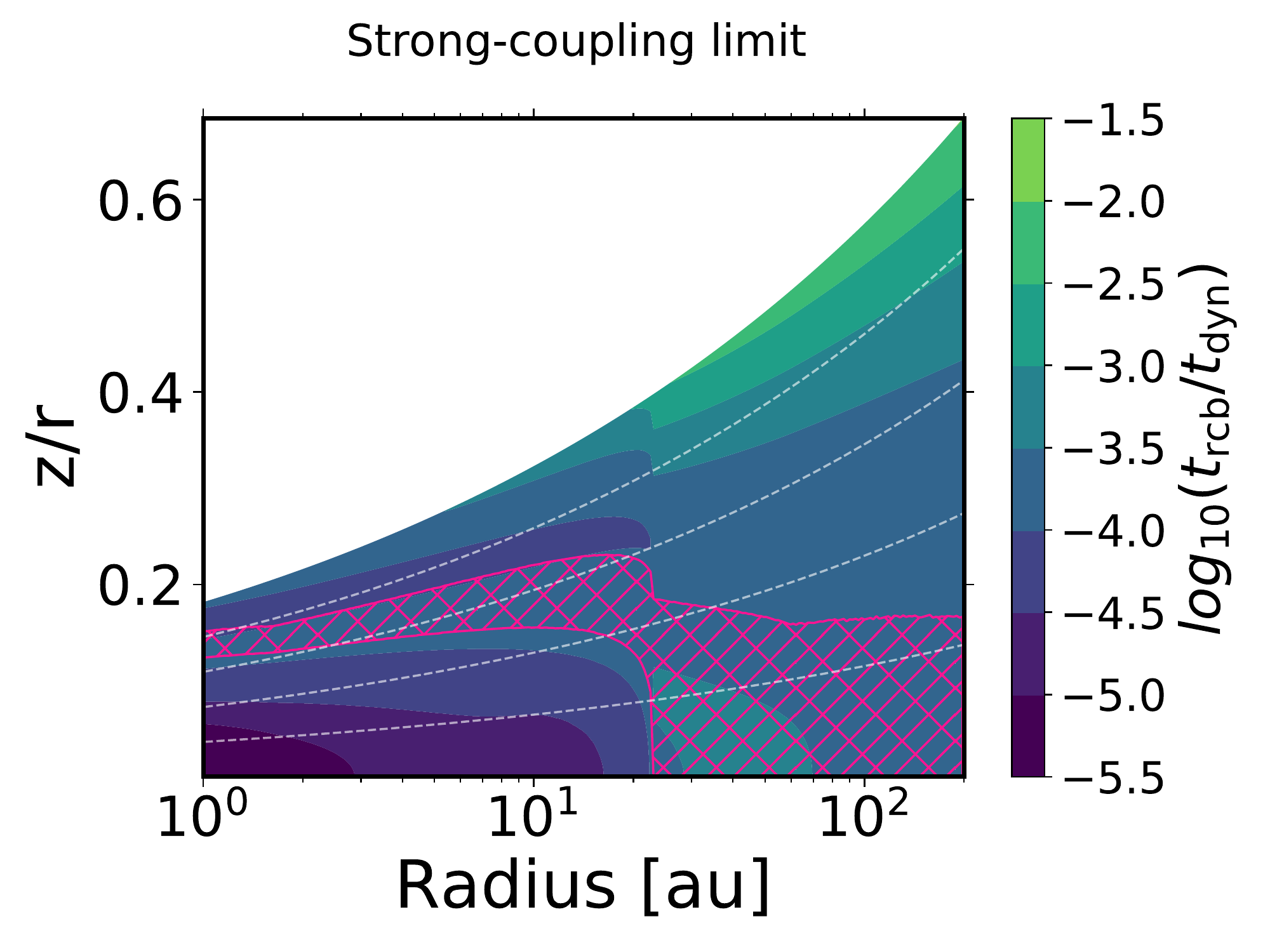}
 \caption{Ratio of the recombination timescale ($t_{\rm{rcb}}$) to the dynamical timescale ($t_{\rm{dyn}}$) as a function of location in the disk, for the fiducial model described in Sect.~\ref{sect:results-fiducial model}. The pink hatched area corresponds to the MRI-active layer. The white dashed lines correspond to the surfaces $z = 1 \, H_{\rm{gas}}$, $z = 2 \, H_{\rm{gas}}$, $z = 3 \, H_{\rm{gas}}$ and $z = 4 \, H_{\rm{gas}}$; from bottom to top, respectively.}
 \label{fig:strong_coupling_limit}
\end{figure}

Since we directly solve for the equilibrium ionization state, we do not have access to the recombination timescale $t_{\rm{rcb}}$. Nonetheless, we can estimate it as follows: Given the first term of Eqs.~\eqref{eq:dnidt} and \eqref{eq:dnedt}, the ionization timescale for ions and free electrons can be respectively written as $t_i = n_i/\left(\zeta n_{\rm{gas}}\right)$ and $t_e = n_e/\left(\zeta n_{\rm{gas}}\right)$. Since $n_i \geq n_e$ (free electrons collide more frequently with grains), $t_i \geq t_e$; hence, the overall relaxation timescale of the gas-phase being given by $t_i$. Assuming local ionization-recombination equilibrium (as it is the case in our semi-analytical chemical model), the ionization timescale ($t_i$) equals the recombination timescale ($t_{\rm{rcb}}$). Therefore, $t_{\rm{rcb}} \approx n_i/\left(\zeta n_{\rm{gas}}\right) \approx 1/\left(s_i u_i\sigma_{\rm dust}n_{\rm dust}P_i + k n_e\right)$; where the second equality is found by equating Eq.~\eqref{eq:dnidt} to zero. Figure~\ref{fig:strong_coupling_limit} shows, for the fiducial model, that the recombination timescale is shorter than the dynamical timescale ($t_{\rm{rcb}}/t_{\rm{dyn}} < 1$) everywhere. As a result, the second requirement of the strong-coupling limit is also justified. 

We can thus conclude that the strong-coupling limit holds in our fiducial disk model, justifying our use of the \citet{2011ApJ...736..144B}'s criterion to describe the effect of ambipolar diffusion on the MRI. We verified that it holds for all the models shown in this present work.

\subsection{The Hall effect} \label{sect:the hall effect}

\begin{figure}
\includegraphics[width=0.50\textwidth]{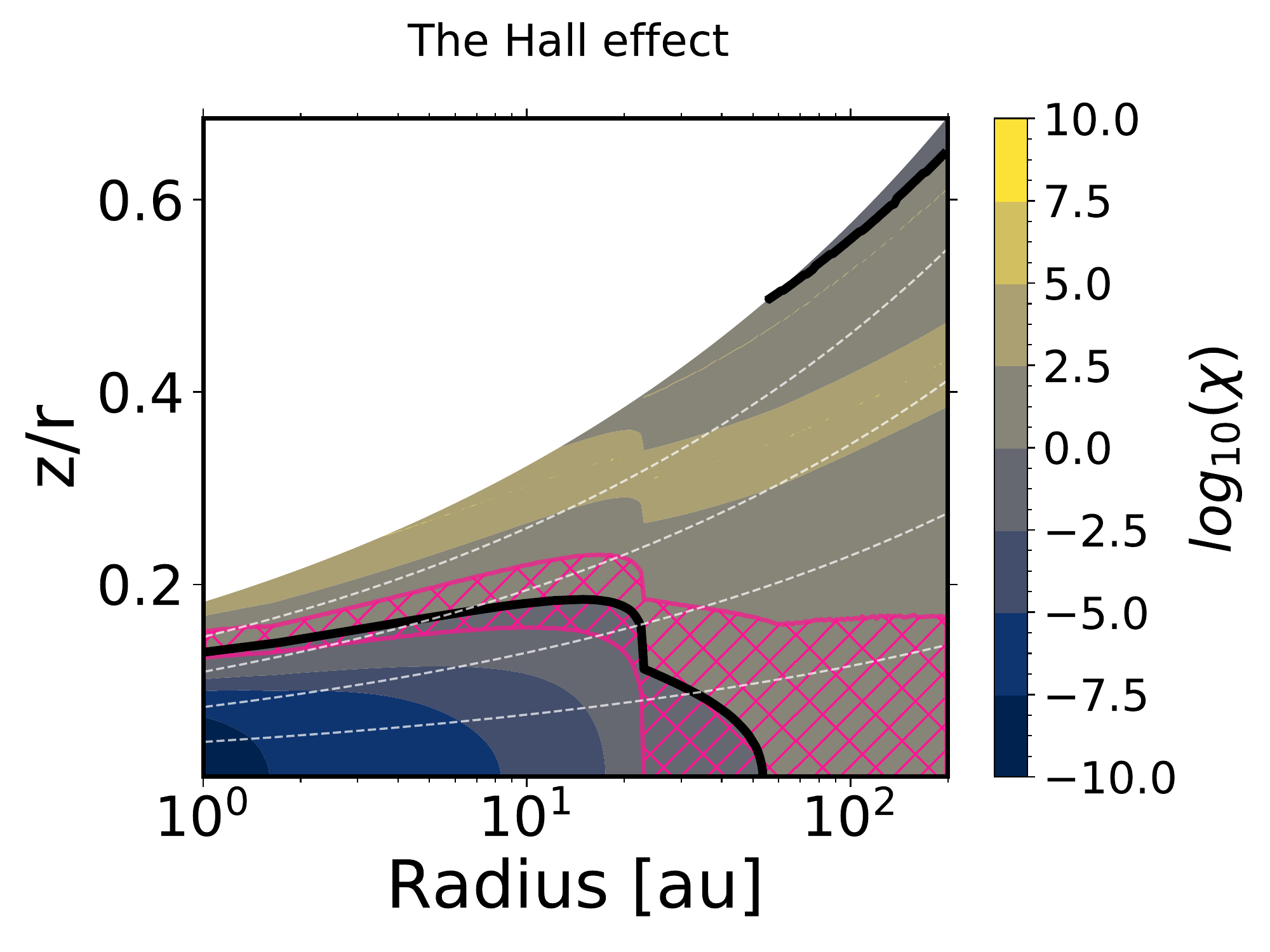}
 \caption{Steady-state Hall Elsasser number $\chi$ as a function of location in the disk, for the fiducial model described in Sect.~\ref{sect:results-fiducial model}. The pink hatched area corresponds to the MRI-active layer. The black solid lines indicates where $\chi = 1$. Everywhere between those lines, $\chi > 1$. The white dashed lines correspond to the surfaces $z = 1 \, H_{\rm{gas}}$, $z = 2 \, H_{\rm{gas}}$, $z = 3 \, H_{\rm{gas}}$ and $z = 4 \, H_{\rm{gas}}$; from bottom to top, respectively.}
 \label{fig:hall_effect}
\end{figure}

The impact of the Hall effect on the MRI is non-trivial: in the presence of a net background vertical magnetic field threading the protoplanetary disk, the non-dissipative character of Hall diffusion implies that the Hall effect depends on whether the field is aligned or anti-aligned with the rotation axis of the disk \citep[see e.g.,][]{1999MNRAS.307..849W, 2001ApJ...552..235B, 2002ApJ...570..314S, 2012MNRAS.422.2737W,2016ApJ...819...68X}.
In Sect.~\ref{sect:behavior of the different accretion layers}, we saw that the Hall effect dominates the non-ideal MHD terms in most of the dead zone, as well as in the inner regions of the MRI-active layer (Fig.~\ref{fig:accretion_layers}(b)). Although we have ignored the Hall effect in our study, we  can investigate a posteriori its potential impact on our results by plotting the Hall Elsasser number $\chi \equiv \frac{\rm{v}_{A}^{2}}{|\eta_{H}|\Omega_K}$ (see Fig.~\ref{fig:hall_effect}), where $\eta_H$ is the Hall magnetic diffusivity \citep[see e.g.,][their Eq.~(30)]{2007Ap&SS.311...35W}.  

If the magnetic field is anti-aligned with the rotation axis of the disk, the Hall effect could substantially reduce the amount of MRI-driven turbulent transport in the MRI-active layer where it dominates the non-ideal MHD terms \citep[see e.g.,][]{2013MNRAS.434.2295K, 2014ApJ...791..137B, 2015ApJ...798...84B,2015MNRAS.454.1117S}. In the fiducial model, the MRI at the mid-plane would thus only develop from $\approx 50 \,$au instead of $\approx 23 \,$au, implying that a significant part of the mid-plane would be laminar driven by non-MRI stresses. A laminar mid-plane would have significant impact on the dust dynamics and the growth process, since lower turbulence implies effective coagulation and less effective fragmentation. It would suggest that the dust at the mid-plane could grow to larger sizes; where the typical grain size reached at a given location in the disk would depend on how much the MRI is suppressed by the Hall effect, as well as how strong the turbulence from non-MRI stresses is.

Conversely, if the magnetic field is aligned with the rotation axis of the protoplanetary disk, the Hall shear instability (HSI) could reactivate the dead zone by producing large-scale and ordered magnetic fields transporting angular momentum radially with little turbulent motion through laminar viscous stresses \citep[see e.g.,][]{2008MNRAS.385.1494K, 2014A&A...566A..56L,2015ApJ...798...84B}. As such, the steady-state pressure maximum at the dead zone outer edge seen in Sect.~\ref{sect:with dust trapping} could thus be removed.  

\section{Additional plots} \label{appendix:Additional plots}

In this section, we present additional plots for completeness.
 
\begin{figure*}
\centering
\includegraphics[width=\textwidth]{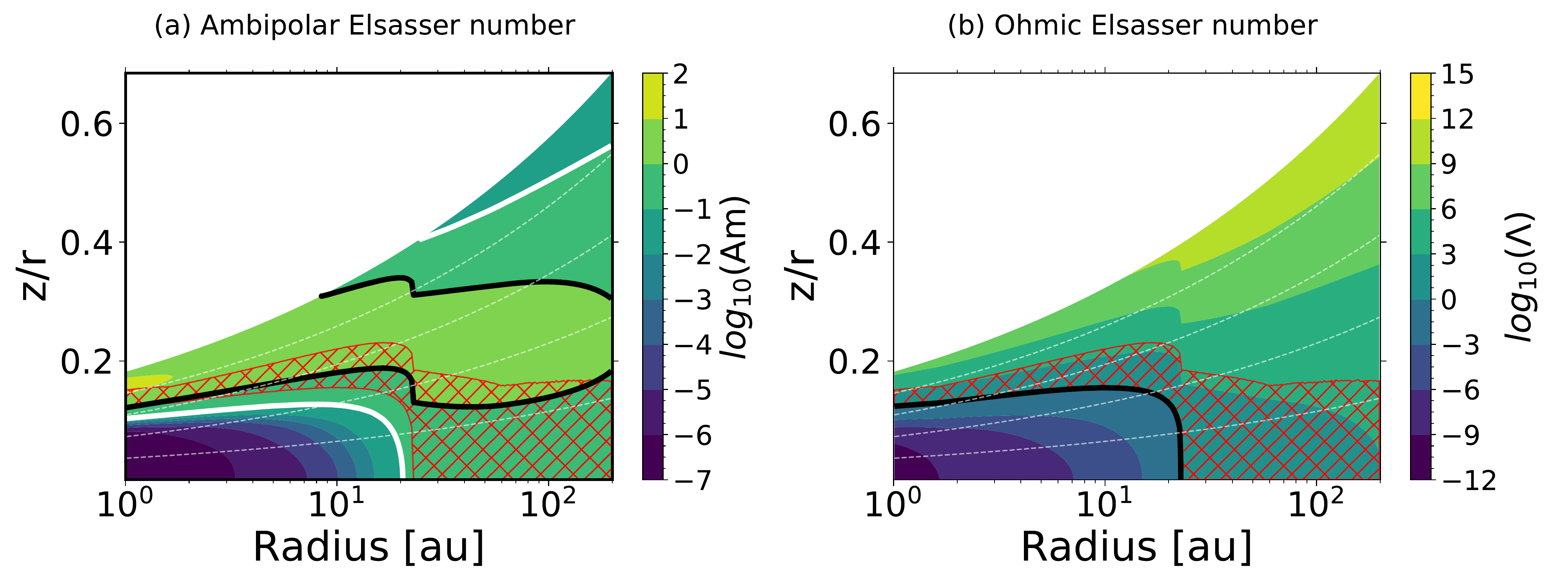}
 \caption{Steady-state ambipolar and Ohmic Elsasser numbers as a function of location in the disk, for the fiducial model described in Sect.~\ref{sect:results-fiducial model}. In both panels, the red hatched area corresponds to the MRI-active layer. Additionally, the white dashed lines correspond to the surfaces $z = 1 \, H_{\rm{gas}}$, $z = 2 \, H_{\rm{gas}}$, $z = 3 \, H_{\rm{gas}}$ and $z = 4 \, H_{\rm{gas}}$; from bottom to top, respectively. \textit{Panel (a)}: Ambipolar Elsasser number $\rm{Am}$. The region within the white solid lines indicates where $\rm{Am} \geq 0.1$. The region within the black solid contour indicates where $\rm{Am} \geq 1$. \textit{Panel (b)}: Ohmic Elsasser number $\Lambda$. The black solid line indicates where $\Lambda = 1$.}
 \label{fig:Elsasser_numbers}
\end{figure*}

\begin{figure}
\includegraphics[width=0.45\textwidth]{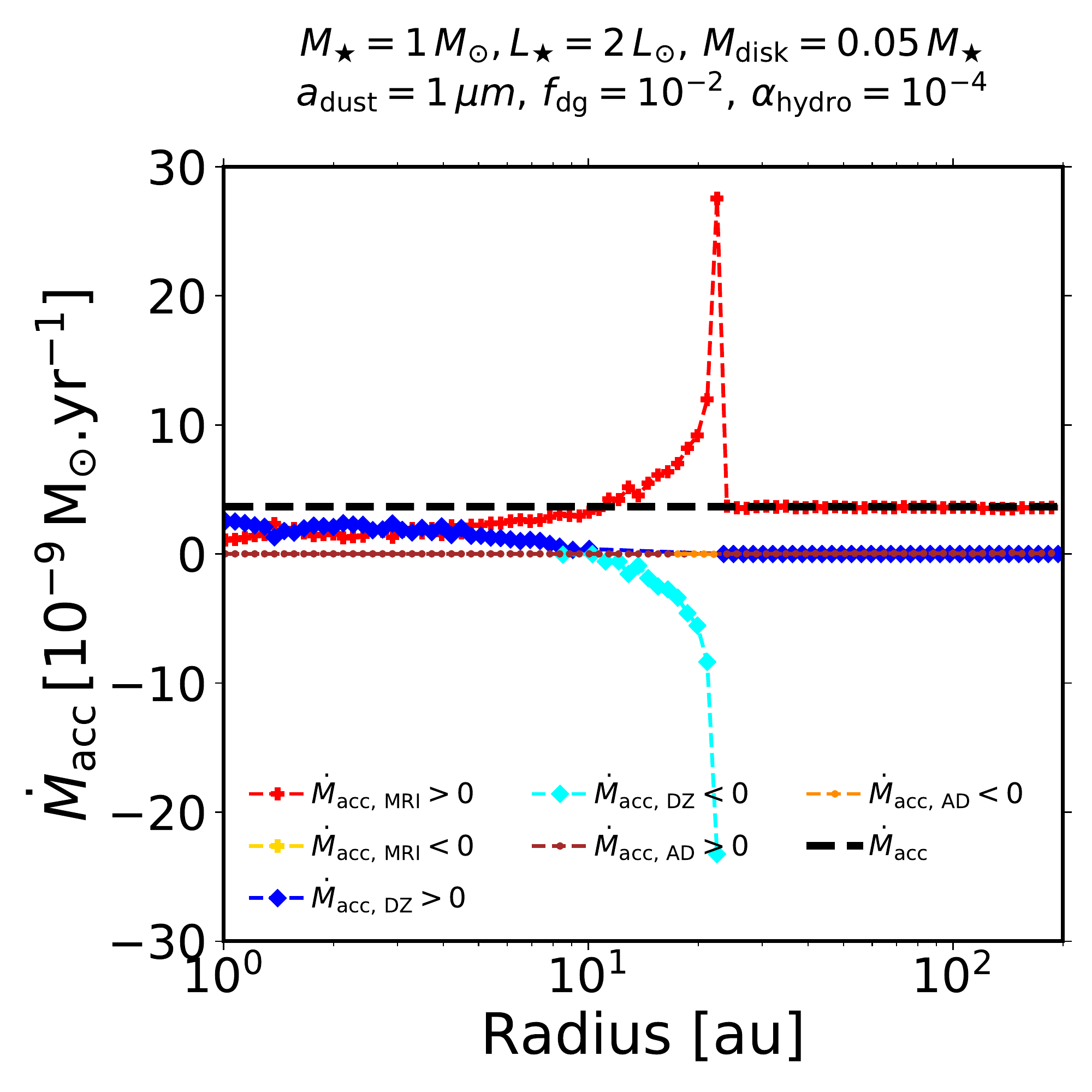}
 \caption{Accretion rates within the individual layers as a function of radius, for the fiducial model described in Sect.~\ref{sect:results-fiducial model}. The overall steady-state inward accretion rate $\dot{M}_{\rm{acc}}$ is defined as $\dot{M}_{\rm{acc}} = \dot{M}_{\rm{acc,\,MRI}} + \dot{M}_{\rm{acc,\,AD}} +\dot{M}_{\rm{acc,\,DZ}}$, where $\dot{M}_{\rm{acc,\,MRI}}$, $\dot{M}_{\rm{acc,\,AD}}$ and $\dot{M}_{\rm{acc,\,DZ}}$ correspond to the radially variable accretion rates within the MRI-active layer, zombie zone and dead zone, respectively. They have been computed using the formula provided in Appendix~B of \citet{2018ApJ...861..144M}. The black dashed line corresponds to $\dot{M}_{\rm{acc}} \approx 3.7 \times 10^{-9} \, M_{\odot}.\rm{yr}^{-1}$, obtained for the fiducial protoplanetary disk model.} 
 \label{fig:accretion_rates_individual_layers}
\end{figure}

\begin{figure}
\includegraphics[width=0.45\textwidth]{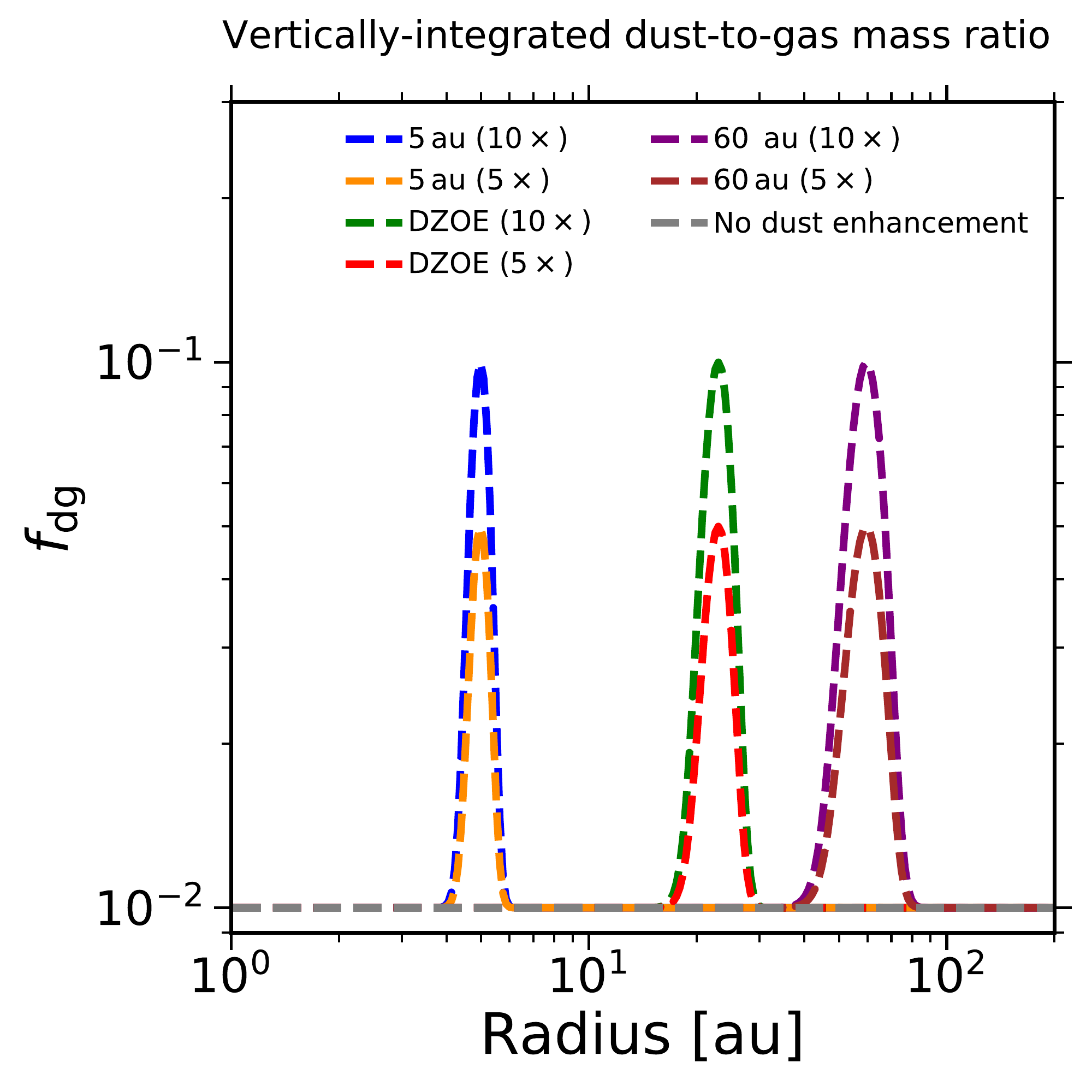}
 \caption{Vertically-integrated dust-to-gas mass ratio profiles for different dust trapping scenarios as function of radius.}
 \label{fig:f_dg}
\end{figure}

\end{appendix}

\end{document}